\documentclass[english]{emulateapj}

\usepackage{textcomp}
\usepackage{amsmath,amssymb,amsfonts}
\usepackage{txfonts}

\usepackage{sidecap}
\usepackage{natbib}

\usepackage{booktabs} % table formatting
\usepackage[normalem]{ulem} % for sout
\usepackage[table]{xcolor} % for font colours
\usepackage{graphicx}	% Including figure files
\usepackage{subfigure}

\usepackage{aas_macros} % for reference shortcuts

\usepackage{xspace}
\newcommand{\rate}[1]{%
  \hypertarget{rate:#1}{\ensuremath{k_{#1}}}\label{rate:#1}\xspace%
}
\newcommand{\rateref}[1]{%
  \hyperlink{rate:#1}{\ensuremath{k_{#1}}}\xspace%
}

\newcommand{\rateh}[1]{%
  \hypertarget{rate:#1}{\ensuremath{#1}}\label{rate:#1}\xspace%
}
\newcommand{\raterefh}[1]{%
  \hyperlink{rate:#1}{\ensuremath{#1}}\xspace%
}

\newcommand{\ratephoton}[1]{%
  \hypertarget{ratephoton:#1}{\ensuremath{\Gamma_\mathrm{#1}}}\label{ratephoton:#1}\xspace%
}
\newcommand{\raterefphoton}[1]{%
  \hyperlink{ratephoton:#1}{\ensuremath{\Gamma_\mathrm{#1}}}\xspace%
}

\newcommand{\abundance}[1]{%
  \hypertarget{abundance:#1}{\ensuremath{\mathrm{A}_\mathrm{#1}}}\label{abundance:#1}\xspace%
}
\newcommand{\abundanceref}[1]{%
  \hyperlink{abundance:#1}{\ensuremath{\mathrm{A}_\mathrm{#1}}}\xspace%
}

\newcommand{\zetat}[1]{%
  \hypertarget{zetat:#1}{\ensuremath{\zeta_\mathrm{t}(\mathrm{#1})}}\label{zetat:#1}\xspace%
}
\newcommand{\zetatref}[1]{%
  \hyperlink{zetat:#1}{\ensuremath{\zeta_\mathrm{t}(\mathrm{#1})}}\xspace%
}

\usepackage{hyperref}

\begin{document}

% \title{\bf{Out-of-equilibrium $\mathrm{OH^+}$, $\mathrm{H_2O^+}$ and $\mathrm{H_3^+}$ distributions inside turbulent interstellar media}}
% 
\title{\bf{Multiphase turbulence \\
as the origin of $\mathrm{OH^+}$, $\mathrm{H_2O^+}$ and H$_3^+$ column density scatter in the local ISM}}

\author{Uri Malamud{$^1$}, Shmuel Bialy{$^1$}, Benjamin Godard{$^2$,$^3$}, David Neufeld{$^4$}, Blakesley Burkhart{$^5$,$^6$}}

\affil{{$^1$}Department of Physics, Technion Israel Institute of Technology, Technion City, Haifa, 3200003, Israel}
\affil{{$^2$}Observatoire de Paris, Université PSL, Sorbonne Université, CNRS, 75014 Paris, France}
\affil{{$^3$}Laboratoire de Physique de l’École Normale Supérieure, ENS, Université PSL, CNRS, Sorbonne Université, 75005 Paris, France}
\affil{{$^4$}Department of Physics \& Astronomy, The Johns Hopkins University, 3400 North Charles Street, Baltimore, MD 21218, USA}
\affil{{$^5$}Department of Physics \& Astronomy, Rutgers, The State University of New Jersey, Piscataway, NJ, USA}
\affil{{$^6$}Center for Computational Astrophysics, Flatiron Institute, Simons Foundation, New York, NY, USA}

\email{urimala@physics.technion.ac.il}

\begin{abstract}

Observations of the reactive ions $\mathrm{OH^+}$, $\mathrm{H_2O^+}$, and H$_3^+$ in the Galactic interstellar medium reveal large sight-line-to-sight-line scatter in their column densities, commonly interpreted as evidence for substantial variations in the cosmic-ray ionization rate (CRIR). We revisit this interpretation using high-resolution three-dimensional
magneto-hydrodynamic simulations of the multiphase ISM with
time-dependent chemistry for H, H$_2$, H$^+$ and electrons, building on the fiducial model of \citet{GodardEtAl-2023}. We find that a single CRIR of $\simeq 2\times10^{-16}\,\mathrm{s^{-1}}$, together with standard Galactic-scale parameters, naturally produces broad column-density distributions for all three tracers in good agreement
with the observed medians and percentile widths, with no fine tuning. Reaching this match requires that the post-processing of $\mathrm{OH^+}$, $\mathrm{H_2O^+}$, and H$_3^+$ retain the time-dependent H$_2$ field generated by the turbulent flow rather than assume chemical equilibrium: turbulence drives long-lived H$_2$ enhancements in the unstable neutral medium where $\mathrm{OH^+}$ and $\mathrm{H_2O^+}$ predominantly reside, and an equilibrium treatment under-predicts their columns substantially. H$_3^+$, which receives most of its column from denser CNM gas closer to equilibrium, is much less affected. Our results caution against interpreting sight-line-to-sight-line scatter as
direct evidence for large CRIR fluctuations, and motivate a shift from independent 1D equilibrium analyses toward 3D dynamical frameworks when inferring ionization conditions in the ISM.

\end{abstract}

\section{Introduction}\label{S:Intro}
A central goal in interstellar medium (ISM) physics is to understand what sets the thermal, chemical, and dynamical state of gas that ultimately forms stars and regulates galaxy evolution. Two environmental drivers are especially influential: the far-ultraviolet (FUV) radiation field, which shapes the atomic-to-molecular transition through photodissociation and photoelectric heating \citep{HollenbachTielens-1999, LePetitEtAl-2006, BialySternberg-2015b, BialySternberg-2016}, and the cosmic-ray ionization rate (CRIR; hereafter $\zetatref{H}$, refers to the total ionization rate per H atom including secondary ionizations), which regulates ionization fractions, ion-neutral coupling to magnetic fields, and initiates the ion-molecule chemistry controlling the abundances of many widely observed tracers \citep{WolfireEtAl-2003, IndrioloMcCall-2012}.

A standard empirical approach to constrain the CRIR has been to observe reactive molecular ions, particularly $\mathrm{OH^+}$, $\mathrm{H_2O^+}$, and H$_3^+$, and interpret their column densities using analytic chemical frameworks or one-dimensional photo-/cosmic-ray-dissociation region (PDR+CRDR) models. In this traditional approach, each sight line is treated as an independent cloud, modeled in isolation with its own dedicated set of parameters. The result is not a single, coherent description of the ISM, but rather a collection of separate cloud models, each yielding its own inferred CRIR. Observational studies have reported considerable sight-line-to-sight-line variation in these inferred values, not only between the Galactic center and disk \citep{LePetitEtAl-2016, NeufeldWolfire-2017}, where large-scale gradients are theoretically expected, but even within the local Galactic neighborhood \citep{IndrioloEtAl-2007, IndrioloMcCall-2012, IndrioloEtAl-2015, ObolentsevaEtAl-2024}. The traditional interpretation has been that this large scatter in molecular abundances reflects genuine spatial variation in the CRIR across the Galaxy, implying significant differences in cosmic-ray source distributions or diffusion properties even on local scales.

However, this interpretation rests on a critical omission: it neglects the fact that the real ISM is multiphase and highly turbulent. Turbulence in a multiphase medium drives energy and mass exchanges between the two stable states of the ISM, and maintains a large fraction of the gas in the unstable state at intermediate temperatures \citep{AuditHennebelle-2005,HennebelleAudit-2007,HennebelleFalgarone-2012}. Fluctuations in density and temperature in turn produce fluctuations in molecular abundances, even when the CRIR is spatially uniform \citep{Imara_2016,BialyEtAl-2017}. In addition, turbulent mixing can transport molecular gas across CNM/WNM interface faster than the local chemistry re-equilibrates, sustaining molecular abundances out of chemical equilibrium \citep{LesaffreEtAl-2007}. This physical insight motivates a new paradigm: rather than modeling each sight line independently, one can describe a large number of observables in the ISM with a self-consistent magnetohydrodynamic (MHD) framework, and treat observations as statistical samplings of sight lines through this one inhomogeneous, multiphase, dynamic medium. This shift is enabled both conceptually and practically: conceptually, by the growing understanding that turbulence and multiphase structure are intrinsic to the ISM and cannot be ignored when interpreting abundance variations; and practically, by advances in computational power that now permit the construction of large-scale MHD simulations incorporating the key physical ingredients: thermal heating and cooling (producing a self-consistent multiphase ISM with cold neutral medium, CNM, warm neutral medium, WNM, and the thermally unstable intermediate phase, UNM), time-dependent chemistry (self-consistently evolving H, H$_2$, H$^+$, and electron abundances), and both subsonic and supersonic turbulence \cite{2021PASP..133j2001B}.

This paradigm, formally introduced by \cite{BellomiEtAl-2020}, carries a fundamentally different and more stringent goal. Rather than fitting each observation separately with a tailored model, a global simulation can naturally produce a continuous spectrum of density, temperature and shielding conditions, from diffuse warm gas to cold dense clouds, along with the full distribution of cloud properties. The challenge is then to study if any rigorous set of immutable parameters (defined by global scale parameters) exists, that can account for the observed scatter in molecular ion columns across many independent sight lines. This is a considerably harder and more powerful constraint than the traditional approach, and also enables studying the dependence of these statistical properties on Galactic scale parameters.

The components of this paradigm developed gradually, through a series of progressively more complete studies. \citet{ValdiviaEtAl-2016,BialyEtAl-2017,NickersonEtAl-2018} first demonstrated how turbulent density fluctuations affect the H\,{\sc i}-to-H$_2$ transition. \citet{BialyEtAl-2019} extended this analysis to $\mathrm{OH^+}$, $\mathrm{H_2O^+}$, and ArH$^+$ using isothermal turbulent box simulations with equilibrium post-processing chemistry, and showed that turbulent density fluctuations alone can produce substantial scatter in observed column densities. However, as those authors acknowledged, two important limitations remained: the simulations were isothermal and therefore could not capture the multiphase nature of the ISM, and the chemistry was assumed to be in equilibrium, a questionable assumption for H$_2$ in dynamic environments.

\cite{BellomiEtAl-2020} addressed both of these shortcomings by incorporating non-isothermal turbulent box simulations with fully time-dependent chemistry of H, H$_2$, H$^+$, and electrons, demonstrating that this more complete model successfully reproduces the observed statistics of H and H$_2$ column densities. The H-H$_2$ transition was found to mostly depend on G$_0$ and the mean total density ($n_\mathrm{H}$) in the simulation box, hence the density and mean distance of OB association and the galactic midplane density, and to weakly depend on turbulence and magnetic field as long as below or at equipartition. \citet{GodardEtAl-2023} then further demonstrated that this same simulation can account for the long-puzzling overabundance problem of CH$^+$, which culminated in a single set of model parameters, following Galactic scale structure constraints: (a) the midplane H\,{\sc i} density at a Galactocentric distance of 8.5\,kpc (corrected for the volume filling factor of fully ionized gas); (b) the ambient radiation field fixed to the local interstellar radiation field (ISRF) in the solar neighborhood; (c) the turbulent energy density required to maintain the vertical structure of the gas; (d) the mean magnetic field strength adopted from Galactic surveys, and assumed to be in approximate equipartition with the turbulent energy density; and (e) the characteristic irradiation scale taken to correspond to the typical separation between OB associations, which sets the spatial frequency of stellar feedback sources. 

Hence, with a fixed CRIR, and a constrained set of model parameters, \cite{GodardEtAl-2023} had managed to reproduce a broad set of Galactic observables: (a) the observed H and H$_2$ column density distributions; (b) the probability distribution function of thermal pressure inferred from fine-structure excitation of carbon in the CNM; (c) the velocity dispersion deduced from H\,{\sc i} emission spectra at high Galactic latitude; and (d) the observed statistical abundance of CH$^+$ and its line profile distribution.

Here we continue this line of development. We use the same simulation presented by \citet{GodardEtAl-2023}, run with the \textsc{ramses} code with identical parameters (turbulence driving, CRIR, FUV field strength $G_0$, and numerical setup). Building on this established and validated baseline, we now ask whether this same simulation can simultaneously explain the observed distribution of $\mathrm{OH^+}$, $\mathrm{H_2O^+}$ and H$_3^+$ to hydrogen column density ratios. We show that a large portion of the observed scatter in these molecular ions in the local ISM, arises naturally from the density and temperature fluctuations inherent to a multiphase turbulent medium, and that a single CRIR close to the canonical value $\zetatref{H}) \sim 2\times10^{-16}$\,s$^{-1}$ \citep{IndrioloEtAl-2007}, consistent with the recent direct constraint from CR-excited vibrational H$_2$ emission \citep{BialyEtAl-2025}, is sufficient to account for the bulk of the observed distribution. While spatial variations in the CRIR may still exist in the real ISM due to the discrete nature of cosmic-ray sources and finite diffusion lengths, our results suggest that such variations need not be invoked as the dominant explanation for the observed scatter. We also stress that our 3D framework, while better matched to the structure of the real ISM, carries limitations of its own, such as finite resolution, reduced chemical networks and a uniform ionization rate, which must equally be accounted for when comparing models with observations.

The paper is structured as follows. In Section \ref{S:Observations} we describe the studies used to assemble our observational sample and provide the adopted sight lines and their associated column densities of $\mathrm{OH^+}$, $\mathrm{H_2O^+}$, H$_3^+$, and hydrogen. In Section \ref{S:Methods} we present our theoretical methods, including our numerical model, the details concerning time-dependent chemistry and the post-processing chemical network used for obtaining the trace molecules. We also describe our line-of-sight reconstruction technique for synthesizing a distribution of column densities from the simulations for direct comparison with the observed distribution. In Section \ref{S:Results} we present our main results. We first analyze the non-equilibrium behavior of electrons and molecular hydrogen and identify the ISM conditions under which equilibrium assumptions are most problematic (Section \ref{SS:Equilibrium}). We then compute the abundances of $\mathrm{OH^+}$, $\mathrm{H_2O^+}$ and H$_3^+$ in post-processing, using the time-dependent H$_2$ and electron fields generated by the simulation as input. As a benchmark, we contrast these predictions with the values that would be obtained if H$_2$ and electrons were also assumed to be in equilibrium, and compare both to observations (Section \ref{SS:TraceResponse}). We summarize our conclusions in Section \ref{S:Conclusions}, emphasizing the importance of time-dependent chemistry, and showing that the \citet{GodardEtAl-2023} framework naturally produces much of the observed scatter in tracer-to-hydrogen columns even with a single local CRIR.

\section{Observational samples}\label{S:Observations}
Our observational comparison combines two qualitatively different kinds of absorption measurements, a distinction rooted in the molecular structure of the tracers. $\mathrm{OH^+}$ and $\mathrm{H_2O^+}$ possess permanent electric dipole moments and therefore have pure rotational transitions, whose ground-state lines lie at sub-millimeter wavelengths and are inaccessible from the ground owing to atmospheric water vapor, requiring space-borne spectroscopy. $\mathrm{H_3^+}$, in contrast, is a symmetric ion with no permanent dipole moment and hence no allowed pure rotational spectrum. It is instead observed through its $\nu_2$ vibration-rotation band near 3.7 micron, from the ground, in a near-infrared atmospheric window. For the latter, nearby hot stars are sufficiently bright background sources (typically within 1-2 kpc). For the former, one requires a much brighter background continuum generated by compact massive star‐forming regions, which can be as far as $\sim$10 kpc. Hence, $\mathrm{OH^+}$ and $\mathrm{H_2O^+}$ absorption is often decomposed into multiple velocity intervals, typically extending across spiral arms in the disk.

The trace molecules have been the target of many observational campaigns in recent decades. In this work, we compile published column densities for each molecule, incorporating distances to the relevant background sources, reported molecular column densities, and the corresponding atomic or total hydrogen column densities. The latter are integrated either up to the background source or else over defined velocity intervals.

\begin{table*}[t]
\centering
\small
\begin{tabular}{lccccc}
\toprule
Source & $v_{\mathrm LSR}$ (km s$^{-1}$) &
$N({\mathrm OH}^+)$ ($10^{13}$ cm$^{-2}$) &
$N({\mathrm H_2O}^+)$ ($10^{13}$ cm$^{-2}$) &
$N({\mathrm H})$ ($10^{21}$ cm$^{-2}$) &
$d$ (kpc) \\
\midrule
M$-0.13-0.08$ & $[-50,-39]$ & $5.11\pm1.42$ & $0.68\pm0.38$ & $1.91\pm0.18$ & $5.2$ \\
M$-0.13-0.08$ & $[-39,-14]$ & $>10.79$ & $2.82\pm1.06$ & $3.24\pm0.41$ & $3.8$ \\
M$-0.02-0.07$ & $[-61,-47]$ & $>8.14$ & $1.43\pm0.39$ & $3.19\pm0.06$ & $5.2$ \\
M$-0.02-0.07$ & $[-47,-37]$ & $3.96\pm0.5$ & $0.39\pm0.22$ & $0.67\pm0.04$ & $3.8$ \\
M$-0.02-0.07$ & $[-37,-23]$ & $7.18\pm1.1$ & $1.41\pm0.38$ & $1.72\pm0.06$ & $3.8$ \\
M$-0.02-0.07$ & $[-23,-13]$ & $6.16\pm0.98$ & $1.09\pm0.29$ & $1.99\pm0.04$ & $3.8$ \\
Sgr B2(M) & $[-57,-33]$ & $34.83\pm6.97$ & $7.63\pm1.19$ & $1.25\pm0.51$ & $5.2$ \\
Sgr B2(M) & $[-33,-2]$ & $60.33\pm12.07$ & $14.59\pm2.29$ & $5.89\pm1.04$ & $3.8$ \\
Sgr B2(N) & $[-60,-31]$ & $54.59\pm10.92$ & $12.71\pm2.54$ & $1.48\pm0.58$ & $5.2$ \\
Sgr B2(N) & $[-31,-7]$ & $44.94\pm8.99$ & $10.64\pm2.13$ & $3.99\pm0.62$ & $3.8$ \\
W28A & $[18,28]$ & $0.73\pm0.05$ & $<0.03$ & $3.17\pm0.43$ & $1.3$ \\
W31C & $[12,24]$ & $7.81\pm0.31$ & $1.99\pm0.15$ & $4.95\pm1.04$ & $4.5$ \\
W31C & $[24,36]$ & $9.97\pm0.46$ & $2.11\pm0.15$ & $>2.81$ & $3.6$ \\
W31C & $[36,43]$ & $8.37\pm0.36$ & $1.91\pm0.11$ & $>1.91$ & $4.3$ \\
W31C & $[43,60]$ & $2.37\pm0.14$ & $0.71\pm0.14$ & $1.84\pm0.29$ & $4.7$ \\
W33A & $[-4,18]$ & $0.85\pm0.14$ & $0.27\pm0.07$ & $>3.19$ & $0.6$ \\
W33A & $[18,25]$ & $0.60\pm0.06$ & $0.19\pm0.02$ & $>0.60$ & $2.4$ \\
G029.96$-00.02$ & $[-10,6]$ & $1.57\pm0.10$ & $0.23\pm0.09$ & $1.40\pm0.23$ & $0.1$ \\
G029.96$-00.02$ & $[6,17]$ & $2.25\pm0.10$ & $0.59\pm0.07$ & $3.45\pm0.16$ & $0.7$ \\
G029.96$-00.02$ & $[17,28]$ & $1.21\pm0.07$ & $0.16\pm0.06$ & $1.42\pm0.16$ & $1.5$ \\
G029.96$-00.02$ & $[28,38]$ & $0.24\pm0.05$ & $<0.05$ & $0.17\pm0.15$ & $2.2$ \\
G029.96$-00.02$ & $[38,45]$ & $0.53\pm0.04$ & $0.05\pm0.04$ & $0.23\pm0.10$ & $2.7$ \\
G029.96$-00.02$ & $[45,50]$ & $0.69\pm0.04$ & $0.06\pm0.03$ & $0.65\pm0.07$ & $3.0$ \\
G029.96$-00.02$ & $[50,56]$ & $0.99\pm0.05$ & $0.12\pm0.03$ & $2.81\pm0.09$ & $3.2$ \\
G029.96$-00.02$ & $[56,65]$ & $2.57\pm0.10$ & $0.23\pm0.05$ & $1.62\pm0.13$ & $3.7$ \\
G029.96$-00.02$ & $[65,73]$ & $3.74\pm0.16$ & $0.68\pm0.06$ & $2.32\pm0.12$ & $4.1$ \\
G029.96$-00.02$ & $[73,79]$ & $2.11\pm0.08$ & $0.40\pm0.04$ & $1.69\pm0.01$ & $4.4$ \\
G029.96$-00.02$ & $[79,88]$ & $1.26\pm0.06$ & $0.09\pm0.05$ & $2.58\pm0.13$ & $4.8$ \\
G034.3+00.15 & $[-12,7]$ & $2.51\pm0.13$ & $0.26\pm0.03$ & $1.34\pm0.18$ & $0.2$ \\
G034.3+00.15 & $[7,18]$ & $2.95\pm0.11$ & $0.55\pm0.02$ & $2.05\pm0.16$ & $0.9$ \\
G034.3+00.15 & $[18,36]$ & $2.69\pm0.12$ & $0.31\pm0.03$ & $>3.64$ & $1.8$ \\
G034.3+00.15 & $[36,44]$ & $1.71\pm0.07$ & $0.21\pm0.01$ & $2.44\pm0.21$ & $2.5$ \\
G034.3+00.15 & $[44,52]$ & $3.67\pm0.13$ & $0.84\pm0.07$ & $3.61\pm0.36$ & $3.0$ \\
W51e & $[-4,11]$ & $3.25\pm0.08$ & $0.61\pm0.07$ & $1.94\pm0.31$ & $0.5$ \\
W51e & $[11,16]$ & $1.10\pm0.02$ & $0.18\pm0.02$ & $0.84\pm0.10$ & $0.9$ \\
W51e & $[16,21]$ & $0.84\pm0.02$ & $0.08\pm0.02$ & $0.88\pm0.09$ & $1.3$ \\
W51e & $[21,33]$ & $1.85\pm0.05$ & $0.18\pm0.04$ & $1.42\pm0.19$ & $1.6$ \\
W51e & $[33,42]$ & $0.96\pm0.03$ & $0.20\pm0.03$ & $0.84\pm0.12$ & $2.7$ \\
W51e & $[62,75]$ & $2.04\pm0.05$ & $0.63\pm0.05$ & $>1.54$ & $5.4$ \\
DR21C & $[3,21]$ & $9.53\pm0.83$ & $1.47\pm0.17$ & $>7.58$ & $1.2$ \\
DR21(OH) & $[3,25]$ & $9.36\pm0.60$ & $1.75\pm0.09$ & $>7.59$ & $1.2$ \\
NGC~7538~IRS1 & $[-17,-3]$ & $2.28\pm0.17$ & $0.27\pm0.05$ & $2.96\pm0.03$ & $0.7$ \\
NGC~7538~IRS1 & $[-3,18]$ & $2.50\pm0.23$ & $0.37\pm0.07$ & $1.99\pm0.05$ & $0.1$ \\
W3~IRS5 & $[-28,-8]$ & $2.83\pm0.13$ & $0.26\pm0.07$ & $1.20\pm0.26$ & $1.5$ \\
W3~IRS5 & $[-8,10]$ & $3.71\pm0.15$ & $0.51\pm0.07$ & $0.97\pm0.21$ & $0.7$ \\
W3(OH) & $[-25,-8]$ & $3.09\pm0.10$ & $0.50\pm0.06$ & $1.39\pm0.03$ & $1.5$ \\
W3(OH) & $[-8,9]$ & $3.56\pm0.11$ & $0.66\pm0.06$ & $0.95\pm0.03$ & $0.7$ \\
G327.30$-0.6$ & $[-28,-15]$ & $2.92\pm0.12$ & $0.43\pm0.03$ & $0.90\pm0.04$ & $1.2$ \\
G327.30$-0.6$ & $[-15,-8]$ & $1.48\pm0.05$ & $0.23\pm0.02$ & $0.97\pm0.02$ & $0.9$ \\
G327.30$-0.6$ & $[-8,6]$ & $1.59\pm0.07$ & $0.33\pm0.04$ & $2.90\pm0.04$ & $0.2$ \\
NGC 6334 I & $[1,14]$ & $1.84\pm0.07$ & $0.31\pm0.03$ & $3.94\pm0.09$ & $1.4$ \\
NGC 6334 I(N) & $[1,12]$ & $1.88\pm0.06$ & $0.27\pm0.03$ & $4.42\pm0.06$ & $1.4$ \\
\bottomrule
\end{tabular}
\caption{Observational sample, based on data from \cite{IndrioloEtAl-2015}. Columns: Source name, velocity intervals, associated column densities ($N$($\mathrm{OH^+}$), $N$($\mathrm{H_2O^+}$) and atomic hydrogen $N$(H)) and heliocentric distance.}
\label{tab:Indriolo2015_table5}
\end{table*}

\begin{table*}[t]
\centering
\small
\begin{tabular}{lccccc}
\toprule
Source &
$N({\mathrm OH}^+)$ ($10^{13}$ cm$^{-2}$) &
$N({\mathrm H_2O}^+)$ ($10^{13}$ cm$^{-2}$) &
$N({\mathrm H})$ ($10^{21}$ cm$^{-2}$) &
$d$ (kpc) \\
\midrule
W28A & $0.73\pm0.05$ & $<0.03$ & $3.17\pm0.43$ & $1.28$ \\
W31C & $28.52\pm1.27$ & $6.72\pm0.55$ & $>11.51$ & $4.95$ \\
W33A & $1.45\pm0.20$ & $0.46\pm0.09$ & $>3.79$ & $2.40$ \\
G029.96$-00.02$ & $17.16\pm0.85$ & $2.61\pm0.57$ & $18.34\pm1.35$ & $5.26$ \\
G034.3+00.15 & $13.53\pm0.56$ & $2.17\pm0.16$ & $>13.08$ & $3.80$ \\
W51e & $10.04\pm0.25$ & $1.88\pm0.23$ & $>7.46$ & $5.41$ \\
DR21C & $9.53\pm0.83$ & $1.47\pm0.17$ & $>7.58$ & $1.50$ \\
DR21(OH) & $9.36\pm0.60$ & $1.75\pm0.09$ & $>7.59$ & $1.50$ \\
W3~IRS5 & $6.54\pm0.28$ & $0.77\pm0.14$ & $2.17\pm0.47$ & $1.83$ \\
W3(OH) & $6.65\pm0.21$ & $1.16\pm0.12$ & $2.34\pm0.06$ & $2.04$ \\
G327.30$-0.6$ & $5.99\pm0.24$ & $0.99\pm0.09$ & $4.77\pm0.10$ & $3.30$ \\
NGC~6334~I & $1.84\pm0.07$ & $0.31\pm0.03$ & $3.94\pm0.09$ & $1.35$ \\
NGC~6334~I(N) & $1.88\pm0.06$ & $0.27\pm0.03$ & $4.42\pm0.06$ & $1.35$ \\
\bottomrule
\end{tabular}
\caption{Same as Table \ref{tab:Indriolo2015_table5} but with velocity intervals integrated to the source.}
\label{tab:Indriolo2015_table5_integrated}
\end{table*}

For $\mathrm{OH^+}$ and $\mathrm{H_2O^+}$ observations, we use the \textit{Herschel} data compiled by \cite{IndrioloEtAl-2015}, which provides information about observed column densities over defined velocity intervals, corresponding to what these authors identify to be separate absorption components. Column densities have measured values with defined errors, while some are given as limits. In most cases, limit column densities are lower limits. We use these data with the following caveats. The CRIR is believed to be much higher near the galactic center than in the disk, so we exclude velocity intervals near the galactic center. We also exclude any velocity intervals that may be too close to the background sources (indicating a velocity range within 5 km s$^{-1}$ of the background source systemic velocity), where the assumption of a low excitation temperature may be invalid and in turn the measured column densities may be unreliable. Finally, we exclude the velocity intervals for the source W49N, whose distances have two alternatives under the galactic rotation model used by \cite{IndrioloEtAl-2015}, hence making it difficult to ascertain the true distance to each of its velocity intervals. After these exclusions, we have (Table \ref{tab:Indriolo2015_table5}) 52 valid velocity intervals, in 17 distinct sources. The column densities towards each source (in Table \ref{tab:Indriolo2015_table5_integrated}) are obtained by integrating all valid velocity intervals, but if even one is a lower limit, the integrated sum is also a lower limit. For the integrated values, we however exclude M$-0.13-0.08$, M$-0.02-0.07$, Sgr B2(M) and Sgr B2(N), as these sources are in the galactic center. Since $\mathrm{OH^+}$ and $\mathrm{H_2O^+}$ preferentially trace mostly atomic/transitional gas, \cite{IndrioloEtAl-2015} provide atomic hydrogen columns, specifically, several of which are lower limits.

For $\mathrm{H_3^+}$ observations, we use the (\textit{VLT} and \textit{IRTF}) data from \cite{ObolentsevaEtAl-2024}, in combination with the (\textit{VLT}, \textit{UKIRT}, \textit{Keck}, \textit{KPNO} and \textit{Gemini South}) data from \cite{IndrioloEtAl-2025}. As previously noted, here the sources are located at close heliocentric distances (up to 2.47 kpc), and these studies provide only the integrated column densities towards the sources. In Table \ref{tab:Obolentseva2024_table1_Indriolo2025_table4} we have in total 37 sources and their associated column densities, the majority of which have measured values (not limits) with defined errors. Since $\mathrm{H_3^+}$ forms predominantly in denser gas, it is associated with more molecular gas. Therefore, in Table \ref{tab:Obolentseva2024_table1_Indriolo2025_table4} we list the total hydrogen column as provided by \cite{ObolentsevaEtAl-2024} and \cite{IndrioloEtAl-2025}. 

We note that the fiducial simulation is intended to represent \emph{local} ISM conditions, so distant sight lines are treated separately later in our study as potentially probing different environments.

\begin{table}[t]
\centering
\small
\begin{tabular}{lccc}
\toprule
Source &
$N(\mathrm{H_3^+})$ ($10^{13}$ cm$^{-2}$) &
$N_{\mathrm{H}}$ ($10^{21}$ cm$^{-2}$) &
$d$ (pc) \\
\midrule
HD 24398 & $6.3\pm0.5$ & $1.66^{+0.63}_{-0.32}$ & $230$ \\
HD 24534 & $7.3\pm0.9$ & $2.21^{+0.18}_{-0.16}$ & $820$ \\
HD 41117 & $5.3\pm2.0$ & $3.66^{+1.28}_{-0.76}$ & $1300$ \\
HD 73882 & $9.0\pm0.5$ & $3.99^{+0.80}_{-0.59}$ & $1000$ \\
HD 110432 & $5.2\pm0.2$ & $1.63^{+0.36}_{-0.22}$ & $420$ \\
HD 154368 & $9.4\pm1.3$ & $3.93^{+0.55}_{-0.43}$ & $1080$ \\
HD 210839 & $7.6\pm1.2$ & $3.02^{+0.27}_{-0.23}$ & $810$ \\
HD 23180 & $2.73\pm0.66$ & $1.61^{+0.39}_{-0.32}$ & $301$ \\
HD 281159 & $5.16\pm2.06$ & $4.86^{+3.40}_{-1.92}$ & $347$ \\
HD 167971 & $5.98\pm0.82$ & $5.40^{+4.37}_{-2.30}$ & $1680$ \\
HD 170740 & $3.35\pm0.54$ & $2.86^{+0.84}_{-0.63}$ & $250$ \\
HD 179406 & $4.55\pm1.04$ & $2.77^{+0.89}_{-0.66}$ & $287$ \\
HD 203374 & $6.81\pm1.16$ & $2.52^{+0.33}_{-0.29}$ & $780$ \\
HD 206165 & $5.13\pm0.87$ & $2.73^{+1.63}_{-0.85}$ & $996$ \\
HD 206267 & $6.92\pm1.71$ & $3.44^{+0.96}_{-0.71}$ & $690$ \\
HD 207198 & $6.31\pm2.21$ & $3.26^{+0.70}_{-0.57}$ & $980$ \\
HD 216532 & $8.74\pm2.76$ & $4.92^{+3.19}_{-1.80}$ & $920$ \\
HD 216898 & $7.51\pm1.51$ & $6.71^{+3.88}_{-2.28}$ & $870$ \\
HD 224151 & $8.8\pm1.39$ & $2.92^{+0.47}_{-0.41}$ & $910$ \\
HD 22951 & $<2.82$ & $1.67^{+0.43}_{-0.34}$ & $369$ \\
HD 37367 & $<4.53$ & $2.16^{+0.77}_{-0.56}$ & $1274$ \\
HD 37903 & $<1.30$ & $3.14^{+0.59}_{-0.48}$ & $471$ \\
HD 42087 & $<2.18$ & $3.12^{+0.88}_{-0.69}$ & $2470$ \\
HD 47129 & $<4.90$ & $1.91^{+0.74}_{-0.51}$ & $1271$ \\
HD 48099 & $<4.76$ & $1.80^{+0.62}_{-0.45}$ & $1289$ \\
HD 102065 & $<3.51$ & $0.98^{+0.42}_{-0.24}$ & $200$ \\
HD 145502 & $<2.07$ & $1.57^{+0.62}_{-0.44}$ & $170$ \\
HD 147933 & $<1.53$ & $7.20^{+1.48}_{-1.23}$ & $170$ \\
HD 149757 & $<2.70$ & $1.42^{+0.11}_{-0.10}$ & $135$ \\
HD 184915 & $<3.10$ & $1.20^{+0.34}_{-0.26}$ & $498$ \\
HD 192639 & $<9.25$ & $3.07^{+0.73}_{-0.57}$ & $2130$ \\
HD 199579 & $<1.84$ & $1.74^{+0.43}_{-0.34}$ & $940$ \\
HD 217035 & $<21.0$ & $4.59^{+1.22}_{-0.95}$ & $720$ \\
HD 217312 & $<6.28$ & $4.25^{+0.88}_{-0.72}$ & $600$ \\
HD 21859 & $<9.20$ & $1.34^{+0.29}_{-0.21}$ & $78$ \\
HD 148184 & $<1.90$ & $2.36^{+0.59}_{-0.39}$ & $153$ \\
HD 149404 & $<14.1$ & $3.91^{+1.15}_{-0.71}$ & $1320$ \\
\bottomrule
\end{tabular}
\caption{Observational sample, based on data from \cite{ObolentsevaEtAl-2024} and \cite{IndrioloEtAl-2025}. Columns: source name, column densities ($N(\mathrm{H_3^+})$ and total hydrogen $N_{\mathrm{H}}$) and heliocentric distance.}
\label{tab:Obolentseva2024_table1_Indriolo2025_table4}
\end{table}

For completion, we note that complementary $\mathrm{OH^+}$ measurements toward reddened background stars are available from near-UV absorption studies, including the EDIBLES survey \citep{BacallaEtAl-2019}. These observations probe a different sample of local diffuse sight lines. However, we do not include them in the primary comparison here, for a number of reasons. First, the near-UV data provide $\mathrm{OH^+}$ only, and therefore do not allow a joint $\mathrm{OH^+}$-$\mathrm{H_2O^+}$ comparison in the same observational sample. Second, the EDIBLES analysis is formulated primarily in terms of the total hydrogen column density, \citep{BacallaEtAl-2019} providing only a few sight lines with explicit $N(\mathrm{H})$, whereas the \textit{Herschel} sample of \cite{IndrioloEtAl-2015} provides the atomic hydrogen column $N(\mathrm{H})$ used in our study for comparison with $\mathrm{OH^+}$ and $\mathrm{H_2O^+}$. Because $\mathrm{OH^+}$ and $\mathrm{H_2O^+}$ preferentially trace mostly atomic/transitional gas, converting the near-UV $N_{\rm H}$ values into homogeneous $N(\mathrm{H})$ columns associated with the $\mathrm{OH^+}$ bearing material would require additional assumptions. Third, the \cite{IndrioloEtAl-2015} atomic hydrogen columns are assigned to the same velocity intervals as $\mathrm{OH^+}$ and $\mathrm{H_2O^+}$, whereas the far-UV columns give source-integrated columns instead, which is partially incompatible with our analysis of the sample, as outlined in Section \ref{SS:LOS_reconstruction}. For all these reasons, we do not combine the near-UV $\mathrm{OH^+}$ data with the main sample.

\section{Methods}\label{S:Methods}
Our modeling procedure has three layers. First, our MHD simulations self-consistently evolve the gas dynamics, thermal state, and the time-dependent abundances of H, H$_2$, H$^+$, PAH charge states and electrons (Section \ref{SS:Overview}). Second, using the resulting cell-by-cell values of the various abundances, density, temperature and shielding conditions, we compute the abundances of trace molecular ions $\mathrm{OH^+}$, $\mathrm{H_2O^+}$ and H$_3^+$ in post-processing, assuming instantaneous steady state (Section \ref{SS:Chemistry}). Third, we reconstruct synthetic lines of sight through the simulation volume and compare the resulting column-density distributions with the observational samples (Section \ref{SS:LOS_reconstruction}).

\subsection{Simulations - overview and input parameters}\label{SS:Overview}
The setup used here is identical to the fiducial simulation of \cite{GodardEtAl-2023}, utilizing the MHD code \textit{RAMSES} \citep{Teyssier-2002,FromangEtAl-2006}. Here we briefly describe the main features of the code. The reader is referred to \cite{GodardEtAl-2023} for further information and detailed references.

We simulate the magnetized and partially ionized gas inside a $L$=200 pc box with periodic boundary conditions, illuminated from all sides by the standard isotropic UV radiation field \citep{MathisEtAl-1983} scaled by the parameter $G_0$. The total ionization rate of H by cosmic ray particles is denoted by $\zetatref{H})$. A homogeneous initial magnetic field $B$ is included. Mechanical energy is injected in the gas at large scale. The amplitude of this turbulent forcing and the relative power injected in compressive modes are controlled by the forcing strength $F$ and compressive ratio $\xi$, as defined in \cite{GodardEtAl-2023}. In our current setup, adaptive mesh refinement in \textit{RAMSES} is deactivated, and the number of grid cells used is straightforward - $K^3$, where $K$ is the linear mesh size. We compared setups with $K$ = 32, 64, 128, 256, 512 and 1024, to check for convergence (Appendix \ref{Appendix:C}), while the main results and plots presented in the study are based on the highest resolution.

The above mentioned parameters are summarized in Table \ref{tab:standard_parameters}. These parameters are tightly constrained by a series of independent constraints, as specified briefly in Section \ref{S:Intro}, while the full justification and details are provided by the predecessor study of \cite{GodardEtAl-2023}, reproducing a broad set of galactic observables.

\begin{table}[htb!]
  \centering
  \caption{Standard values of the simulation parameters.}
  \label{tab:standard_parameters}
  \begin{tabular}{l l l l}
    \hline\hline
    Parameter & Symbol & Value & Unit \\
    \hline
    Box size & $L$ & 200 & pc \\
    Mean density & $n_\mathrm{H}$ & 1.5 & cm$^{-3}$ \\
    UV radiation & $G_0$ & 1 & Mathis field \\
    CR ionization rate & $\zetat{H}$ & $2\times10^{-16}$ & s$^{-1}$ \\
    Mesh size & $K$ & $64, 128, 256, 512, 1024$ & \\
    Forcing strength & $F$ & $1.5\times10^{-3}$ & kpc\,Myr$^{-2}$ \\
    Compressive ratio & $\xi$ & 0.1 & \\
    Initial magnetic field & $B$ & 3.8 & $\mu$G \\
    \hline
    Abundance of (total) PAHs & $\abundance{PAH}$ & $6 \times 10^{-7}$ & \\
    Abundance of oxygen & $\abundance{O}$ & $3.2 \times 10^{-4}$ & \\
    Abundance of carbon & $\abundance{C}$ & $1.4 \times 10^{-4}$ & \\
    \hline
  \end{tabular}
\end{table}

Heating is induced by the photo-electric effect, cosmic ray particles, the formation of H$_2$ and its photo-destruction. Cooling is induced by the Lyman-$\alpha$ line, the fine structure lines of OI and CII, the recombination of electrons onto grains and the radiative cooling induced by the collisional excitation of H$_2$ rovibrational levels (see appendix B in \cite{BellomiEtAl-2020} for full details).

The out-of-equilibrium chemical evolution is computed in the simulation, as follows: (a) the formation of H$_2$ onto grains and its photo-destruction by UV photons, accounting for H$_2$ self-shielding and dust attenuation using the method of \cite{ValdiviaEtAl-2016}; (b) H$^+$ forms by ionization of H, and recombines on PAHs, accounting for dust attenuation. PAHs ionization states are determined by photo-detachment, photoionization and by recombination with free electrons, with the total abundance of PAHs ($\abundanceref{PAH}$) being held constant (Table \ref{tab:standard_parameters}).

These processes are described by a set of chemical reactions, fully listed in Table \ref{tab:reactions_RAMSES}. The \textit{RAMSES} simulations solve these time-dependent reactions and output the following fractional abundances: $x$(H), $x$(H$_2$), $x$(PAH$^-$), $x$(PAH$^0$), $x$(PAH$^+$) and $x$(e) (the latter including a post-processing correction specified in Appendix \ref{Appendix:A}), in addition to the temperature and density in each cell. These are used in the following section in order to calculate, in post-processing, the abundances of the studied trace molecules.

\begin{table*}[]
  \centering
  \caption{\textit{RAMSES} chemical network.}
  \label{tab:reactions_RAMSES}
  \begin{tabular}{l l l l}
    \hline\hline
    Reaction & Rate coefficient & Units & Rate Source\\
    \hline
    \rowcolor{gray!20}
    \multicolumn{4}{c}{H$^+$, H and PAHs} \\
    \hline
    H + CR $\rightarrow$ H$^+$ + e & $\zetatref{H})=2\times10^{-16}$ & s$^{-1}$ & \cite{IndrioloEtAl-2007} \\
    PAH$^-$ + photon $\rightarrow$ PAH$^0$ + e & $1.83\times10^{-8} \mathrm{e}^{-\tau} G\mathrm{_0}$ & s$^{-1}$ & \cite{LePetitEtAl-2006} \\
    PAH$^0$ + photon $\rightarrow$ PAH$^+$ + e & $1.17\times10^{-8} \mathrm{e}^{-\tau} G\mathrm{_0}$ & s$^{-1}$ & \cite{LePetitEtAl-2006} \\
    H$^+$ + PAH$^-$ $\rightarrow$ H + PAH$^0$ & $8.3\times10^{-7} (T/100~\mathrm{K})^{-0.5}$ & cm$^3$ s$^{-1}$ & \cite{DrainSutin-1987}\\
    H$^+$ + PAH$^0$ $\rightarrow$ H + PAH$^+$ & $3.1\times10^{-8}$ & cm$^3$ s$^{-1}$ & \cite{DrainSutin-1987}\\
    PAH$^0$ + e $\rightarrow$ PAH$^-$ + photon & $1.5625\times10^{-6}$ & cm$^3$ s$^{-1}$ & \cite{DrainSutin-1987} \\
    PAH$^+$ + e $\rightarrow$ PAH$^0$ + photon & $4.0625\times10^{-5} (T/100~\mathrm{K})^{-0.5}$ & cm$^3$ s$^{-1}$ & \cite{DrainSutin-1987} \\
    \hline
    \rowcolor{gray!20}
    \multicolumn{4}{c}{H$_2$ formation / destruction} \\
    \hline
    H + H + grain $\rightarrow$ H$_2$ + grain & $\rateh{R} = 3 \times 10^{-17} \sqrt{T/100~\mathrm{K}} \left( 1 + (T/464~\mathrm{K})^{1.5} \right)^{-1}$ & cm$^3$ s$^{-1}$ & \citep{Jura-1974,LeBourlotEtAl-2012,BronEtAl-2014}\\
    H$_2$ + photon $\rightarrow$ H + H & $\rateh{D} = 3.3 \times 10^{-11} f_\mathrm{att} G_0$ & s$^{-1}$ & \citep{BertoldiDrain-1996,DrainBertoldi-1996}\\
    \hline
    \multicolumn{4}{l}{*$\tau$ is the optical depth, and $f_\mathrm{att}$ incorporates the attenuation by both dust and self-shielding of H$_2$}.
  \end{tabular}
\end{table*}

\subsection{Post-processing: $\mathrm{OH^+}$, $\mathrm{H_2O^+}$ and $\mathrm{H_3^+}$ chemistry}\label{SS:Chemistry}
We compute the abundances of $\mathrm{OH^+}$, $\mathrm{H_2O^+}$ and $\mathrm{H_3^+}$ in post-processing, at equilibrium, while under the constraint of time-dependent H$_2$ and e. This approach is justified since the relevant formation/destruction rates of ion molecules are characteristically around $\sim 10^{-9}$ cm$^3$ s$^{-1}$ (Tables \ref{tab:reactions_OHp}-\ref{tab:reactions_H3p}), implying equilibration timescales of thousands of years for the densities of interest -- several orders of magnitude faster than either the dynamical timescale set by turbulence or the formation of H$_2$ on dust grains (further discussion in Section \ref{S:Results}). Hence, these ion molecules equilibrate extremely quickly once H$_2$ and e are fixed. The chemical network solved is fully provided in Tables \ref{tab:reactions_OHp}-\ref{tab:reactions_H3p}.

Given our chemical reaction network, we write the rate equations for the formation and destruction of $\mathrm{OH^+}$, $\mathrm{H_2O^+}$ and H$_3^+$. Assuming a chemical equalibrium (formation rate=destruction rate) we obtain analytic formulae for the abundances of each of these molecules. For $\mathrm{OH^+}$, we obtain
\begin{equation}\label{eq:x_OHp}
\begin{aligned}
x(\mathrm{OH}^+) = \frac{\epsilon \zetatref{H})}{n_\mathrm{H} \left(\rateref{4} x(\mathrm{H}_2) + \rateref{5} x(\mathrm{e})\right)},
\end{aligned}
\end{equation}
where
\begin{equation}\label{eq:epsilon_of_x_OHp}
\begin{aligned}
\epsilon = \frac{1}{1+Y}
\end{aligned}
\end{equation}
and
\begin{equation}\label{eq:Y_of_x_OHp}
\begin{aligned}
Y = \frac{\left(\rateref{2} + \rateref{3} x(\mathrm{H}_2)\right) \left(\rateref{6} x(\mathrm{e}) + \rateref{7} x(\mathrm{PAH^-}) + \rateref{8} x(\mathrm{PAH^0})\right)}{\rateref{1} \rateref{3} \abundanceref{O} x(\mathrm{H}_2)}.
\end{aligned}
\end{equation}

This expression is identical to the established analytic framework of \cite{HollenbachEtAl-2012}, although several updates to the rate coefficients have been incorporated. Most importantly, the rate coefficient, $k_5$, for the dissociative recombination of $\mathrm{OH^+}$ has been recently measured \citep{KalosiEtAl-2023} for rotationally-cold $\mathrm{OH^+}$ at the Cryogenic Storage Ring (CSR) in Heidelberg and found to be a factor 3-5 larger than that for thermally-populated $\mathrm{OH^+}$ at room temperature. This new value of $k_5$ is more appropriate for low-temperature, low-density astrophysical environments and was therefore adopted in our model: this change significantly reduces the predicted $\mathrm{OH^+}$ abundance. It remains to be seen whether future measurements of the dissociative recombination rate for rotationally-cold $\mathrm{H_2O^+}$, planned at the CSR, will require similar changes to the rate coefficient $k_{11}$.

\begin{table*}[htb]
  \centering
  \caption{Reaction channels for $\mathrm{OH^+}$.}
  \label{tab:reactions_OHp}
  \begin{tabular}{l l l l}
    \hline\hline
    Reaction & Rate coefficient & Units & Rate Source\\
    \hline
    H + CR $\rightarrow$ H$^+$ + e & $\zetatref{H}) = 2\times10^{-16}$ & s$^{-1}$ & \cite{IndrioloEtAl-2007}\\
    H$^+$ + O $\rightarrow$ O$^+$ + H & \rate{1}(Equation \ref{eq:gamma1}) & cm$^3$ s$^{-1}$ & \cite{StancilEtAl-1999}\\
    O$^+$ + H $\rightarrow$ H$^+$ + O & \rate{2}(Equation \ref{eq:gamma2}) & cm$^3$ s$^{-1}$ & \cite{StancilEtAl-1999}\\
    H$_2$ + O$^+$ $\rightarrow$ $\mathrm{OH^+}$ + H & $\rate{3} = 1.3\times10^{-9}$ & cm$^3$ s$^{-1}$ & \cite{KovalenkoEtAl-2018}\\
    $\mathrm{OH^+}$ + H$_2$ $\rightarrow$ $\mathrm{H_2O^+}$ + H & $\rate{4} = 1\times10^{-9}$ & cm$^3$ s$^{-1}$ & \cite{TranEtAl-2018} \\
    $\mathrm{OH^+}$ + e $\rightarrow$ O + H & $\rate{5}$ (Equation \ref{eq:gamma5}) & cm$^3$ s$^{-1}$ & \cite{KalosiEtAl-2023}\\
    H$^+$ + e $\rightarrow$ H & $\rate{6} = 3.5\times10^{-12} (T/300~\mathrm{K})^{-0.75}$ & cm$^3$ s$^{-1}$ &  \cite{HollenbachEtAl-2012}\\
    H$^+$ + PAH$^-$ $\rightarrow$ PAH$^0$ + H & $\rate{7} = 8.3\times10^{-7} (T/100~\mathrm{K})^{-0.5}$ & cm$^3$ s$^{-1}$ & \cite{DrainSutin-1987}\\
    H$^+$ + PAH$^0$ $\rightarrow$ PAH$^+$ + H & $\rate{8} = 3.1\times10^{-8}$ & cm$^3$ s$^{-1}$ & \cite{DrainSutin-1987}\\
    \hline
  \end{tabular}
\end{table*}

\begin{table*}[htb]
  \centering
  \caption{Reaction channels for $\mathrm{H_2O^+}$.}
  \label{tab:reactions_H2Op}
  \begin{tabular}{l l l l}
    \hline\hline
    Reaction & Rate coefficient & Units & Rate Source\\
    \hline
    $\mathrm{OH^+}$ + H$_2$ $\rightarrow$ $\mathrm{H_2O^+}$ + H & $\rate{9} = 1\times10^{-9}$ & cm$^3$ s$^{-1}$ & \cite{TranEtAl-2018} \\
    $\mathrm{H_2O^+}$ + H$_2$ $\rightarrow$ H$_3$O$^+$ + H & $\rate{10} = 9.7\times10^{-10}$ & cm$^3$ s$^{-1}$ & \cite{TranEtAl-2018}\\
    $\mathrm{H_2O^+}$ + e $\rightarrow$ products & $\rate{11} = 4.3\times10^{-7} (T/300~\mathrm{K})^{-0.74}$ & cm$^3$ s$^{-1}$ & \cite{RosenEtAl-2000}\\
    \hline
  \end{tabular}
\end{table*}

\begin{table*}[htb]
  \centering
  \caption{Reaction channels for $\mathrm{H_3^+}$.}
  \label{tab:reactions_H3p}
  \begin{tabular}{l l l l}
    \hline\hline
    Reaction & Rate coefficient & Units & Rate Source\\
    \hline
    H$_2$ + CR $\rightarrow$ H$_2$$^+$ + e & $\zetat{H_2})=\sim 2 \zetatref{H})~(=4\times10^{-16})$ & s$^{-1}$ & \cite{Dalgarno-2006} \\
    H$_2$$^+$ +H$_2$ $\rightarrow$ $\mathrm{H_3^+}$ + H & $2.23 \times 10^{-9}$ & cm$^3$ s$^{-1}$ & \cite{Jimenez-RedondoEtAl-2024}\\
    H$_2$$^+$ +H $\rightarrow$ H$^+$ + H$_2$ & $6.4 \times 10^{-10}$ & cm$^3$ s$^{-1}$ & \cite{NeufeldWolfire-2017}\\
    $\mathrm{H_3^+}$ + e $\rightarrow$ products & $\rate{12} = 6.7\times10^{-8} (T/300~\mathrm{K})^{-0.52}$ & cm$^3$ s$^{-1}$ & \cite{LePetitEtAl-2016}\\
    $\mathrm{H_3^+}$ + O $\rightarrow$ products & $\rate{13} = \sim 1\times10^{-9}$ & cm$^3$ s$^{-1}$ & \cite{HillenbrandEtAl-2022}\\
    \hline
  \end{tabular}
\end{table*}

For $\mathrm{H_2O^+}$ we obtain
\begin{equation}\label{eq:x_H2Op}
\begin{aligned}
x(\mathrm{H}_2\mathrm{O}^+) = \frac{x(\mathrm{OH}^+)}{\rateref{10}/\rateref{9}+\rateref{11}x(\mathrm{e})/\rateref{9}x(\mathrm{H}_2) },
\end{aligned}
\end{equation}
where $x$($\mathrm{OH^+}$) derives Equation \ref{eq:x_OHp}. This expression is identical to the analytic framework of \cite{GerinEtAl-2010}.

For $\mathrm{H_3^+}$ we obtain
\begin{equation}\label{eq:x_H3p}
\begin{aligned}
x(\mathrm{H}_3^+) = \frac{\zetatref{H_2}) x(\mathrm{H}_2)}{n_\mathrm{H} \left( \rateref{12} x(\mathrm{e}) + \rateref{13} \abundanceref{O} \right)} \frac{1}{1+0.3 x(\mathrm{H})/x(\mathrm{H_2})}.
\end{aligned}
\end{equation}

This expression is similar to that obtained by the chemical framework suggested by \cite{Dalgarno-2006}, however with a couple of revisions. Our first revision to the \cite{Dalgarno-2006} expression is the addition of the term $\rateref{13} \abundanceref{O}$, complementing the standard $\rateref{12} x(\mathrm{e})$. Ordinarily, studies may assume that the destruction of $\mathrm{H_3^+}$ is dominated by dissociative recombination with electrons. However, as previously mentioned in Section \ref{SS:Overview}, in dense CNM it cannot be assumed that the electron fractional abundance is derived solely from fully ionized atomic carbon. The fraction of ionized atomic carbon instead lowers with increasing density and shielding. Eventually, as this in turn leads to diminishing electron fractional abundance, destruction by electron dissociative recombination becomes sub-dominant to destruction by $\mathrm{H_3^+}$ reactions with neutral constituents (O, CO, PAH, etc.). Of these neutral constituents, atomic oxygen initially dominates, and therefore we add the contribution of destruction by atomic oxygen ($\rateref{13} \abundanceref{O}$) and neglect other neutral constituents for simplicity.

Our second revision is based on the discussion by \cite{NeufeldWolfire-2017}. Past studies typically assumed that following the formation of H$_2$$^+$, a reaction with H$_2$ subsequently forms $\mathrm{H_3^+}$. However, H$_2$$^+$ can also react with H instead. While the rate for the latter reaction is lower (see Table \ref{tab:reactions_H3p}), the branching ratio is not negligible. The denominator term ($1+ 0.3 x(\mathrm{H})/x(\mathrm{H_2})$) allows for this possibility.

%$x(\mathrm{C}^+)=\max([\mathrm{C}]/S,10^{-7})$.The parameters that enter into $S$ are acquired from \textit{RAMSES}. Since the fraction of ionized atomic hydrogen is negligible in the CNM phase, $x(\mathrm{e})\approx x(\mathrm{C^+})$ is reduced in kind. However, the reduction in $x(\mathrm{e})$ is not indefinite, as eventually other species become the main charge carriers. We account for this by capping $x(\mathrm{C}^+)$ at $10^{-7}$ (see e.g., \cite{BronEtAl-2021,BeslicEtAl-2025}). In practice, the destruction term $k_\mathrm{O} x(\mathrm{O})$ becomes dominant over $k_\mathrm{e} x(\mathrm{e})$ (Equarion \ref{eq:x_H3p}) already when $x(\mathrm{e})$ drops to a values of about $\sim$$10^{-6}$, so that the cap is precautionary but not not mandatory.

\subsection{Line of sight reconstruction}\label{SS:LOS_reconstruction}
Observations span a wide range of LOS lengths and their associated column densities, whereas our simulation consists of a single, 200 pc box, representing a chunk of ISM (neutral) volume. Thus, in order to enable a comparison, we follow the approach of \cite{BialyEtAl-2019,BellomiEtAl-2020,GodardEtAl-2023}, where we stack simulation boxes whenever lengths are larger than the simulation box, to match the overall distribution of LOS lengths. In turn, we can compare the observed column densities and those synthesized from our simulations, via the following steps: (a) choosing a distribution of LOSs; (b) calculating synthesized column densities of trace molecules and hydrogen, either atomic or total; (c) drawing histograms of hydrogen per trace molecules, and compare observed sample to the synthesized sample. A good match between the latter is a pivotal factor in choosing the LOS distribution wisely. Because the observed samples mix source-integrated sight lines and velocity-interval components, we consider two separate reconstruction strategies.

\noindent \textbf{Distance to source:} This algorithm applies to the case where column densities are based on all the intervening gas between us and the background source. From our 200 pc 3D box, we draw random 1D lines of length $L_i$, $i$ denoting a distribution of distances. The distances are obtained from the last column ($d$) in Tables \ref{tab:Indriolo2015_table5_integrated} and \ref{tab:Obolentseva2024_table1_Indriolo2025_table4}, however filtering out distant sources beyond 3 kpc, postulating that the CRIR can be markedly different beyond the local ISM. For $L_i \leq 200$ pc the sight line is a single random chord of length $L_i$ terminating within the box, while for $L_i>$ 200 pc we stack $\lceil L_i/(200\ \mathrm{pc})\rceil$ independent random lines, so as to avoid re-sampling the same structures along a single sight line. We correct for the absence of hot ionized medium (HIM) in our simulations, since only neutral ISM contributes significantly to both trace molecules and neutral hydrogen, multiplying $L_i$ pc by a factor of (1-$\varphi$), where $\varphi$=0.5 is our heuristic for the HIM fraction \citep{BellomiEtAl-2020,GodardEtAl-2023}.

\noindent \textbf{Velocity intervals power-law size distribution:} This algorithm applies to the case where column densities are defined over specific velocity intervals, the last column ($d$) in Table \ref{tab:Indriolo2015_table5} now indicating the distance to the clouds associated with each interval. Unfortunately, there is no information about the physical size of these clouds, necessitating some further assumption. E.g., \cite{BialyEtAl-2019} assumed a constant size. However, it is expected that the absorption features arise from ISM patches of different sizes, smaller clouds typically being more common than large ones. Hence, we adopt a truncated power law mathematical framework for $L_i$, such that $p(L) \propto L^{-\gamma}$ over [$L_\mathrm{min}$, $L_\mathrm{max}$], where $\gamma$ is the power-law exponent coefficient. These lengths apply only to the simulated neutral ISM, while ignoring the negligible contribution of HIM. We explored a grid of $(\gamma, L_{\min}, L_{\max})$ values and selected the combination that best reproduces the observed $N(\mathrm{H})$ distribution.

While the first algorithm is more robust, anchoring $L_i$ in direct distance observations, it nevertheless has two substantial disadvantages for the $\mathrm{OH^+}$ and $\mathrm{H_2O^+}$ data: (a) the observed sample contains a factor of 3 less data points, and (b) the resulting integrated column density ratios are often limits (when at least 1 velocity interval is a limit in either trace or hydrogen column). 

\begin{figure}[]
    \subfigure[$\mathrm{H_3^+}$ distance-to-source] {\label{fig:H3p_DistanceToSource}\includegraphics[scale=0.44]{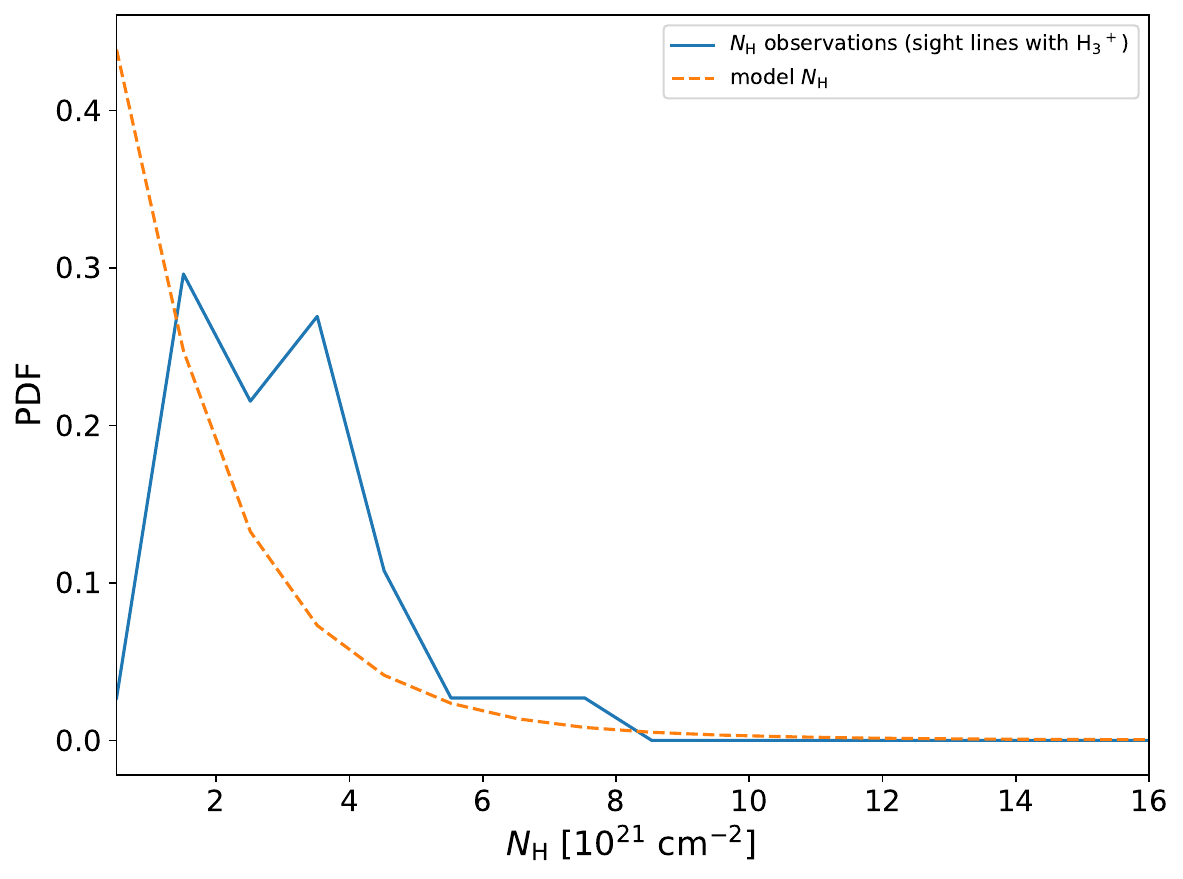}}
 	\subfigure[$\mathrm{OH^+}$ and $\mathrm{H_2O^+}$ distance-to-source] {\label{fig:OHp_H2Op_DistanceToSource}\includegraphics[scale=0.44]{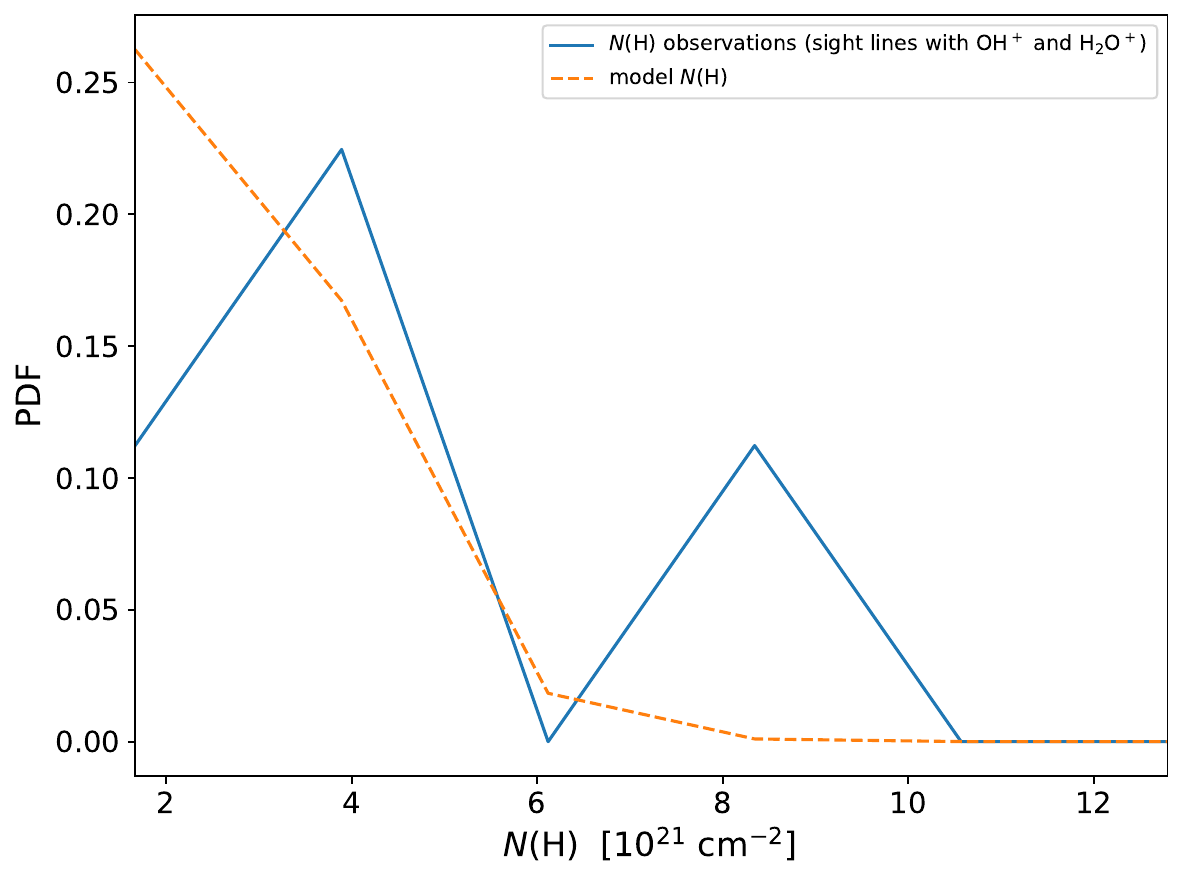}}
    \subfigure[$\mathrm{OH^+}$ and $\mathrm{H_2O^+}$ power-law ($\gamma=0.75$, $L_\mathrm{min}=50$ pc, $L_\mathrm{max}=1600$ pc)] {\label{fig:OHp_H2Op_PowerLaw}\includegraphics[scale=0.44]{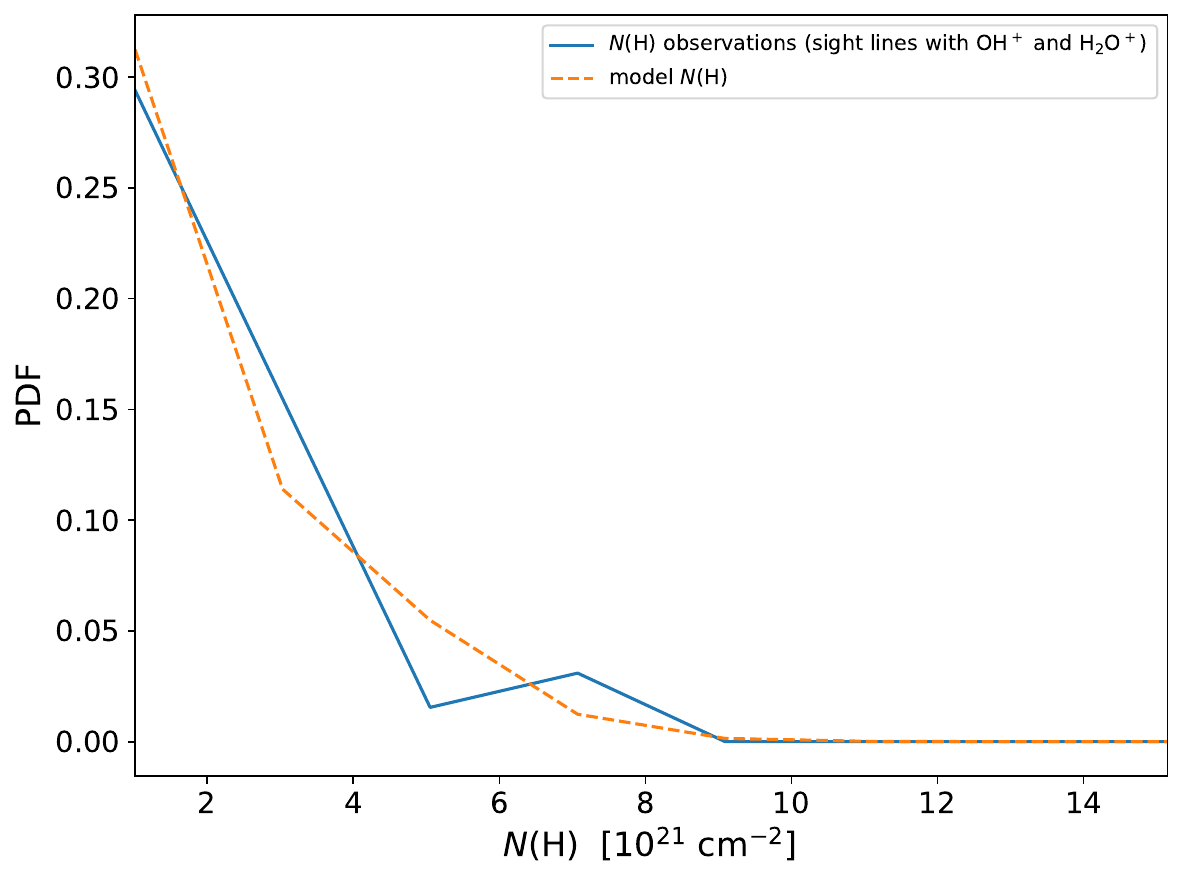}}

    \caption{LOS reconstruction algorithms distance-to-source (panels (a) and (b)) and velocity-intervals-power-law-distribution (panel (c)) are validated by comparing column densities of atomic hydrogen N(H) between model and observations.}
    \label{fig:LOS_Reconstruction_Validation}
\end{figure}

Given the strengths and weaknesses of both algorithms, we henceforth present the reader with both options. Figure \ref{fig:LOS_Reconstruction_Validation} shows a very good agreement between synthesized and observed atomic hydrogen column density probability distribution function (PDF) histograms, for both algorithms. The most notable deviation is however in the case of the distance-to-source algorithm for the integrated velocity intervals of $\mathrm{OH^+}$ and $\mathrm{H_2O^+}$, probably caused due to small number statistics. As indicated in Section \ref{S:Observations}, out of 14 distinct sources, only 7 are within the limit distance of 3 kpc, which decreases the reliability of this histogram. Figure \ref{fig:OHp_H2Op_PowerLaw} shows our second algorithm with the power-law width distribution interpretation, for the individual velocity intervals of $\mathrm{OH^+}$ and $\mathrm{H_2O^+}$. We have tried (by trial and error) several sets of $\gamma$, $L_\mathrm{min}$ and $L_\mathrm{max}$, and the parameter combination we eventually found as giving the closest match with observations is $\gamma=0.75$, $L_\mathrm{min}=50$ pc, $L_\mathrm{max}=1600$ pc.

\section{Results}\label{S:Results}
Our central question is whether the observed scatter in $\mathrm{OH^+}$,
$\mathrm{H_2O^+}$ and H$_3^+$ column densities can be naturally reproduced by the density, temperature and shielding fluctuations of a turbulent multiphase medium. The trace-ion abundances depend on the background fields $x(\mathrm{H_2})$ and $x(\mathrm{e})$, which in a turbulent medium need not equal their local equilibrium values, so we organize the results in two steps. In Section \ref{SS:Equilibrium} we identify where and why $x(\mathrm{H_2})$ and $x(\mathrm{e})$ depart from
equilibrium. As we will show, the strongest departures live in the UNM for $x(\mathrm{H_2})$ and in the WNM for $x(\mathrm{e})$. While both are intrinsically interesting for ISM chemistry, mostly the $x(\mathrm{H_2})$ departure materially affects the trace ions studied here, since $\mathrm{OH^+}$, $\mathrm{H_2O^+}$ and H$_3^+$ form predominantly in UNM and CNM gas where $x(\mathrm{e})$ closely tracks equilibrium. In Section \ref{SS:TraceResponse} we then compute the trace-ion to hydrogen column density ratios and compare them to the observational sample.

\subsection{Revisiting equilibrium in turbulent interstellar media}\label{SS:Equilibrium}
Classic treatment of ISM conditions often entails the assumption of equilibrium, under the supposition that the local microphysical processes in the ISM act much faster than the large scale dynamical evolution, especially in dense regions. While fully time-dependent MHD simulations are computationally expensive, the assumption of equilibrium simplifies matters. It allows to decouple the MHD simulation from the chemical processes, solving the former first, and then compute the chemical state based on straightforward analytic or semi-analytic post-processing calculations. Although convenient, in this section we show that in practice, when calculating the abundances of H$_2$ and e, the assumption of equilibrium is often incorrect, and skews our perception of ISM properties. Instead, we find that it is important to evolve the H$_2$ (and e) chemistry and hydrodynamics together. With this in mind, throughout Sections \ref{SS:Equilibrium} and \ref{SS:TraceResponse}, the time-dependent calculation is our physically faithful prediction, and the equilibrium calculation is presented purely as a benchmark, to isolate the impact of dropping the equilibrium assumption.
% it is important to evolve the H$_2$ (and e) chemistry and hydrodynamics together. 

As outlined in Section \ref{SS:Chemistry}, the equilibrium state in neutral ISM is largely controlled by molecular hydrogen and electrons, since once these species are fixed, the trace molecules reach equilibrium rapidly. The time-dependent abundances of H, H$^+$, H$_2$ and e are computed self-consistently by \textit{RAMSES}. To quantify the effect of the equilibrium assumption, we additionally solve the same chemical rate equations (\textit{RAMSES}'s chemical network; Table \ref{tab:reactions_RAMSES}) in post-processing under chemical equilibrium, using the density and temperature output by \textit{RAMSES} as input. 

For H$_2$, we make the simplifying assumption that $x(\mathrm{H^+})<<1$, a highly judicious assumption in neutral ISM essentially across all thermal phases, and especially in the relatively dense regions where H$_2$ forms. Under equilibrium, we can equate the rate of H$_2$ formation ($\raterefh{R}$, per encounter) and the rate of H$_2$ destruction ($\raterefh{D}$, per particle) such that $\raterefh{R} n_\mathrm{H} n(\mathrm{H}) = \raterefh{D} n(\mathrm{H_2})$ \footnote{H$_2$ is also destroyed by cosmic rays, so in principle $\raterefh{D}$ should include a cosmic-ray destruction term. However, this process is significant only in dense, shielded gas, which is already predominantly molecular ($x({\rm H_2}) \approx 0.5$), affecting only the residual atomic H abundance \citep{GoldsmithEtAl-2007, SternbergEtAl-2024}. We therefore neglect it both in the current implementation of time-dependent chemistry in \textit{RAMSES} and in our post-process calculation.}. If $x(\mathrm{H^+})<<1$, it follows that $n(\mathrm{H}) \simeq \left( n_\mathrm{H} - 2n(\mathrm{H_2}) \right)$, and we get
\begin{equation}\label{eq:n_H2}
\begin{aligned}
x(\mathrm{H_2}) = \left( 2 + \raterefh{D}/(n_\mathrm{H} \raterefh{R}) \right)^{-1}.
\end{aligned}
\end{equation}

Inside CNM, the dissociation rate satisfies $\raterefh{D}<<(n_\mathrm{H} \raterefh{R})$, and in that limit all hydrogen is molecular ($x$(H$_2$)=1/2).

Unlike H$_2$, the equilibrium fractional abundance of electrons cannot be obtained through a simple analytic expression. As previously mentioned, $x$(e) = $x$(C$^+$) + $x$(H$^+$). Therefore, we obtain $x$(e) in equilibrium by calculating the equilibrium value of $x$(H$^+$), solving the chemical network implemented in \textit{RAMSES}, shown in Table \ref{tab:reactions_RAMSES}. The equilibrium solution is computed numerically, in post-processing, by finding the common root of the coupled nonlinear steady-state equations. I.e., at each ($n_\mathrm{H}$, $T$, $\zetatref{H}$, e$^{-\tau}$), we solve a corresponding system for ($x$(H), $x$(H$^+$), $x$(PAH$^-$), $x$(PAH$^0$), $x$(PAH$^+$)) by enforcing steady state (net formation/destruction rates vanish) and hydrogen/PAH number conservation.

\begin{figure*}[]
    \subfigure[$x$(H$_2$)] {\label{fig:EqVsNonEq_H2}\includegraphics[scale=0.46]{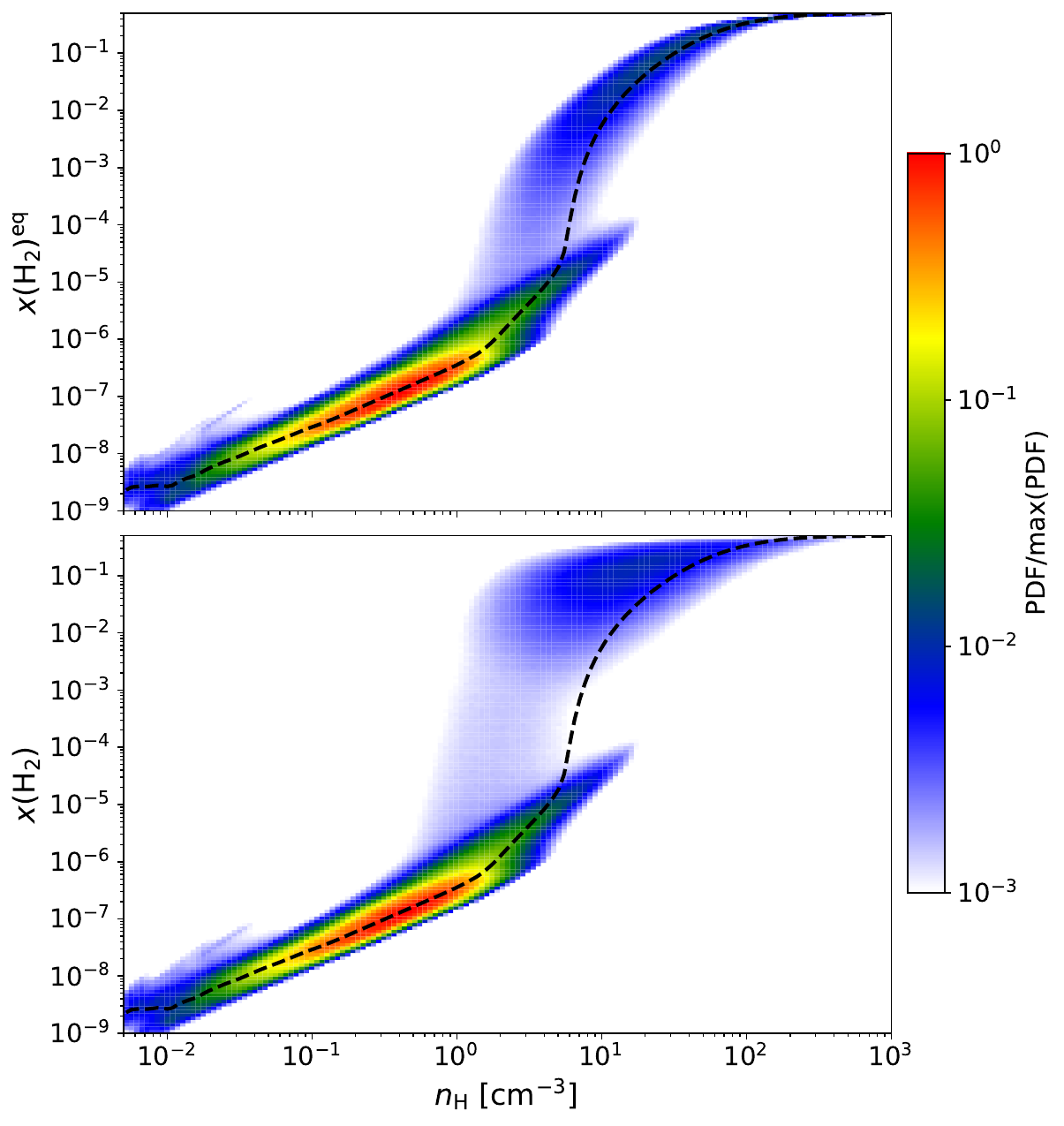}}
    \subfigure[$x$(e)] {\label{fig:EqVsNonEq_e}\includegraphics[scale=0.46]{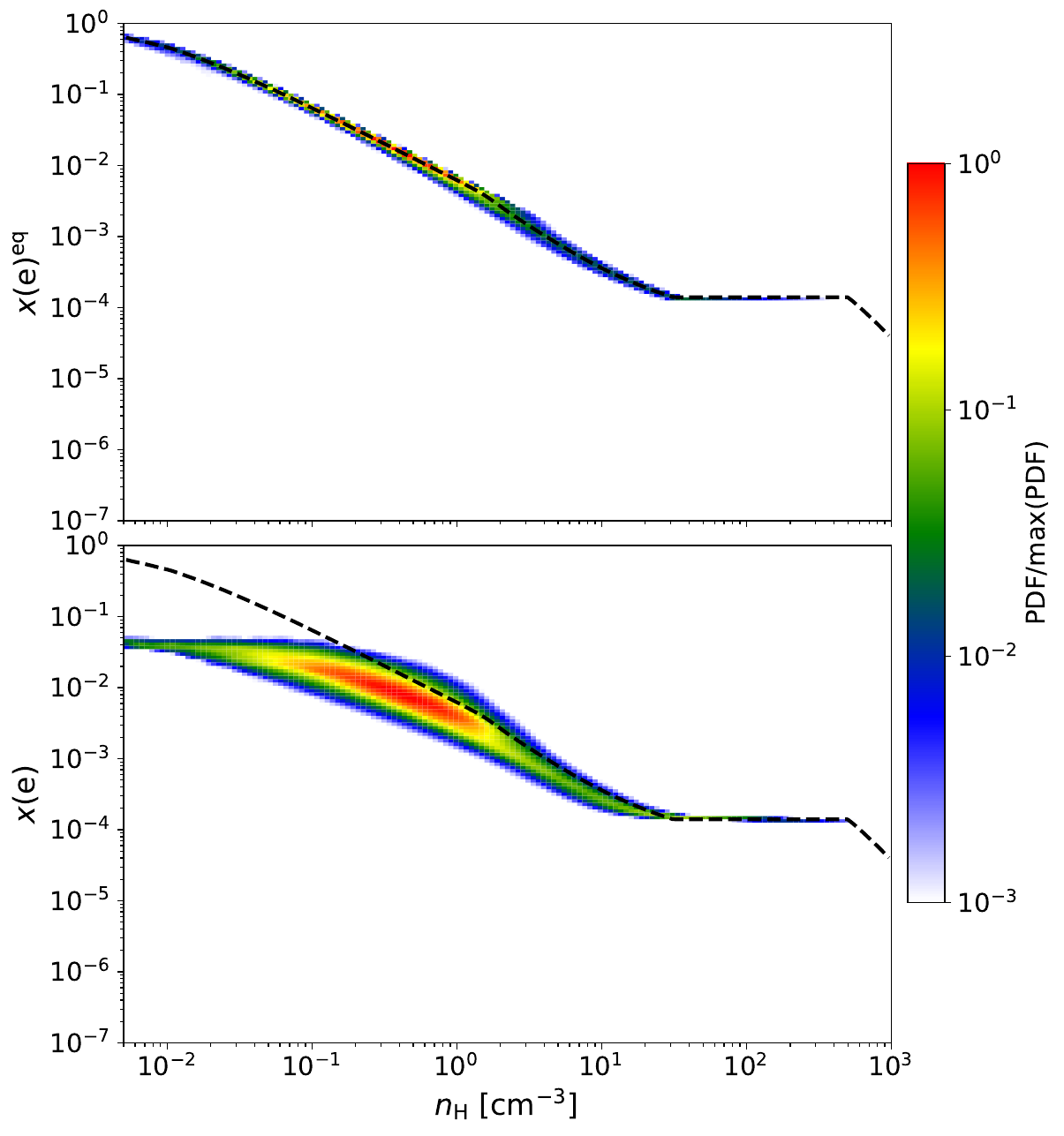}}
    \caption{2D volume-weighted histograms of the molecular hydrogen
    (panel (a)) and electron fractional (panel (b)) abundances versus
    total density, comparing the equilibrium benchmark (top) with the
    time-dependent calculation (bottom). The equilibrium median is shown
    in both.}
    \label{fig:EqVsNonEq_H2_and_e}
\end{figure*}

Figure \ref{fig:EqVsNonEq_H2_and_e} shows the volume-weighted abundances of H$_2$ and e as functions of the total density $n_\mathrm{H}$, for the equilibrium calculation (top) and the time-dependent abundances obtained directly from RAMSES (bottom). The dashed curves in all panels show the
equilibrium median abundances versus $n_\mathrm{H}$. Comparing these curves with the time-dependent distributions reveals where and by how much the two realizations diverge. Across most of the density range, the time-dependent abundances closely track the equilibrium trend, but in specific regimes they depart by several orders of magnitude. Specifically, (a) the time-dependent H$_2$ abundance is significantly higher than equilibrium at $n_\mathrm{H} \sim 10^0$--$10^1$ cm$^{-3}$, corresponding to the UNM regime, and (b) the time-dependent electron abundance is markedly lower than equilibrium at $n_\mathrm{H} \lesssim 10^{-1}$ cm$^{-3}$, corresponding to the diffuse WNM. In both cases the scatter is also enhanced in these regions, so the time-dependent abundances span a noticeably wider range of values than their equilibrium counterparts.

The origin of these departures can be understood by comparing two timescales: the chemical relaxation timescale $\tau_\mathrm{chem}$, on which the gas would reach local equilibrium, and the turbulent mixing timescale $\tau_\mathrm{turb}$, on which gas parcels are advected between phases. When $\tau_\mathrm{chem} \ll \tau_\mathrm{turb}$, chemistry tracks equilibrium. When $\tau_\mathrm{chem} \gtrsim \tau_\mathrm{turb}$, advection imprints the chemical state of the source phase onto the destination phase, driving the gas out of equilibrium.

The characteristic turbulent timescale is the turn-over time. For our entire simulation, this timescale is $\sim 0.5\,L/\sigma_v \approx 10$ Myr, with a  velocity dispersion $\sigma_v = 8$ km s$^{-1}$ and box size $L = 200$ pc. However, the chemical and turbulent timescales should be compared at the scale on which turbulent mixing actually takes place, which ranges from entire
CNM clouds down to thin sheet-like CNM boundary interfaces. In our simulations these mixing scales lie in the range $\sim 0.1$--$10$ pc. Since the turbulent timescale scales with size as $l^{1/2}$ \citep{McKeeOstriker-2007}, we obtain $\tau_\mathrm{turb} \approx 0.3$--$3$ Myr at these scales.

For electrons, the chemical timescale is set by the balance between cosmic-ray ionization and recombination, $\tau_\mathrm{chem} \sim x(\mathrm{e})/\zetatref{H}$. In the CNM, recombination is efficient, $x(\mathrm{e})$ is small, and this timescale is short compared to $\tau_\mathrm{turb}$. Therefore, $x(\mathrm{e})$ remains close to its equilibrium value. In the WNM, the equilibrium ionization fraction is $x(\mathrm{e}) \sim$ few $\times
10^{-2}$, set by cosmic-ray ionization of H balanced by radiative recombination, giving $\tau_\mathrm{chem} \sim 10$ Myr (see Table \ref{tab:reactions_RAMSES}), substantially longer than $\tau_\mathrm{turb}$. Gas parcels are therefore advected from the CNM, where $x(\mathrm{e})$ is small, into the WNM faster than they can be re-ionized to the local equilibrium value, and the WNM electron abundance remains under-ionized relative to equilibrium, as seen in Figure \ref{fig:EqVsNonEq_e}. The process is not symmetric. Although turbulence can also advect WNM parcels into the CNM, at CNM densities the recombination timescale is comparable to (or shorter than) $\tau_\mathrm{turb}$, so the over-ionized parcels quickly relax toward local equilibrium. CNM electrons therefore remain close to chemical equilibrium, whereas WNM electrons are driven away from it.

\begin{figure*}[]
	\begin{center}
		\includegraphics[scale=0.49]{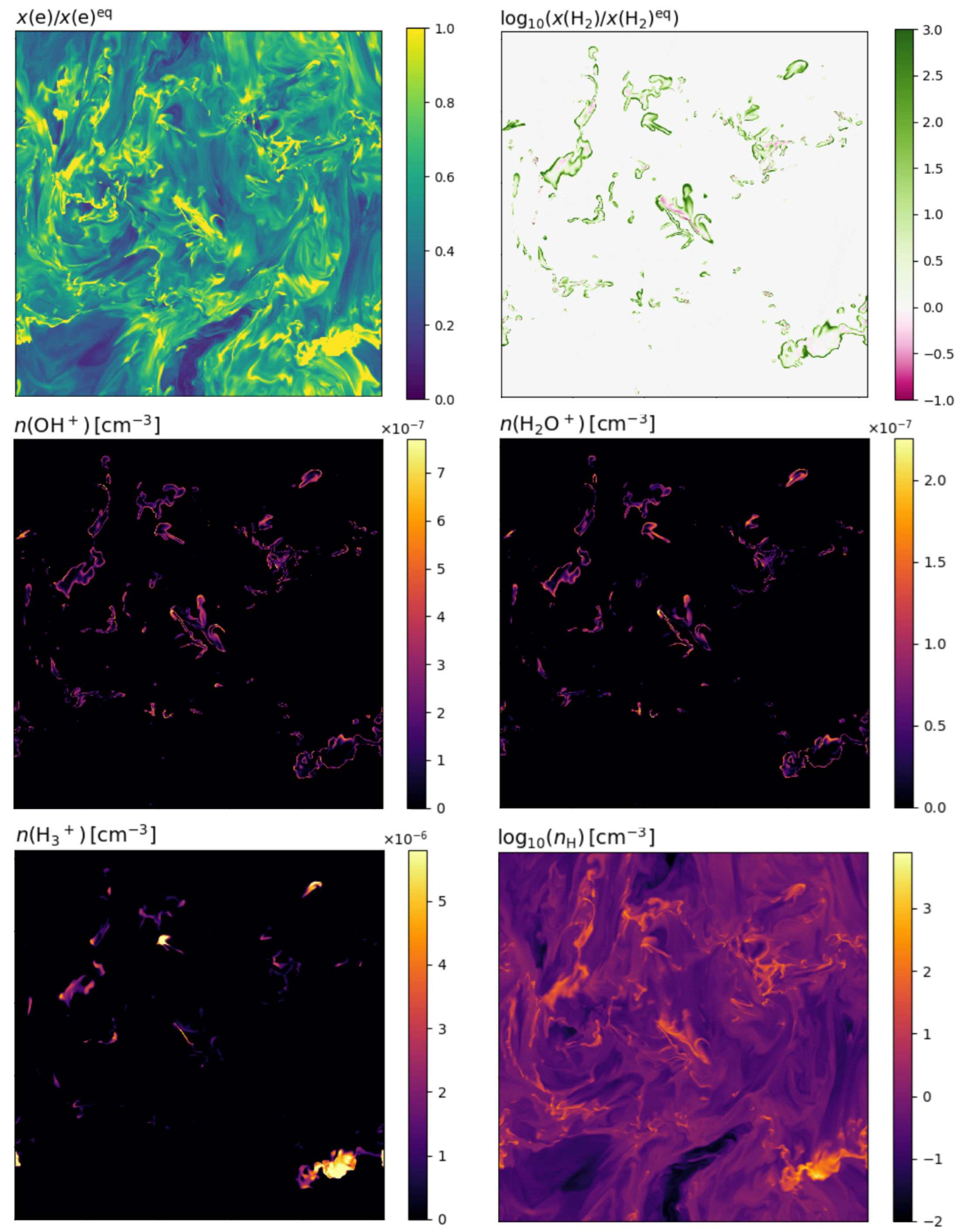}
		\caption{A 2D slice through the middle 2 the 3D simulation box (resolution $K=1024$). The top panels show the ratio between the equilibrium and time-dependent $x$(H$_2$) and $x$(e), respectively, the middle and bottom panels show the trace molecule density and the total density, the latter providing context for interpreting the various ISM phases.} \label{fig:spatial_distribution}
	\end{center}
\end{figure*}

The mechanism for H$_2$ is analogous: turbulence advects gas from the
CNM into more diffuse phases faster than the local chemistry can
re-equilibrate, so the destination phase inherits the chemical state of
the source. The regime where this drives the strongest departures is,
however, different. For H$_2$ it is the UNM rather than the WNM, as
we now show. The H$_2$ chemical timescale is $\tau_\mathrm{chem} =
1/(\raterefh{D} + 2\raterefh{R}n_\mathrm{H})$ \citep{BialyEtAl-2019},
set by the faster of the dissociation and formation rates. In the CNM,
strong shielding suppresses $\raterefh{D}$, formation dominates, and
$\tau_\mathrm{chem} \simeq 1/(2\raterefh{R}n_\mathrm{H})$. With
$\raterefh{R} \approx 3 \times 10^{-17}$ cm$^3$ s$^{-1}$, this gives
$\tau_\mathrm{chem} \approx 0.5$--$5$ Myr at $n_\mathrm{H} \approx
10^3$--$10^2$ cm$^{-3}$, comparable to $\tau_\mathrm{turb}$ at the
mixing scales of CNM clouds. The CNM is therefore close to equilibrium with $x(\mathrm{H_2})$ a few tens of percent below the asymptotic equilibrium value. In the diffuse WNM, by contrast, self-shielding is ineffective and the
photodissociation rate $\raterefh{D} \simeq D_0 \approx 3.3 \times 10^{-11}$ s$^{-1}$ (the free-space rate) dominates, giving $\tau_\mathrm{chem} \simeq 1/D_0 \sim 10^3$ yr, vastly shorter than $\tau_\mathrm{turb}$. The WNM therefore tracks equilibrium. The strongest departures occur in the UNM. When a CNM parcel is advected into the UNM, it carries a large initial H$_2$ fraction that provides strong self-shielding (combined with partial and typically very minor geometrical shielding from the surrounding CNM), strongly suppressing $\raterefh{D}$ through a small attenuation factor $f_\mathrm{att} \lesssim 10^{-4}$ and giving $\tau_\mathrm{diss} \sim 6$ Myr, longer than $\tau_\mathrm{turb}$. The enhanced H$_2$ fraction therefore persists out of equilibrium as the parcel mixes into the more diffuse medium. Turbulence can also drive gas parcels from the WNM into the UNM, but the WNM H$_2$ fraction is negligible to begin with, so the UNM H$_2$ budget is completely dominated by CNM-advected parcels.

Figure \ref{fig:spatial_distribution} shows the spatial distribution of these departures in a 2D slice of the simulation box: the equilibrium-to-time-dependent ratio for H$_2$ and e (top panels), the trace molecule densities (further explored in Section \ref{SS:TraceResponse}), and the total density providing phase context (bottom right). The H$_2$ over-abundance forms a clearly visible boundary layer wrapping each CNM cloud, with ratios reaching several orders of magnitude (capped at $10^3$ for color-scheme clarity). The WNM electron under-abundance is similarly evident, while the CNM is nearly indistinguishable from equilibrium\footnote{Interactive figures allowing an exploration of slices vs height inside the simulation box is available in the following \href{https://github.com/UriMalamud/OUT_OF_EQUILIBRIUM_ISM/tree/main}{GitHub repo}}.

In summary, Figure \ref{fig:EqVsNonEq_H2_and_e} shows that the equilibrium assumption breaks down most strongly in the UNM for H$_2$ and in the WNM for electrons, where the chemical relaxation timescale greatly exceeds the turbulent mixing time and the abundances depart from equilibrium by up to several orders of magnitude. The CNM H$_2$ abundance carries a milder, tens-of-percent offset, reflecting the marginal regime in which $\tau_\mathrm{chem} \approx \tau_\mathrm{turb}$.

While both departures are intrinsically interesting for ISM chemistry, their relevance to the trace molecules studied in the next section is asymmetric. $\mathrm{OH^+}$, $\mathrm{H_2O^+}$, and H$_3^+$ form predominantly in the UNM and CNM, where electron abundances closely track equilibrium. The strong WNM $x(\mathrm{e})$ departure therefore has little impact on their column densities, whereas the UNM $x(\mathrm{H_2})$ enhancement directly boosts their formation. The remainder of this paper therefore focuses on the role of time-dependent $x(\mathrm{H_2})$, with $x(\mathrm{e})$ evolved self-consistently in the simulation but expected to matter primarily for tracers formed in more diffuse gas. With this in mind, the equilibrium-vs-time-dependent comparisons in the next section are predominantly a probe of the H$_2$ departures.

\subsection{Response of trace molecules to equilibrium departures}\label{SS:TraceResponse}

\begin{figure*}[]
    \subfigure[Equilibrium, distance-to-source] {\label{fig:2D_N_OHp_ratio_eq-N_H-DS}\includegraphics[scale=0.46]{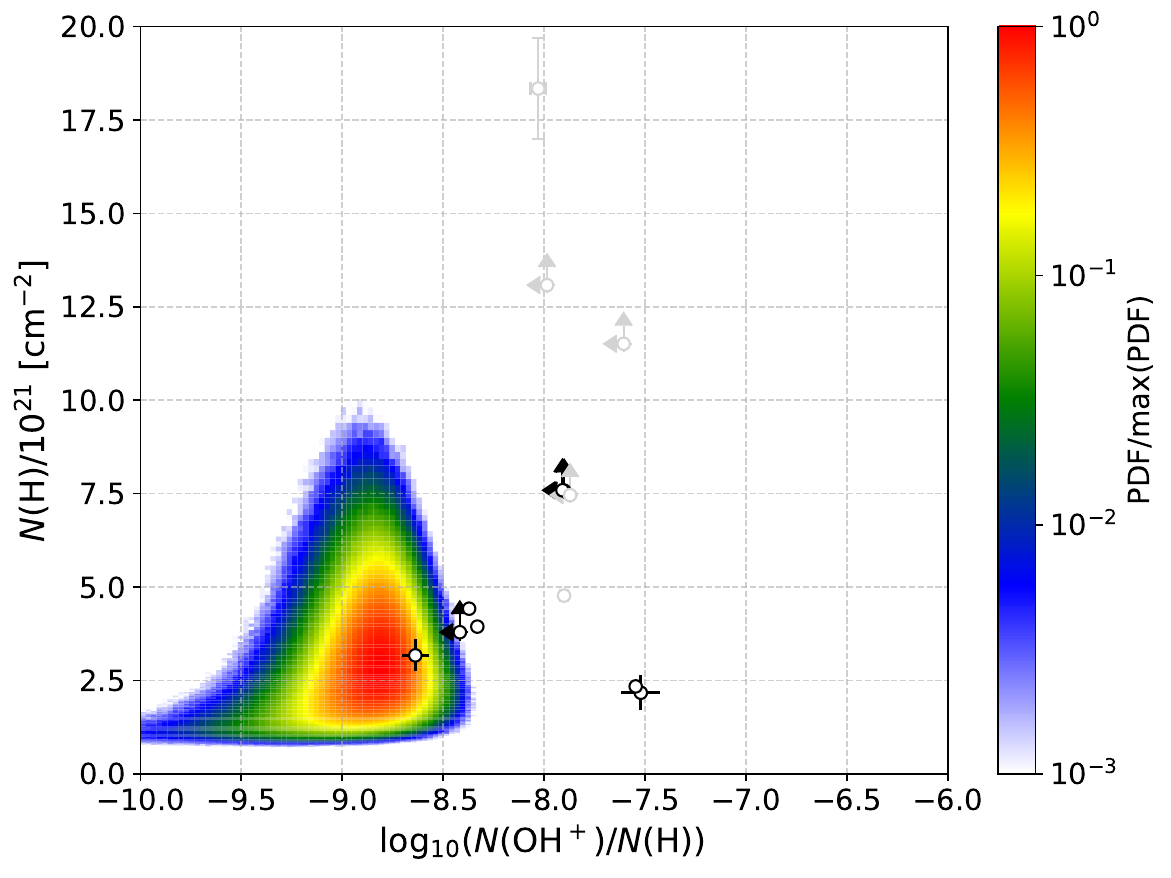}}
    \subfigure[Equilibrium, velocity-intervals] {\label{fig:2D_N_OHp_ratio_eq-N_H-VI}\includegraphics[scale=0.46]{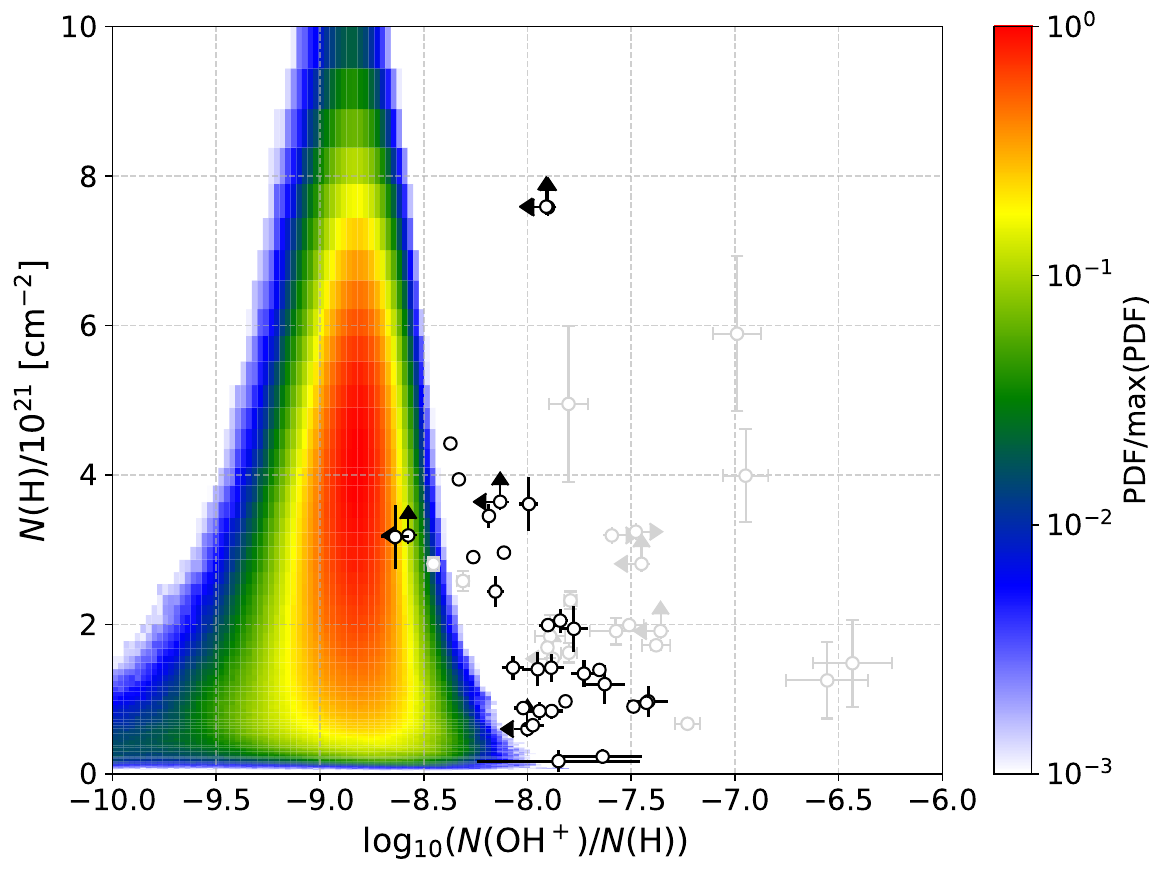}}
    \subfigure[Time-dependent, distance-to-source] {\label{2D_N_OHp_ratio-N_H-DS}\includegraphics[scale=0.46]{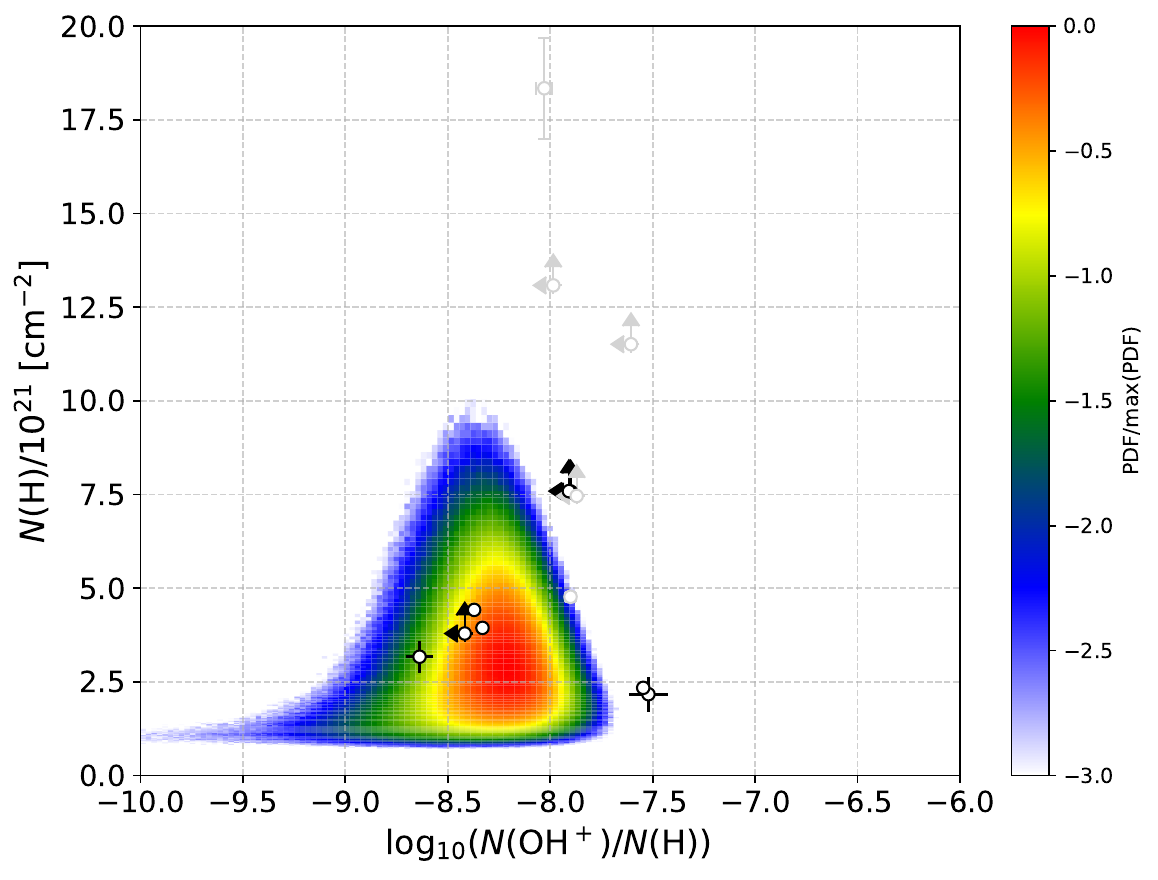}}
 	\subfigure[Time-dependent, velocity-intervals] {\label{fig:2D_N_OHp_ratio-N_H-VI}\includegraphics[scale=0.46]{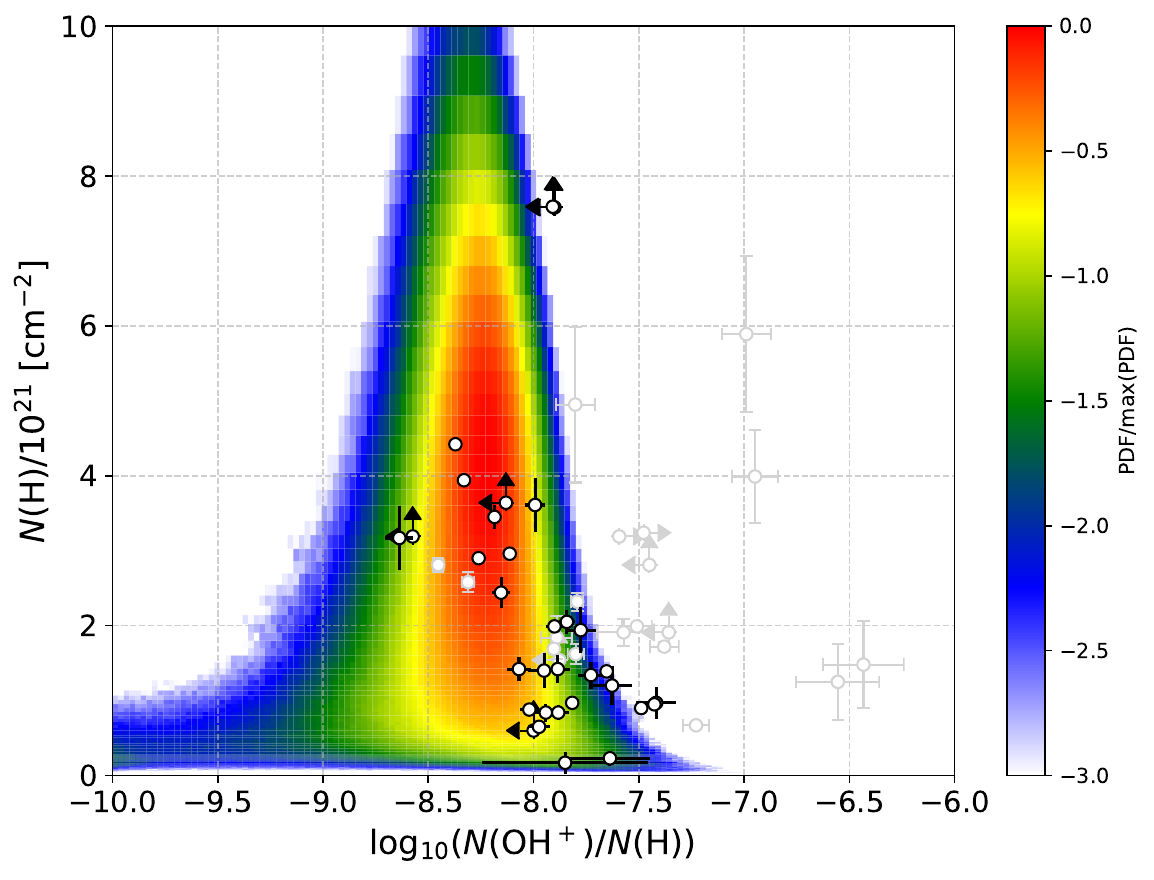}}

    \caption{$\mathrm{OH^+}$ comparison between observed and simulation-synthesized column density distributions, showing $N(\mathrm{H})$ as a function of log$_{10}$($N$($\mathrm{OH^+}$)/$N$(H)). Circles denote integrated values toward observed \textbf{sources} (left plots), or associated with \textbf{velocity intervals} (right plots), grayed if their heliocentric distances exceeds 3 kpc. 
    % The simulation-induced statistics compare equilibrium (top plots) and Time-dependent (bottom plots) realizations.
    The simulation-induced statistics compare the equilibrium benchmark (top plots) and the time-dependent calculation (bottom plots).
    }
    \label{fig:OHp_column_densities}
\end{figure*}

\begin{figure*}[]
    \subfigure[Equilibrium, distance-to-source] {\label{fig:2D_N_H2Op_ratio_eq-N_H-DS}\includegraphics[scale=0.46]{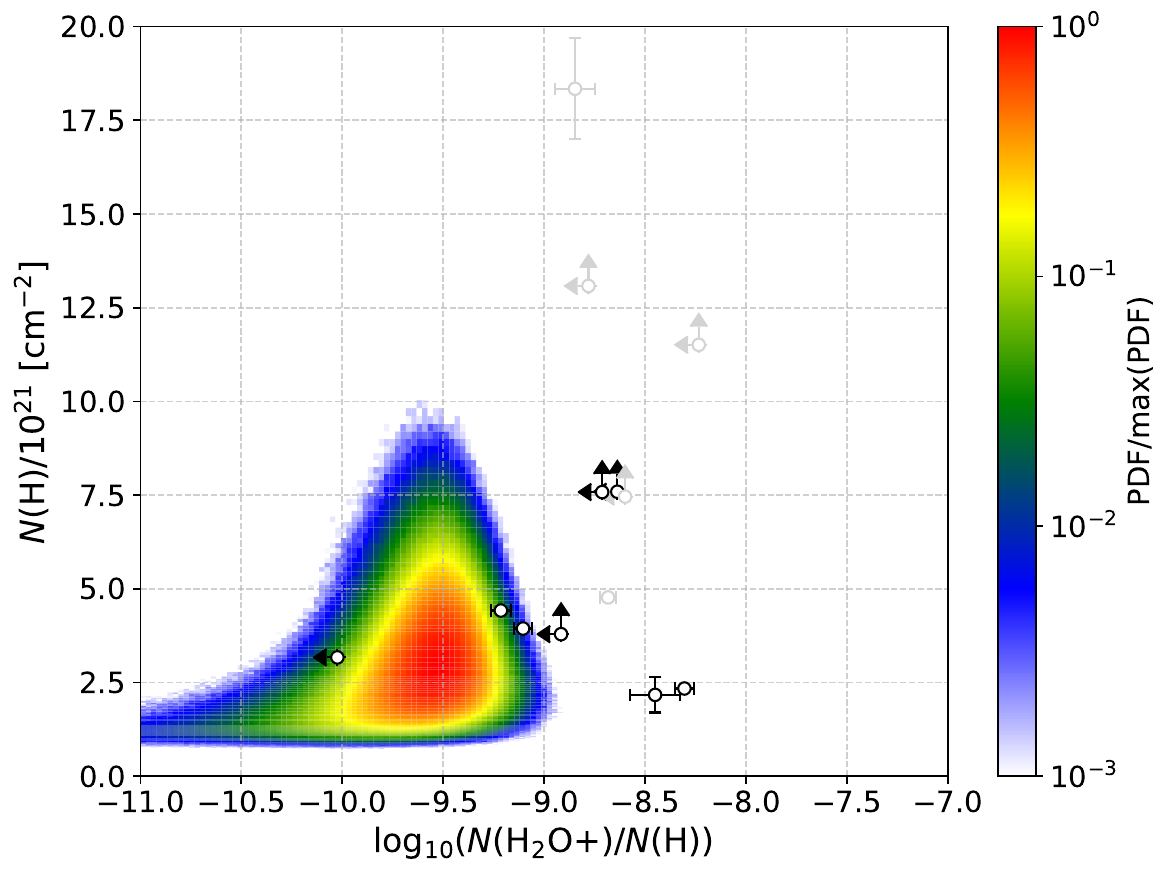}}
    \subfigure[Equilibrium, velocity-intervals] {\label{fig:2D_N_H2Op_ratio_eq-N_H-VI}\includegraphics[scale=0.46]{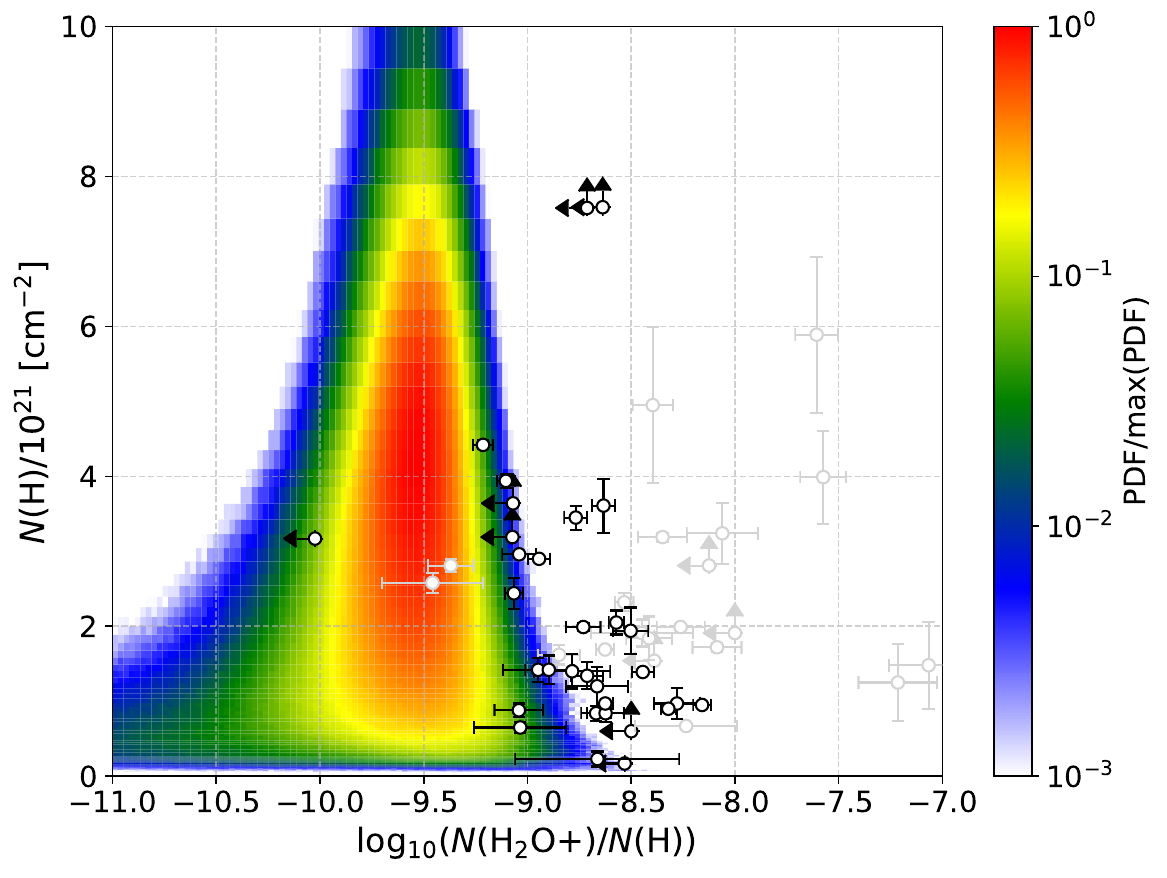}}
    \subfigure[Time-dependent, distance-to-source] {\label{fig:2D_N_H2Op_ratio-N_H-DS}\includegraphics[scale=0.46]{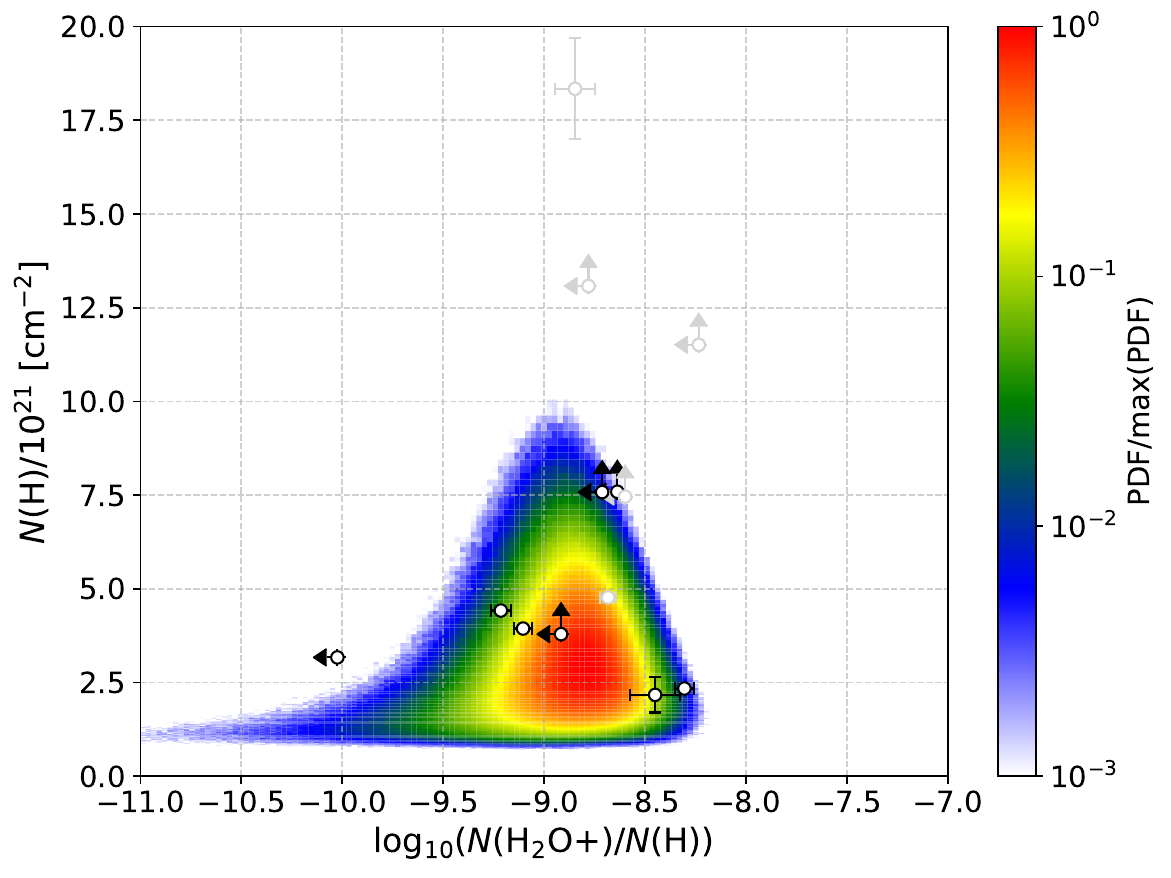}}
 	\subfigure[Time-dependent, velocity-intervals] {\label{fig:2D_N_H2Op_ratio-N_H-VI}\includegraphics[scale=0.46]{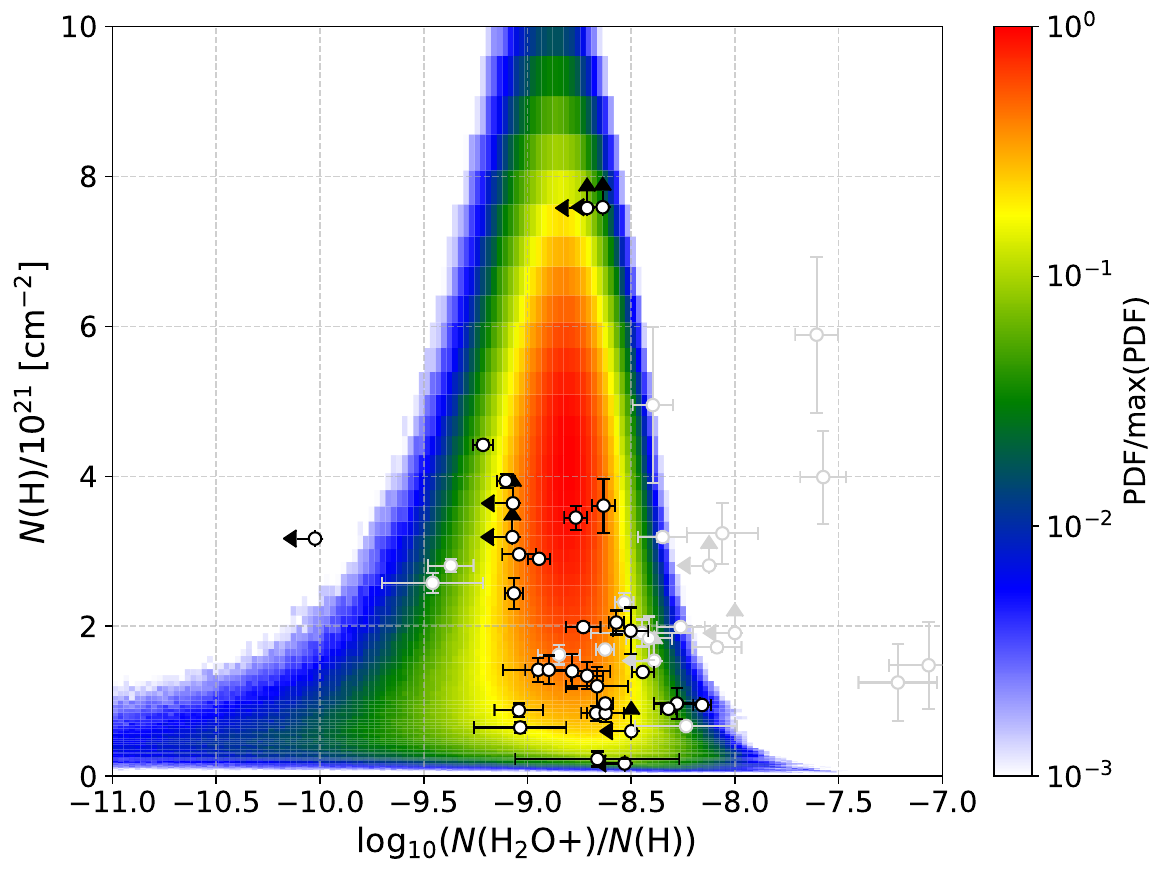}}

    \caption{Same as Figure \ref{fig:OHp_column_densities} but for $\mathrm{H_2O^+}$.}
    \label{fig:H2Op_column_densities}
\end{figure*}

\begin{figure}[]
    \subfigure[Equilibrium, distance-to-source] {\label{fig:2D_N_H3p_ratio_eq-NH-DS}\includegraphics[scale=0.46]{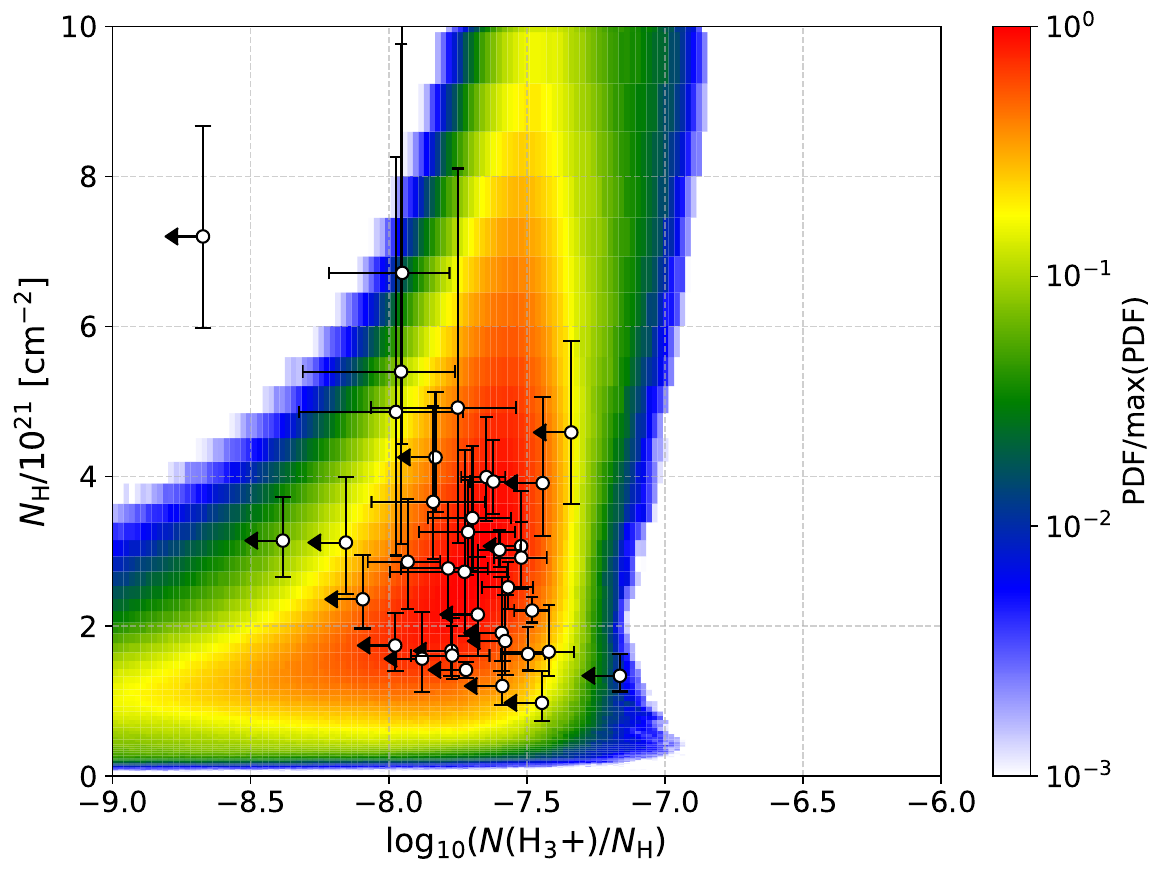}}
    \subfigure[Time-dependent, distance-to-source] {\label{fig:2D_N_H3p_ratio-NH-DS}\includegraphics[scale=0.46]{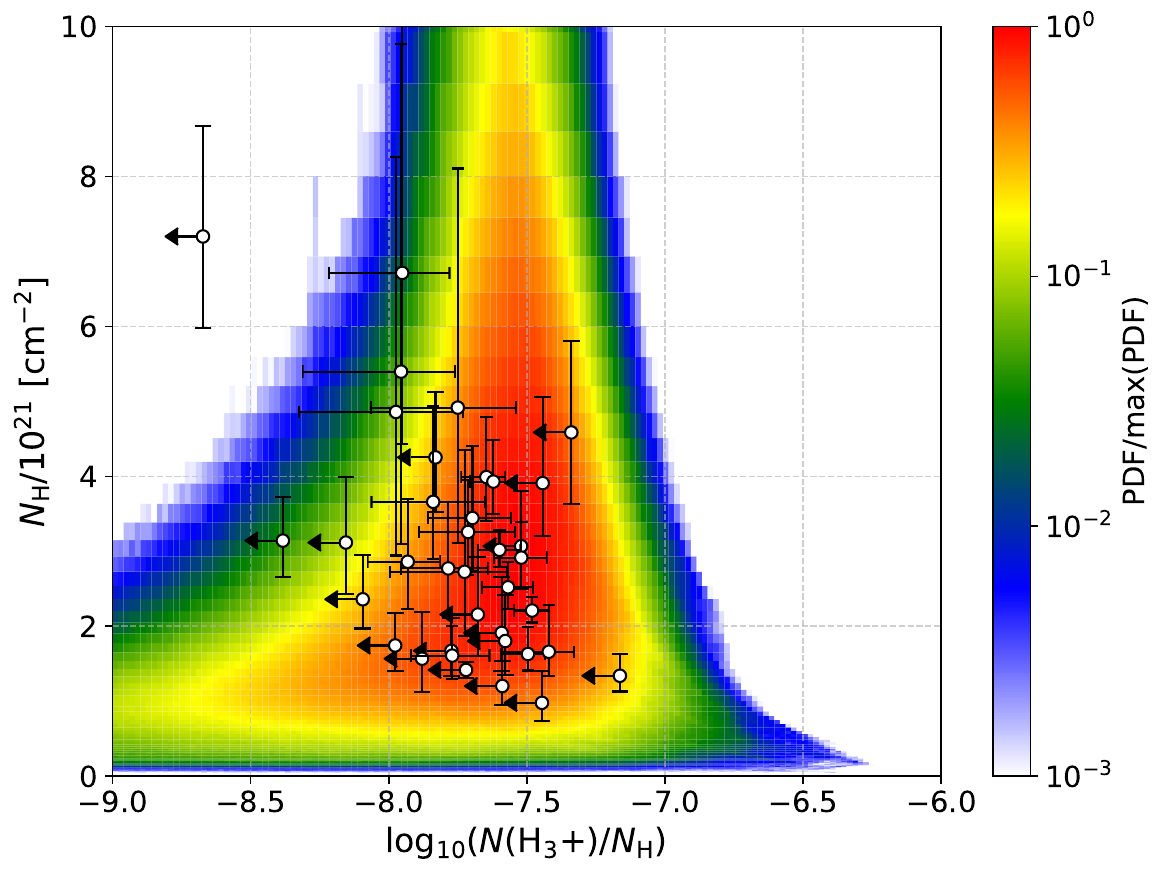}}

    \caption{Same as Figure \ref{fig:OHp_column_densities} but for $\mathrm{H_3^+}$ (observed data includes only integrated values towards the sources).}
    \label{fig:H3p_column_densities}
\end{figure}

Figures \ref{fig:OHp_column_densities}--\ref{fig:H3p_column_densities} show, for each trace species, the joint distribution of the trace-to-hydrogen column density relative abundance. The colored 2D map is the PDF predicted by our simulation (see \S\ref{SS:LOS_reconstruction}), with the time-dependent calculation in the bottom panels and the equilibrium benchmark in the top panels. Circles mark individual observations: black points are local sight lines ($d<3$ kpc), expected to probe conditions comparable to those in our simulation, while gray points are more distant sight lines that may sample regions with significantly different $G_0$ or $\zeta$. Each species is shown for both observational integrations, \textit{distance-to-source} (left panels) and \textit{velocity-intervals} (right panels).

A noteworthy caveat concerns the derivation of the atomic hydrogen columns. These values, following \cite{IndrioloEtAl-2015}, are compiled from H\,{\sc i} 21 cm data. In particular, H\,{\sc i} 21 cm absorption is weighted toward colder gas because the optical depth scales inversely with temperature, and therefore it is most sensitive to CNM material. We therefore interpret the tabulated $N(\mathrm{H})$ values as the best available atomic columns associated with the observed $\mathrm{OH^+}$ and $\mathrm{H_2O^+}$ velocity intervals, noting (see Appendix \ref{Appendix:D} for further discussion) that undetected warm H\,{\sc i} would shift the observed column ratios to somewhat lower values. For the H$_3^+$ comparison we use the total hydrogen column density, $N_{\rm H}$, reported in the corresponding H$_3^+$ studies. These values are primarily based on far-UV/Ly$\alpha$ and H$_2$ absorption measurements toward the background stars, rather than on H\,{\sc i} 21 cm absorption, and therefore are not subject to the same CNM-weighting bias.

\begin{figure}[]
	\begin{center}
		\includegraphics[scale=0.46]{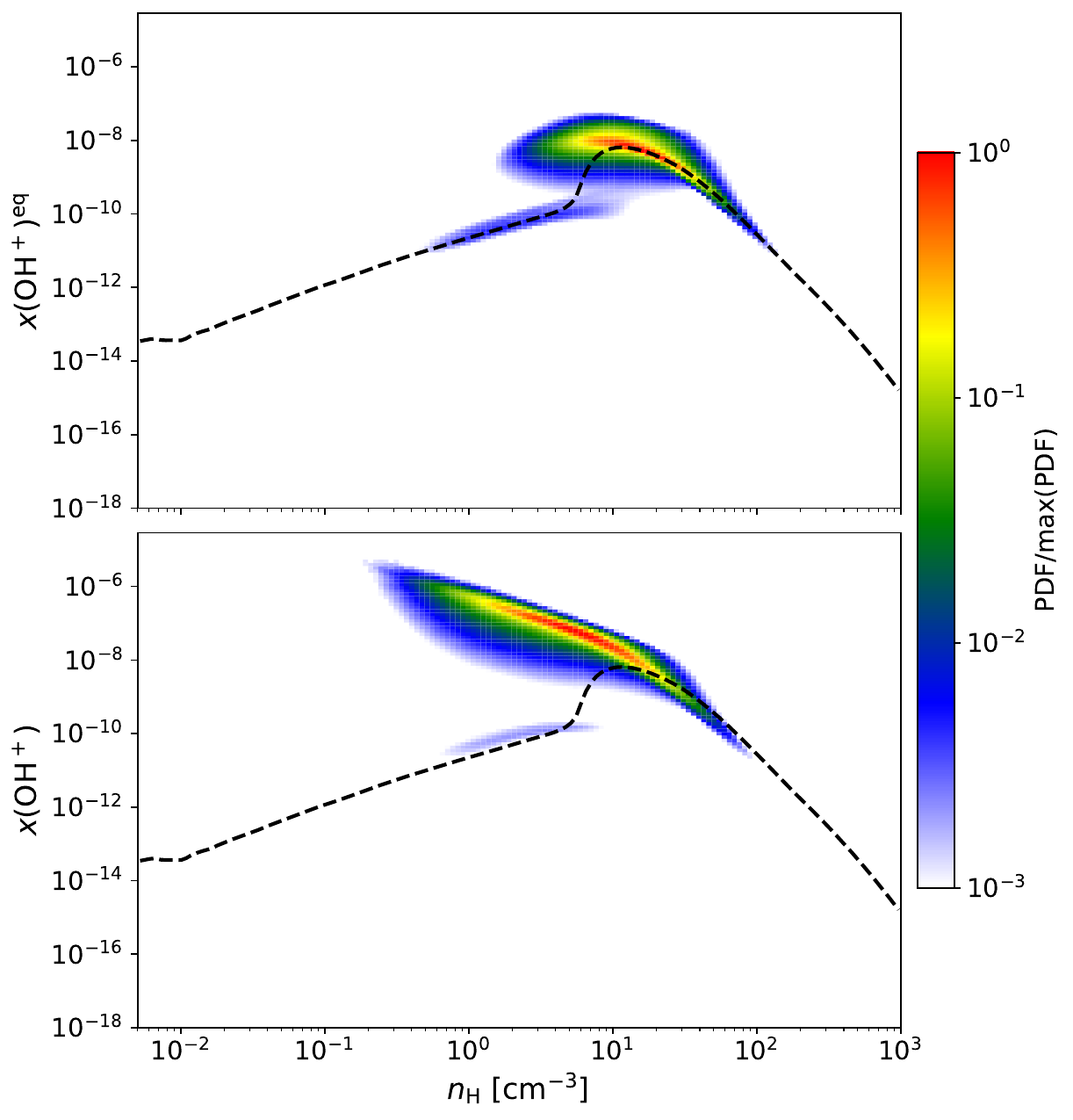}
		% \caption{$\mathrm{OH^+}$ fractional abundance as a function of total density, in (top) and out (bottom) of equilibrium, weighted by $n$($\mathrm{OH^+}$).}
        \caption{$\mathrm{OH^+}$ fractional abundance as a function of total density, for the equilibrium benchmark (top) and the time-dependent calculation (bottom), weighted by $n$($\mathrm{OH^+}$).}
		\label{fig:2D_nH-x_OHp_eq_vs_noneq-n_OHpWeighted}
	\end{center}
\end{figure}

\begin{figure}[]
	\begin{center}
		\includegraphics[scale=0.46]{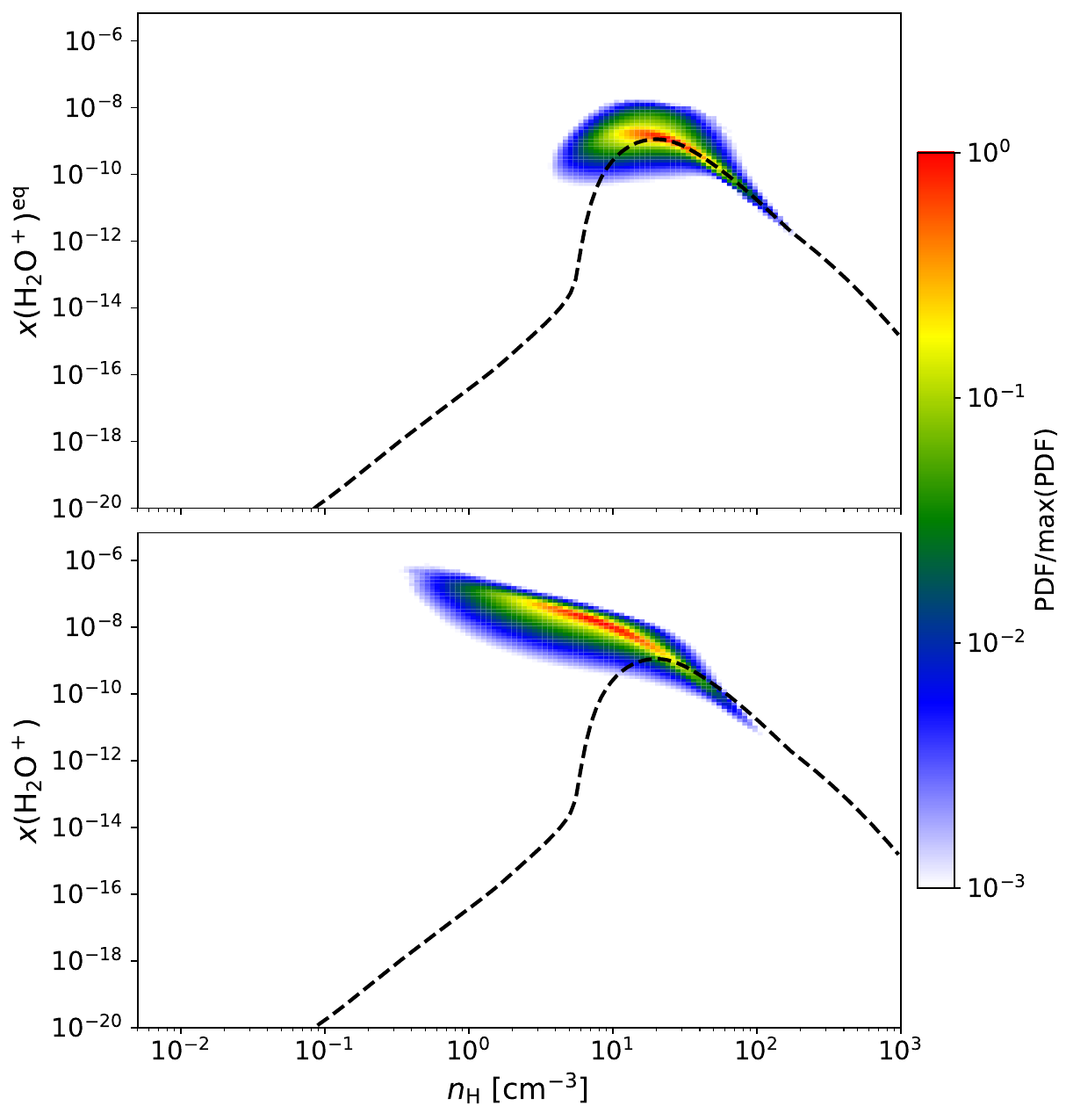}
		% \caption{$\mathrm{H_2O^+}$ fractional abundance as a function of total density, in (top) and out (bottom) of equilibrium, weighted by $n$($\mathrm{H_2O^+}$).}
        \caption{$\mathrm{H_2O^+}$ fractional abundance as a function of total density, for the equilibrium benchmark (top) and the time-dependent calculation (bottom), weighted by $n$($\mathrm{H_2O^+}$).}
		\label{fig:2D_nH-x_H2Op_eq_vs_noneq-n_H2OpWeighted}
	\end{center}
\end{figure}

\begin{figure}[]
	\begin{center}
		\includegraphics[scale=0.46]{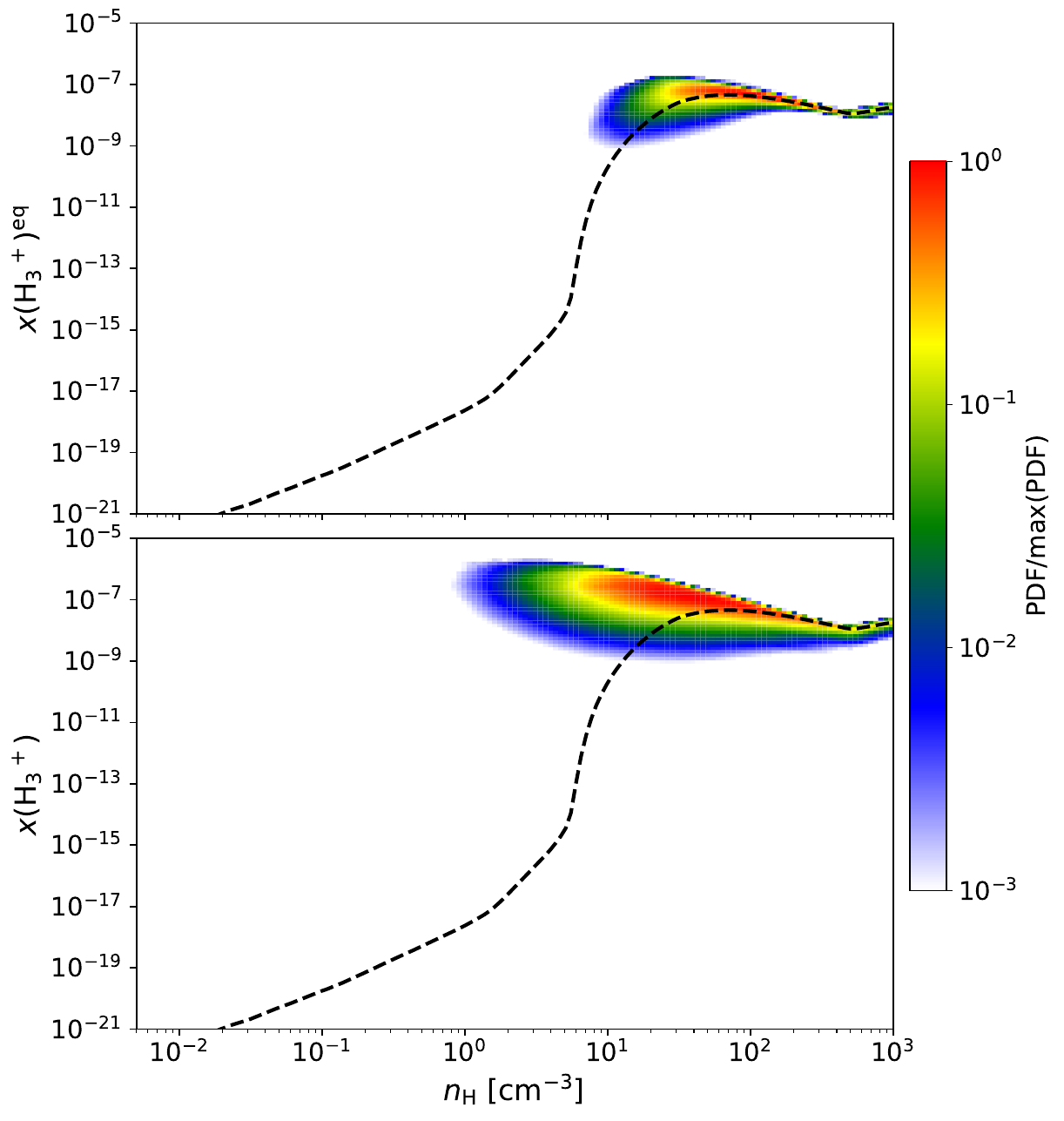}
		% \caption{$\mathrm{H_3^+}$ fractional abundance as a function of total density, in (top) and out (bottom) of equilibrium, weighted by $n$($\mathrm{H_3^+}$).}
        \caption{$\mathrm{H_3^+}$ fractional abundance as a function of total density, for the equilibrium benchmark (top) and the time-dependent calculation (bottom), weighted by $n$($\mathrm{H_3^+}$).}
		\label{fig:2D_nH-x_H3p_eq_vs_noneq-n_H3pWeighted}
	\end{center}
\end{figure}

Across all three species and both integration choices, four robust trends emerge. First, the time-dependent calculation systematically predicts higher trace abundances than the equilibrium benchmark, and with a wider intrinsic scatter. Second, the dominant mode of the time-dependent model PDF (the red--light-green region, hereafter the "1-dex band") encloses nearly all local observations, leaving only a handful of outliers in its low-probability tails. The 1-dex band is not arbitrary: as we will show in Figures \ref{fig:2D_nH-x_OHp_eq_vs_noneq-n_OHpWeighted}-\ref{fig:2D_nH-x_H3p_eq_vs_noneq-n_H3pWeighted}, it corresponds to the gas in which most of each species actually resides. Third, those outliers are almost exclusively distant sight lines ($d>3$ kpc, gray circles), consistent with our identification of the simulation as a local-ISM benchmark; in Figure \ref{fig:H3p_column_densities}, where all sources lie within this distance, no such outliers appear. Fourth, the equilibrium benchmark, by contrast, fails to enclose a large fraction of the observed points within its 1-dex band, particularly for $\mathrm{OH^+}$ and $\mathrm{H_2O^+}$.

% Across all three species and both integration choices, three robust trends emerge. First, the time-dependent calculation systematically predicts higher trace abundances than the equilibrium benchmark, and with a wider intrinsic scatter. Second, the dominant mode of the time-dependent model PDF (the red--light-green region, hereafter the "1-dex band") encloses nearly all local observations, leaving only a handful of outliers in its low-probability tails. The 1-dex band is not arbitrary: as we will show in Figures~\ref{fig:2D_nH-x_OHp_eq_vs_noneq-n_OHpWeighted}--\ref{fig:2D_nH-x_H3p_eq_vs_noneq-n_H3pWeighted}, it corresponds to the gas in which most of each species actually resides. Third, those outliers are almost exclusively distant sight lines ($d>3$ kpc, gray circles), consistent with our identification of the simulation as a local-ISM benchmark; in Figure~\ref{fig:H3p_column_densities}, where all sources lie within this distance, no such outliers appear. The equilibrium benchmark, by contrast, fails to enclose a large fraction of the observed points within its 1-dex band, particularly for $\mathrm{OH^+}$ and $\mathrm{H_2O^+}$.

The origin of these trends is made explicit in Figures~\ref{fig:2D_nH-x_OHp_eq_vs_noneq-n_OHpWeighted}-\ref{fig:2D_nH-x_H3p_eq_vs_noneq-n_H3pWeighted}, which show 2D PDFs of the fractional abundance of each trace ion as a function of $n_\mathrm{H}$, weighted by the species density. The 1-dex bands therefore trace the gas in which most of each species actually resides: $\mathrm{OH^+}$ and $\mathrm{H_2O^+}$ form predominantly in the UNM and diffuse CNM, while H$_3^+$ extends further into the denser CNM (in agreement with \cite{IndrioloEtAl-2012}). The equilibrium median (overlaid on the time-dependent distributions in the bottom panels) lies up to two orders of magnitude below the time-dependent abundance over the range $n_\mathrm{H} \sim 1$-$20$ cm$^{-3}$, with the exact range depending on the species. This is precisely the density regime where Section \ref{SS:Equilibrium} identified the largest non-equilibrium departures in $x(\mathrm{H_2})$, so the overlap between non-equilibrium gas and trace-ion-bearing gas occurs in the UNM and at the diffuse-CNM boundary.

The consequences for the predicted column densities follow directly. For $\mathrm{OH^+}$ and $\mathrm{H_2O^+}$, which form predominantly in the UNM and diffuse CNM, the UNM-scale enhancement translates into a substantial increase in the integrated abundance, accounting for the rightward shift between the equilibrium and time-dependent column density distributions in Figures \ref{fig:OHp_column_densities} and \ref{fig:H2Op_column_densities}. For H$_3^+$ the situation is different: most of the column arises in the denser CNM, where deviations from equilibrium are intrinsically small. The same UNM-scale enhancement that drives the $\mathrm{OH^+}$ and $\mathrm{H_2O^+}$ response is present for H$_3^+$ as well, but it operates on only a small fraction of the H$_3^+$ mass and therefore contributes only modestly to the integrated column. The net impact on $N(\mathrm{H_3^+})$ is correspondingly small, consistent with the closer agreement between the equilibrium and time-dependent $N(\mathrm{H_3^+})/N_\mathrm{H}$ predictions in Figure \ref{fig:H3p_column_densities}. This agreement is especially good when comparing sight lines with large $N_\mathrm{H}$ specifically (higher on the y-axis in Figure \ref{fig:H3p_column_densities}), with their logarithmic trace-to-hydrogen ratio varying by $sim$0.1 dex only, versus $\sim$0.3 dex for the lowest $N_\mathrm{H}$ sight lines. Not surprising, this comes from the fact that the large $N_\mathrm{H}$ sight lines preferentially sample a larger fraction of dense CNM, where, as previously discussed, deviations from equilibrium are smallest. 

Small differences between time-dependent versus steady-state simulations were also demonstrated for $\mathrm{H_3^+}$ in the recent study of \cite{HoEtAl-2026}, which showed that the inferred CRIR under the steady-state assumption is a factor of $\sim$3 higher than in the time dependent case. These authors also demonstrated a dependence on the line-of-sight column density, showing that the this factor decreases as $N_\mathrm{H}$ increases.

Figure \ref{fig:spatial_distribution} depicts these differences in terms of spatial distribution in the simulation. A linear color scheme is preferred for the trace molecule densities (middle row and lower left panels), in order to emphasize where most of their mass actually forms. The densest CNM regions (see total density) correspond with substantial $\mathrm{H_3^+}$ formation, whereas $\mathrm{OH^+}$ or $\mathrm{H_2O^+}$ are typically generated in thin shells around dense CNM (e.g., see the dense CNM cloud in the bottom right corner of the box), and occasionally as filamentary regions within clouds of UNM-like conditions (e.g., see the connected clouds in the central part of the box).

%\begin{figure*}[]
%	\begin{center}
%		\includegraphics[scale=0.415]{TraceMolecules_spatial_distribution.png}
%		\caption{Same simulation slice as in Figure \ref{fig:EqVsNonEq_spatial_distribution}, but for trace molecules and total density, and a resolution of $K=1024$.}
%		\label{fig:TraceMolecules_spatial_distribution}
%	\end{center}
%\end{figure*}

\begin{figure*}[]
    \subfigure{\label{fig:1D_PDF_OHp_ratio}\includegraphics[scale=0.45]{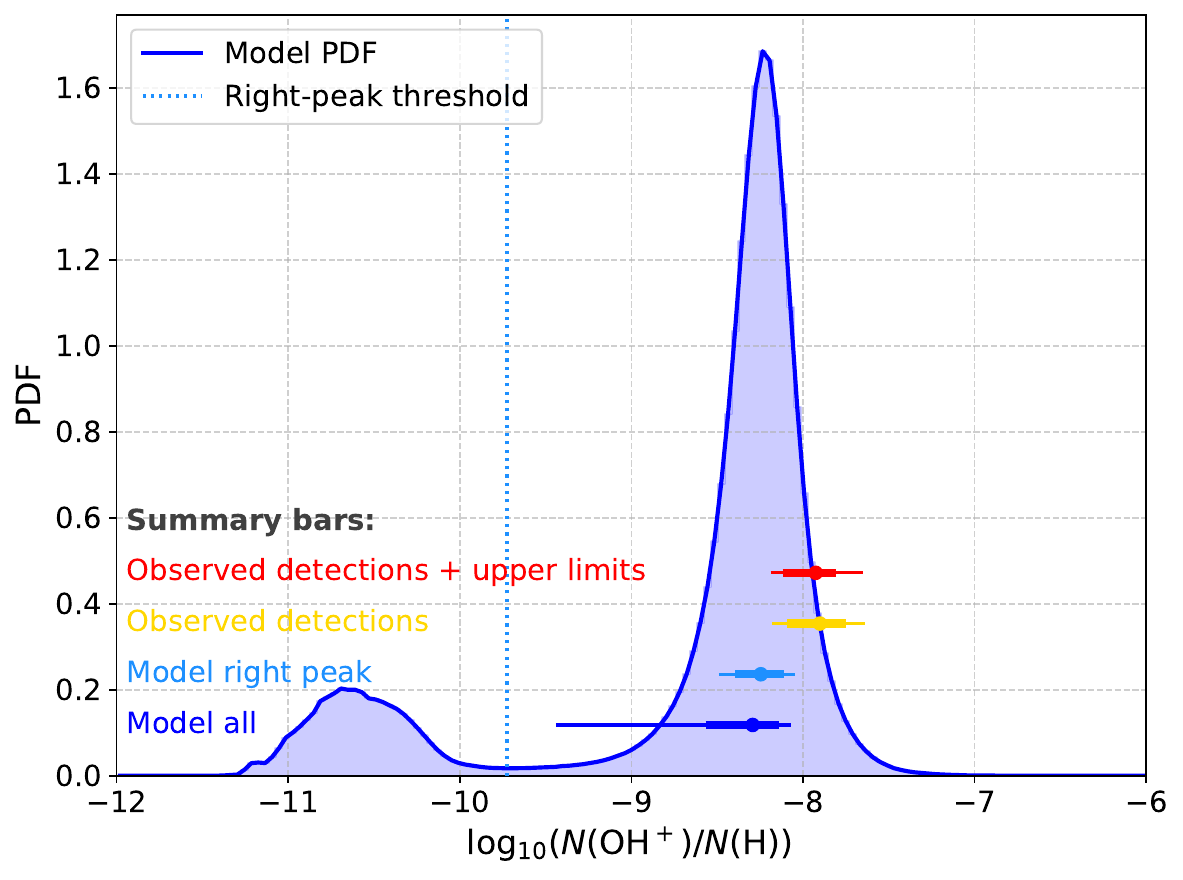}}
 	\subfigure{\label{fig:1D_CDF_OH_ratio}\includegraphics[scale=0.45]{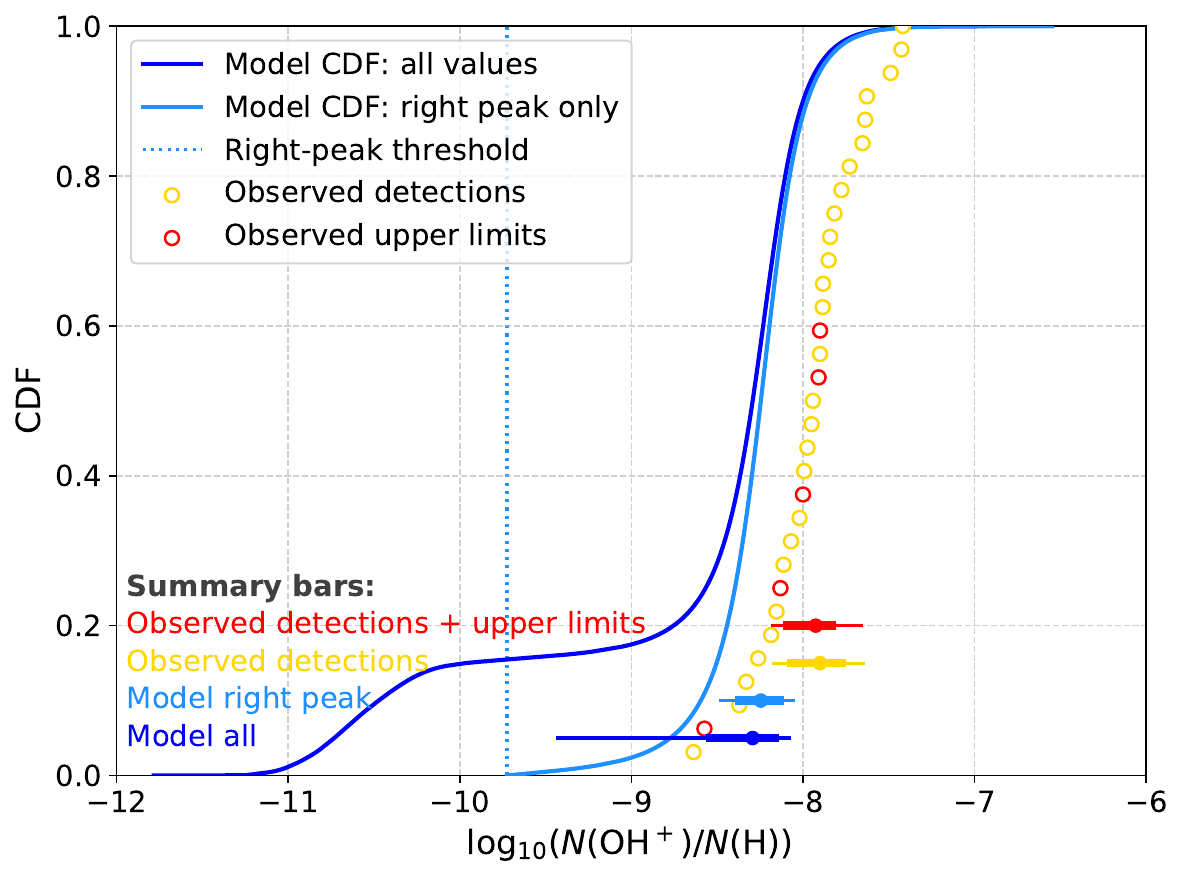}}

    \subfigure{\label{fig:1D_PDF_H2Op_ratio}\includegraphics[scale=0.45]{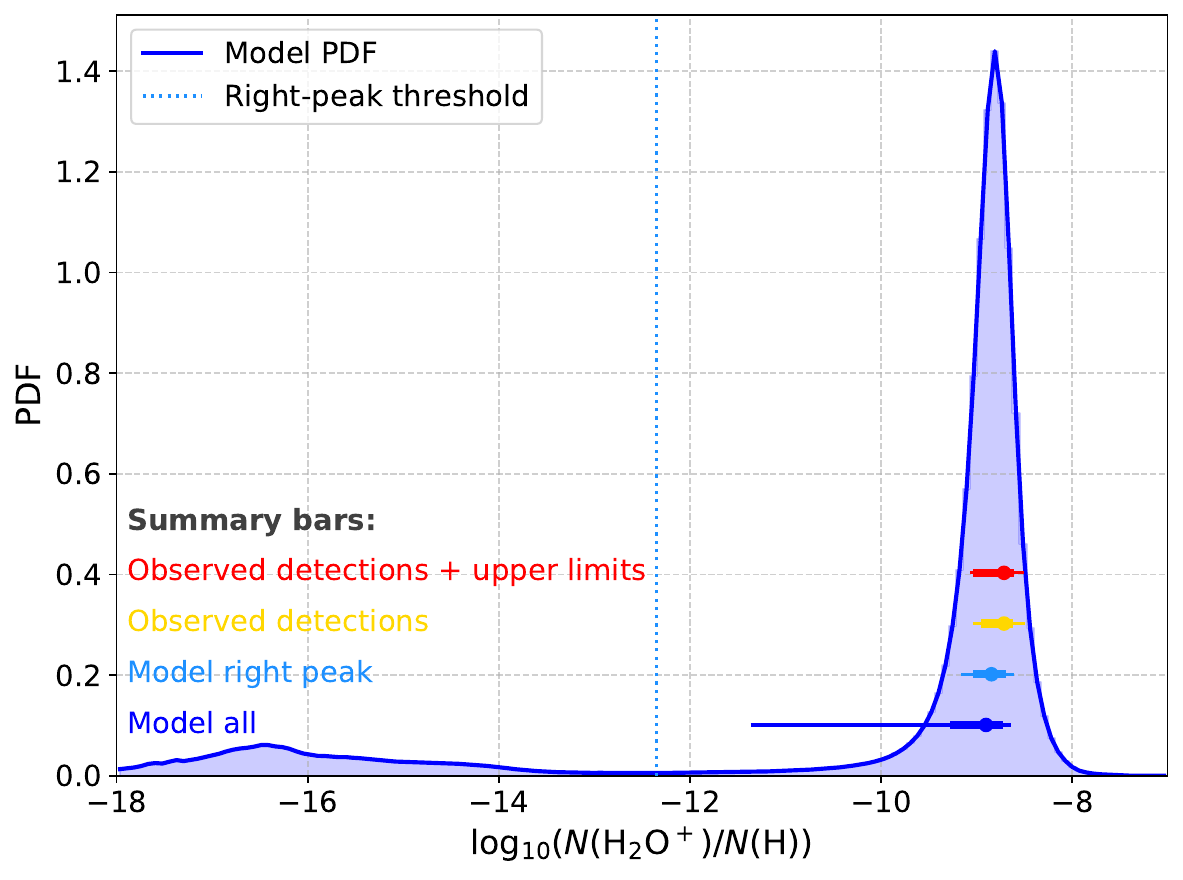}}
 	\subfigure{\label{fig:1D_CDF_H2Op_ratio}\includegraphics[scale=0.45]{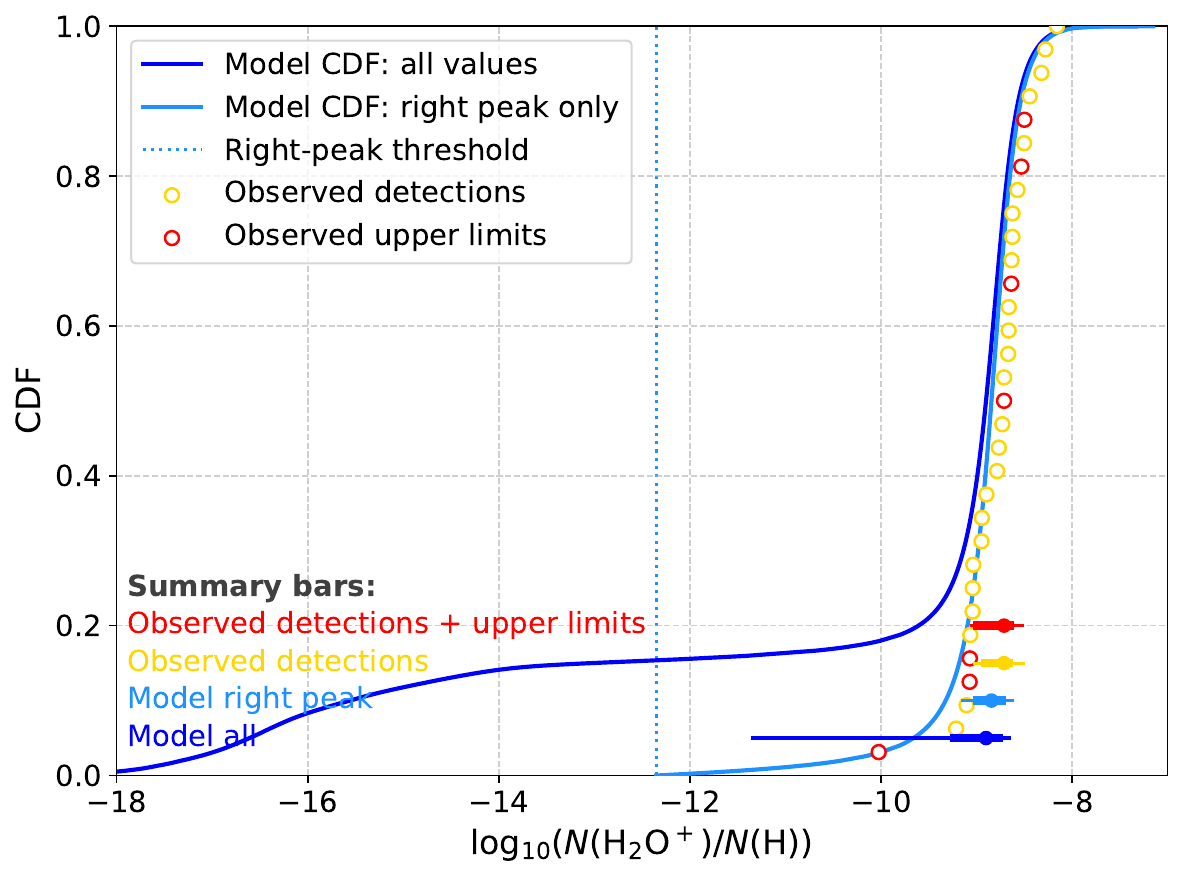}}

    \subfigure{\label{fig:1D_PDF_H3p_ratio}\includegraphics[scale=0.45]{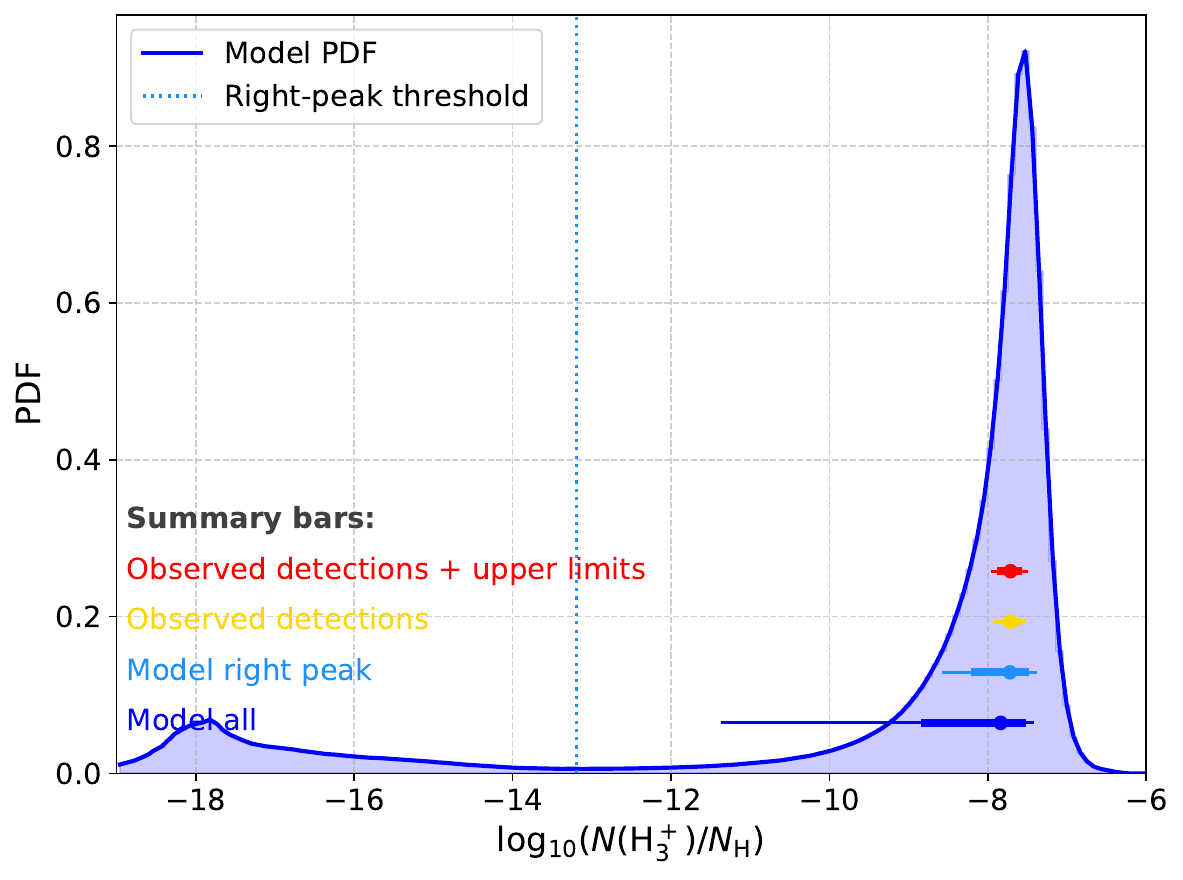}}
 	\subfigure{\label{fig:1D_CDF_H3p_ratio}\includegraphics[scale=0.45]{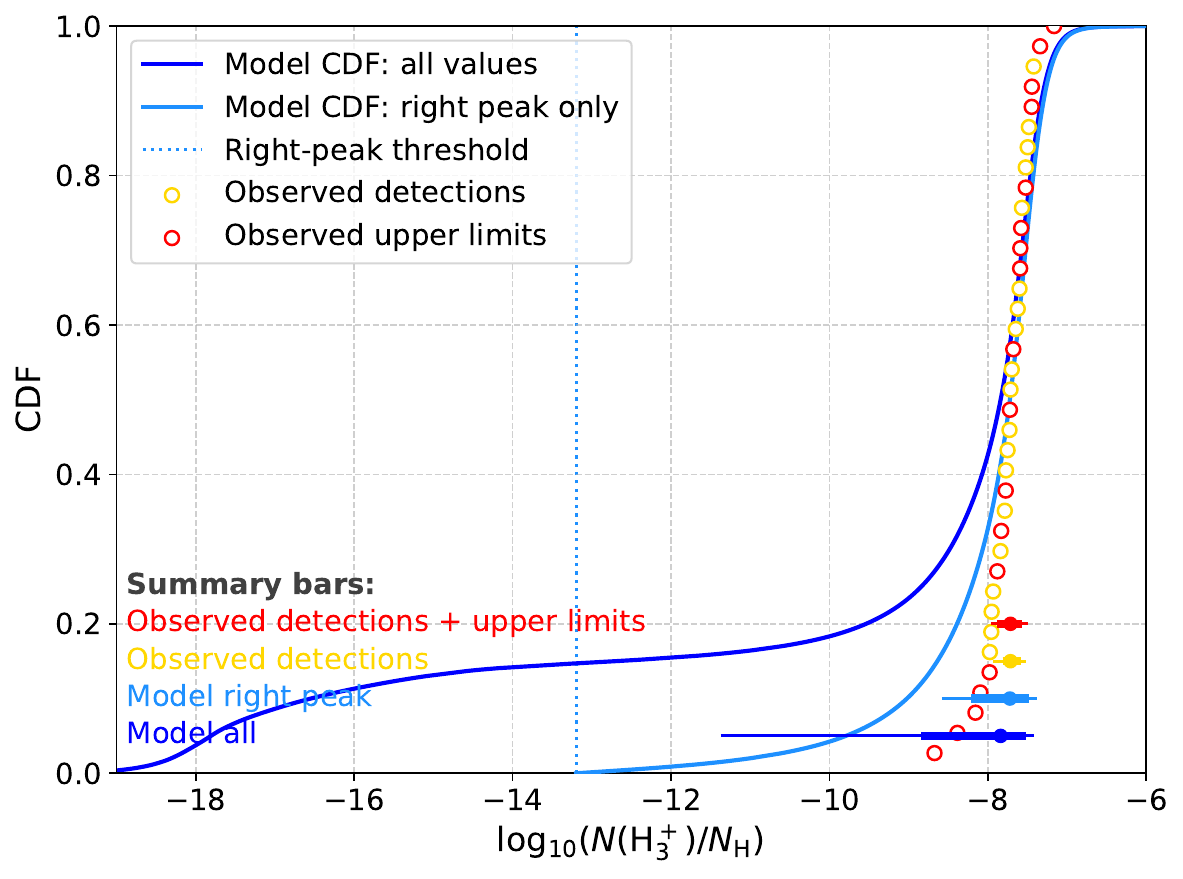}}

    \caption{
    1D PDFs (left panels) and CDFs (right panels) of the logarithmic column density ratios for $\mathrm{OH^+}$, $\mathrm{H_2O^+}$ and H$_3^+$, comparing the time-dependent model with local-ISM observations. The model distributions (blue) are shown both for the full bimodal population and for the high-ratio (right-hand) peak only (light blue), with the separation indicated by the vertical dotted line. The $\mathrm{OH^+}$ and $\mathrm{H_2O^+}$ model is based on the larger velocity-interval (Table \ref{tab:Indriolo2015_table5_integrated}) sample. Observations are shown as detections (yellow) and detections combined with upper limits (red). Statistical summaries ("status bars") indicate the median (point), 25–75 percentiles (thick line), and 16–84 percentiles (thin line), shown separately for the full model, right-peak model, detections only, and detections plus limits.
}
    \label{fig:1D-ModelVersusObservationAbundances}
\end{figure*}

In Figure \ref{fig:1D-ModelVersusObservationAbundances} we compare the time-dependent model predictions for the trace molecules with their corresponding observations in the local ISM, using simplified 1D representations of the column density ratios relative to atomic hydrogen. The $\mathrm{OH^+}$ or $\mathrm{H_2O^+}$ model is based on the velocity-interval LOS reconstruction, due to its much larger statistical sample. The left panels show the (log-scale) probability distribution functions of these ratios. Unlike in Figures \ref{fig:OHp_column_densities}–\ref{fig:H3p_column_densities}, we display here the full range of the model distributions. The resulting bimodal (double-peaked) structure is consistent with previous studies (e.g. \cite{BialyEtAl-2019}), and reflects distinct chemical regimes within interstellar clouds.

The right panels show the corresponding cumulative distribution functions. While the contribution of the low-ratio (left) peak is clearly visible in the model, it is not similarly represented in the observational data, which are inherently biased against low-ratio detections due to instrument sensitivity (see dedicated discussion in Appendix \ref{Appendix:D}). The dotted vertical line marks the threshold used to separate the two peaks.

To quantify the agreement between the model and the observations, we include summary "status bars" that characterize the statistical distributions. Each bar indicates the median (point), the interquartile range (25–75 percentile; thick line), and the 16–84 percentile range (thin line). To account for the observational bias, we show two model status bars: one representing the full (bimodal) distribution, and another restricted to the right-hand peak only. The latter provides a more appropriate comparison to the observations, which preferentially sample higher ratios.

For the observational data, we also present two status bars: one including only detections, and another that additionally incorporates upper limits. Including upper limits shifts the distribution toward lower ratios, representing a conservative estimate of the contribution from undetected low-ratio sight lines (see Appendix \ref{Appendix:D})). As more sensitive observations might become available in the future, the observational distribution is therefore expected to shift further to the left.

The interquartile (25–75) and 16–84 percentile ranges shown by the summary bars in Figure \ref{fig:1D-ModelVersusObservationAbundances} are also listed in Table \ref{tab:ratio_summary_statistics}. We find that the widths of the right-peak model distributions are in good agreement with those inferred from the observations, particularly when upper limits are included, and for $\mathrm{OH^+}$ and $\mathrm{H_2O^+}$. The percentile ranges of the model closely match those of the data, despite the model assuming a single, spatially uniform CRIR. This indicates that the observed sight-line-to-sight-line dispersion in the trace-to-hydrogen column ratios can be largely reproduced by variations in the underlying multiphase structure of the ISM.

Traditionally, such variations in observed column density ratios have been interpreted as evidence for fluctuations in the CRIR. However, our model results show that structural variations alone can produce a comparable (or even larger) spread in the ratios, introducing a degeneracy between CRIR variations and multiphase gas structure. This suggests that at least part of the inferred CRIR variability may instead reflect differences in the physical and chemical conditions along individual lines of sight, and raises the possibility that the CRIR is more homogeneous than commonly assumed.

\begin{table}
\centering
\caption{Summary statistics for the logarithmic column density ratios $N$($\mathrm{OH^+}$)/$N(\mathrm{H})$, $N$($\mathrm{H_2O^+}$)/$N(\mathrm{H})$ and $N$($\mathrm{H_3^+}$)/$N_\mathrm{H}$. The table lists the median, interquartile range (25--75 percentile), and 16--84 percentile range for the model and observational samples shown in Figure \ref{fig:1D-ModelVersusObservationAbundances}.}
\label{tab:ratio_summary_statistics}
\begin{tabular}{lccc}
\hline
Trace molecule & Median & 25--75 percentile & 16--84 percentile \\
\hline
$\mathrm{OH^+}$: Model full & $-8.29$ & $-8.57$--$-8.14$ & $-9.44$--$-8.07$ \\
Model right peak & $-8.25$ & $-8.40$--$-8.11$ & $-8.49$--$-8.04$ \\
Obs. detections & $-7.90$ & $-8.09$--$-7.75$ & $-8.18$--$-7.64$ \\
Obs. det. + upper lim. & $-7.93$ & $-8.12$--$-7.81$ & $-8.19$--$-7.65$ \\
\hline
$\mathrm{H_2O^+}$: Model full & $-8.90$ & $-9.28$--$-8.72$ & $-11.36$--$-8.64$ \\
Model right peak & $-8.84$ & $-9.03$--$-8.69$ & $-9.16$--$-8.61$ \\
Obs. detections & $-8.71$ & $-8.95$--$-8.62$ & $-9.04$--$-8.49$ \\
Obs. det. + upper lim. & $-8.71$ & $-9.04$--$-8.61$ & $-9.07$--$-8.50$ \\
\hline
H$_3^+$: Model full & $-7.84$ & $-8.84$--$-7.52$ & $-11.37$--$-7.41$ \\
Model right peak & $-7.72$ & $-8.21$--$-7.48$ & $-8.58$--$-7.38$ \\
Obs. detections & $-7.71$ & $-7.81$--$-7.58$ & $-7.93$--$-7.52$ \\
Obs. det. + upper lim. & $-7.71$ & $-7.88$--$-7.57$ & $-7.96$--$-7.49$ \\
\hline
\end{tabular}
\end{table}

Finally, we comment on the robustness of these results with respect to numerical resolution. As discussed in Appendix \ref{Appendix:C}, the total masses of $\mathrm{OH^+}$, $\mathrm{H_2O^+}$ and H$_3^+$ are not yet fully converged at our highest resolution ($K=1024$), but the deviations from the asymptotic values are constrained to within a factor of $\sim$2. Importantly, the qualitative trends identified here —- namely the enhancement of $\mathrm{OH^+}$ and $\mathrm{H_2O^+}$ in the time-dependent calculation relative to the equilibrium benchmark, and the comparatively weaker response of H$_3^+$ -— are preserved across resolutions. We therefore conclude that the main results reported in this section are not driven by numerical artifacts, although higher resolution will be required for precise quantitative convergence.

In the context of Table \ref{tab:ratio_summary_statistics}, we expect the median model ratios to shift slightly toward lower values with increasing numerical resolution (details in Appendix \ref{Appendix:C}). A similar shift is expected for the observational medians as improved sensitivity reveals additional low-ratio sight lines currently represented by upper limits (details in Appendix \ref{Appendix:D}). In contrast, the model distribution widths are expected to remain largely unchanged (Appendix \ref{Appendix:C})). Our conclusions regarding the dispersion are therefore robust.

\section{Conclusions and future directions}\label{S:Conclusions}
We investigated whether the large observational scatter in the column densities of the reactive molecular ions $\mathrm{OH^+}$, $\mathrm{H_2O^+}$, and H$_3^+$ necessarily implies large intrinsic variations in the cosmic-ray ionization rate across Galactic sight lines. Building on the high-resolution multiphase MHD simulations of \citet{GodardEtAl-2023}, which already reproduce a broad set of local-ISM constraints, we post-processed the simulation snapshots with analytic steady-state frameworks for $\mathrm{OH^+}$, $\mathrm{H_2O^+}$ and H$_3^+$ while explicitly retaining the time-dependent $x(\mathrm{H_2})$ and $x(\mathrm{e})$ fields. We then synthesized a distribution of line-of-sight column densities and compared it directly to observational samples compiled from the literature.

Our main conclusions are as follows:

\begin{enumerate}

\item \textbf{A single local-ISM CRIR can be consistent with a wide range of observed tracer columns.}
Using $\zetatref{H} \simeq 2\times10^{-16}\,\mathrm{s^{-1}}$, together with the fiducial local-ISM parameters adopted in \citet{GodardEtAl-2023}, the simulations naturally produce broad, observationally comparable distributions of the trace-to-hydrogen ratio column densities without requiring order-of-magnitude sight-line-to-sight-line variations in the CRIR. In this picture, much (however not all) of the apparent scatter in "inferred CRIR" arises from sampling different trajectories through a turbulent, multiphase medium rather than from large intrinsic variations in the underlying ionization environment.

\item \textbf{Time-dependent chemistry of $x(\mathrm{H_2})$ is essential for $\mathrm{OH^+}$ and $\mathrm{H_2O^+}$, but only marginal for H$_3^+$.} 
Turbulence advects gas across phase boundaries faster than the chemistry of $\mathrm{H_2}$ and ionization can relax, producing long-lived departures from equilibrium that are largest in the UNM for $x(\mathrm{H_2})$ and in the diffuse WNM for $x(\mathrm{e})$. In particular, $x(\mathrm{H_2})$ is enhanced by orders of magnitude in UNM "skins" around CNM structures.
$\mathrm{OH^+}$ and $\mathrm{H_2O^+}$ reside preferentially in exactly these transitional environments, so the time-dependent calculation predicts substantially higher columns than the equilibrium benchmark, populating regions of $N(\mathrm{H})$--$N(i)/N(\mathrm{H})$ space that the benchmark cannot reach. H$_3^+$, in contrast, draws most of its column from the denser CNM where equilibrium is closer to valid, so the time-dependent and equilibrium predictions for H$_3^+$ differ much less. This makes H$_3^+$ a relatively cleaner CRIR diagnostic, although still not a clean one: multiphase structure alone produces substantial scatter in $N(\mathrm{H_3^+})/N_\mathrm{H}$ even at fixed CRIR.

% Although $x(\mathrm{e})$ is also suppressed in WNM/UNM gas recently processed through colder phases, this effect is increasingly important only for trace molecules that are preferentially produced in more diffuse environments than $\mathrm{OH^+}$ and $\mathrm{H_2O^+}$.

\item \textbf{Geometry and LOS definitions matter.}
We explored two LOS reconstruction strategies: (i) integrating to observed source distances (relevant for H$_3^+$ and for integrated $\mathrm{OH^+}$ / $\mathrm{H_2O^+}$ columns), and (ii) interpreting velocity intervals as a distribution of path lengths, modeled here by a truncated power-law. Both approaches can reproduce the observed hydrogen column distributions, and once this is satisfied, the same underlying 3D simulation produces tracer-to-hydrogen column distribution ratios broadly consistent with the observational scatter. This underscores that converting observed columns into local physical conditions cannot be done robustly without accounting for how the multiphase medium is sampled.

\item \textbf{The most significant outliers plausibly trace non-local environments.}
Observed points that fall well outside the dominant probability (so-called '1-dex band') of the synthetic distributions are predominantly associated with long-distance ($\gtrsim 3\,\mathrm{kpc}$) sight lines in the observational sample, which are more likely to intersect regions with genuinely different environmental parameters (e.g., elevated CRIR or $G_0$) than the conditions represented by the fiducial simulation. This provides a natural explanation for why the largest apparent CRIR excursions inferred from $\mathrm{OH^+}$ and $\mathrm{H_2O^+}$ often occur in the most distant lines of sight.
\end{enumerate}

Taken together, our results support a picture in which much of the observed dispersion in $\mathrm{OH^+}$, $\mathrm{H_2O^+}$ and H$_3^+$ columns can arise from \emph{multiphase structure + turbulence + time-dependent chemistry} rather than from large intrinsic CRIR fluctuations within the local ISM. In this framework, the commonly used practice of interpreting each sight line with an independent 1D equilibrium model can systematically map environmental diversity and dynamical history into an apparent spread in the CRIR.

Table \ref{tab:tracer_sensitivity_summary} summarizes, in compact form, the dominant physical origin of each tracer in our simulations and its sensitivity to the two key time-dependent ingredient $x(\mathrm{H_2})$.

\begin{table}[h!]
\centering
\small
\caption{Summary of dominant formation environments and non-equilibrium sensitivities for the three tracer ions.}
\label{tab:tracer_sensitivity_summary}
\begin{tabular}{lcc}
\toprule
Tracer & Dominant phase(s) contributions & Sensitivity to non-eq \\
\midrule
$\mathrm{OH^+}$     & UNM / diffuse CNM interfaces & High \\
$\mathrm{H_2O^+}$ & UNM / diffuse CNM interfaces & High \\
H$_3^+$    & Diffuse-to-dense CNM & Low--Moderate \\
\bottomrule
\end{tabular}
\end{table}

\subsection*{Implications and near-term extensions}
Our analysis suggests several immediate implications for observational inference and for future modeling:

\begin{itemize}

\item \textbf{Joint use of $\mathrm{OH^+}$, $\mathrm{H_2O^+}$ and H$_3^+$ can help break degeneracies.}
Because $\mathrm{OH^+}$ and $\mathrm{H_2O^+}$ are most sensitive to transition-layer physics while H$_3^+$ is sensitive to deeper molecular material, combining these tracers provides leverage to disentangle variations in the CRIR from variations in phase structure and the time-dependent chemistry.

% \item \textbf{Broader chemistry and radiative transfer can further refine results.}
% The main driver behind the good correspondence between modeled and observed column distributions is the time-dependent behavior of hydrogen and electrons, captured in the simulations and propagated into the trace molecule abundances. Including additional destruction partners and more detailed treatments of carbon chemistry and shielding -- either in post-processing or directly in the simulation -- would sharpen the quantitative predictions inside the CNM. H$_3^+$ might be particularly sensitive to such modeling improvements.

\item \textbf{Broader chemistry, radiative transfer and higher resolution can further refine results.}
The main driver behind the good correspondence between modeled and observed column distributions is the time-dependent behavior of hydrogen and electrons, captured in the simulations and propagated into the trace molecule abundances. Including additional destruction partners and more detailed treatments of carbon chemistry and shielding -- either in post-processing or directly in the simulation -- would sharpen the quantitative predictions inside the CNM, with H$_3^+$ likely the most sensitive of the three tracers to such improvements. Further gains will also come from increased numerical resolution: as discussed in Appendix \ref{Appendix:C}, the present $K=1024$ simulations are converged to within a factor of $\sim 2$ of the asymptotic mass, but full convergence will require higher resolutions accessible in future work.

\item \textbf{Spatial variations of the CRIR within the simulation.}  
In the current work, where multiple new ideas were presented, we opted for simplicity in depicting the CRIR, and adopted a uniform value throughout the computational domain. However, physically, the CRIR is expected to vary with environment: it may be attenuated in dense CNM due to column-dependent shielding, geometrically modulated in CNM skins and comparatively less attenuated in highly diffuse WNM gas. Such effects would introduce an intrinsic spatial distribution of the CRIR within the simulation itself. Importantly, this would \emph{add} to the scatter in trace molecule column densities, potentially leading to an even better match between model and observation. Incorporating physically motivated CR attenuation into the 3D simulation framework represents a further step for future follow-up studies.

% \item \textbf{Resolution requirements.}
% The results shown in this study are based on our highest resolution simulation, with a $K=1024$ mesh size. This resolution surpasses the highest resolution used by \cite{GodardEtAl-2023} in a predecessor study (previously, $K_\mathrm{max}=512$), representing an important step forward in our numerical modeling efforts. However, in Appendix \ref{Appendix:C} we report that this numerical resolution is still limited, and insufficient for complete convergence of all the trace molecules. We were nevertheless able to constrain the extent of this limitation to within a factor of 2 (compared to the asymptotic masses expected at infinite resolution). Completely resolving the formation of trace molecules $\mathrm{OH^+}$, $\mathrm{H_2O^+}$ and H$_3^+$ therefore remains an important but feasible task to look forward to in future follow-up studies, as we anticipate notable advancements in high performance capabilities in the coming years.
\end{itemize}

In summary, high-resolution 3D multiphase ISM simulations that accurately capture turbulence and time-dependent hydrogen/electron chemistry provide a natural explanation for much of the observed scatter in $\mathrm{OH^+}$, $\mathrm{H_2O^+}$ and H$_3^+$ columns in the local Galaxy. These results caution against interpreting the observed dispersion as direct evidence for large local variations in the CRIR, and motivate a shift toward inference approaches that explicitly account for multiphase structure and chemical non-equilibrium.

\section{Acknowledgments}\label{S:Acknowledgments}
We wish to thank anonymous reviewer for an extremely detailed review, which greatly improved the quality of our manuscript. We wish to thank Alexei Ivlev, Munan Gong, Kedron Silsbee and Ka Wai Ho for helpful discussions related to this project. S.B. acknowledges support from the ISF grant number 2071540, the GIF grant number I-1568-303.7/2024, the NSF-BSF grant number 2023761, and the Alon Fellowship prize for junior faculty. B.B. acknowledges NSF grant AST-2407877. B.B. is grateful for the generous support by the David and Lucile Packard Foundation and the Alfred P. Sloan Foundation. B.B. thanks the Center for Computational Astrophysics (CCA) of the Flatiron Institute and the Mathematics and Physical Sciences (MPS) division of the Simons Foundation for support. The Flatiron Institute is supported by the Simons Foundation. D.A.N. gratefully acknowledges support from the NASA/APRA program ``Continued Laboratory Studies of Dissociation Recombination with Cold Molecular Ions'' through a subcontract from Columbia University.

\newpage
\bibliographystyle{plainnat}
% \bibliography{example} % if your bibtex file is called example.bib
\bibliography{bibfile} % if your bibtex file is called example.bib

\clearpage
\appendix

\section{Appendix A: Electron fractional abundance modifications}\label{Appendix:A}
The \cite{GodardEtAl-2023} \textit{RAMSES} version used in this study takes the electron abundance to be $x(\mathrm{e}) = x(\mathrm{C}^+) + x(\mathrm{H}^+)$, and further assumes that carbon is fully ionized, so that $x(\mathrm{C}^+) = \abundanceref{C}$. The first relation holds across all neutral phases, but the second breaks down in denser CNM, where $x(\mathrm{C}^+)$ falls below $\abundanceref{C}$ as ionized carbon undergoes charge exchange with PAHs and is incorporated into CO and other molecules. To avoid overestimating $x(\mathrm{e})$ in this regime, we apply a simple, approximate correction in post-processing.

\begin{table*}[h!]
  \centering
  \caption{C$^+$ formation and destruction.}
  \label{tab:reactions_Cp}
  \begin{tabular}{l l l l}
    \hline\hline
    Reaction & Rate coefficient & Units & Rate Source\\
    \hline
    C + photon $\rightarrow$ C$^+$ + e & $\ratephoton{C} = 2.6\times10^{-10} \exp(-3.8 \mathrm{A}_\mathrm{v})$ & s$^{-1}$ & \cite{HeaysEtAl-2017} \\
    C$^+$ + PAH$^-$ $\rightarrow$ C + PAH$^0$ & $\rate{-} = 2.4\times10^{-7} (T/100~\mathrm{K})^{-0.5}$ & cm$^3$ s$^{-1}$ & \cite{WolfireEtAl-2008}\\
    C$^+$ + PAH$^0$ $\rightarrow$ C + PAH$^+$ & $\rate{0} = 8.8\times10^{-9}$ & cm$^3$ s$^{-1}$ & \cite{WolfireEtAl-2008}\\
    \hline
  \end{tabular}
\end{table*}

Table \ref{tab:reactions_Cp} considers the main processes that lead to $x$(C$^+$) reduction. In our simple approximation, the ionization of atomic carbon by photons is balanced by charge exchanges with PAHs. Since $\rateref{0}$ (the reaction rate between $x$(C$^+$) and PAH$^0$) is much lower than $\rateref{-}$, particularly with increasing density (decreasing temperature), and since $x(\mathrm{PAH^-})$ and $x(\mathrm{PAH^0})$ are comparable (few $10^{-7}$), we neglect this reaction for simplicity. The recombination of C$^+$ and electrons is likewise neglected, accounting for the baseline electron abundance ($x(\mathrm{e}) < 10^{-4}$) and the characteristic recombination rate $\sim 10^{-11} - 10^{-12}$ \citep{NaharPradhan-1997,Badnell-2006}. This leads to the following expression for $x$(C$^+$):
\begin{equation}\label{eq:x_Cp}
\begin{aligned}
x(\mathrm{C}^+) = \abundanceref{C} / \left( 1 + \frac{n_\mathrm{H} x(\mathrm{PAH^-}) \rateref{-}}{\mathrm{G_0} \raterefphoton{C}} \right).
\end{aligned}
\end{equation}

% Since $\abundanceref{C}$ is the total elemental abundance of carbon, it follows that $\abundanceref{C}$=$x$(C)+$x$(C$^+$)). 
From Equation \ref{eq:x_Cp}, we implement a simple post-processing correction factor $S$ for the reduction in $x(\mathrm{C}^+)$, such that
\begin{equation}\label{eq:x_Cp_reduction_factor}
\begin{aligned}
S = \frac{n_\mathrm{H} x(\mathrm{PAH^-}) \rateref{-}}{\mathrm{G_0} \raterefphoton{C}}.
\end{aligned}
\end{equation}
 
We adopt a thresholded form of this correction rather than the exact factor $(1+S)^{-1}$: no reduction is applied for $S\leq 1$, and for $S>1$ we use the asymptotic form $x(\mathrm{C}^+)=\abundanceref{C}/S$. This deliberately suppresses the correction when the two competing channels are of comparable strength ($S\sim 1$), and only reduces $x(\mathrm{C}^+)$ once PAH-mediated neutralization clearly dominates, while still recovering the smooth form $\abundanceref{C}/S$ in the strongly suppressed limit ($S\gg 1$). This correction is applied as a post-processing step, affecting the calculations of all trace molecules in Section \ref{SS:Chemistry}, and predominantly that of $\mathrm{H_3^+}$ which forms mostly in dense environments.

The resulting electron fractional abundance, appearing in the bottom-right panel of Figure \ref{fig:EqVsNonEq_H2_and_e} shows how the correction for $x(\mathrm{C}^+)$ modifies our model. The $S>1$ criteria in Equation \ref{eq:x_Cp_reduction_factor} turns on at a density of around $n_\mathrm{H} \sim 5 \times 10^2$ cm$^-3$, marking a turn-off point in which $x(\mathrm{C}^+)$ and $x(\mathrm{e})$ concurrently start to decline. We compare this turn-off point to several other detailed studies that account for dense ISM chemistry (E.g., figure D1 in \cite{Ibanez-MejiaEtAl-2019}; figure 11 in \cite{HuEtAl-2021}; figure 4 in \cite{BisbasEtAl-2023}) and find it to be comparable to our own, generally in the range $n_\mathrm{H} \sim 10^2-10^3$ cm$^-3$. This broad agreement, despite various differences in modeling approaches, further validates our approximation, which fits well in the middle range of past results.

\section{Appendix B: Reaction rates}\label{Appendix:B}
Below we provide additional reaction rates whose expressions are too long for the main text:

\begin{equation}\label{eq:gamma1}
\begin{aligned}
k_1 ={}&\Bigl(1.57\times10^{-9} \left(T/10000~\mathrm{K}\right)^{0.298} \exp\left(\frac{T}{75100~\mathrm{K}}\right) +\\
            &1.62\times10^{-7} \left(T/10000~\mathrm{K}\right)^{1.13} \exp\left(\frac{-T}{19.4~\mathrm{K}}\right) \Bigr) \exp(-227~\mathrm{K}/T) ~cm^3~s^{-1}
\end{aligned}
\end{equation}

\begin{equation}\label{eq:gamma2}
\begin{aligned}
k_2 ={}&2.08\times10^{-9} \left(T/10000~\mathrm{K}\right)^{0.405} +\\             &1.11\times10^{-11} \left(T/10000~\mathrm{K}\right)^{-0.458} ~cm^3~s^{-1}
\end{aligned}
\end{equation}

\begin{equation}\label{eq:gamma5}
\begin{aligned}
k_5 ={}& 1.1 \times 10^{-7} (300~\mathrm{K}/T)^{0.767} + (T/1~\mathrm{K})^{-1.5}\Bigl( 3.46\times 10^{-4}\\
            & \exp(-129~\mathrm{K}/T)
              - 9.16\times 10^{-4} \exp(-1220~\mathrm{K}/T) \\
            & - 1.85\times 10^{-3} \exp(-5900~\mathrm{K}/T)
              + 3.33\times 10^{-2} \\
            & \exp(-43400~\mathrm{K}/T)
\Bigr) ~cm^3~s^{-1}
\end{aligned}
\end{equation}

\section{Appendix C: Resolution convergence requirements}\label{Appendix:C}
\begin{figure*}[]
	\begin{center}
		\includegraphics[scale=0.7]{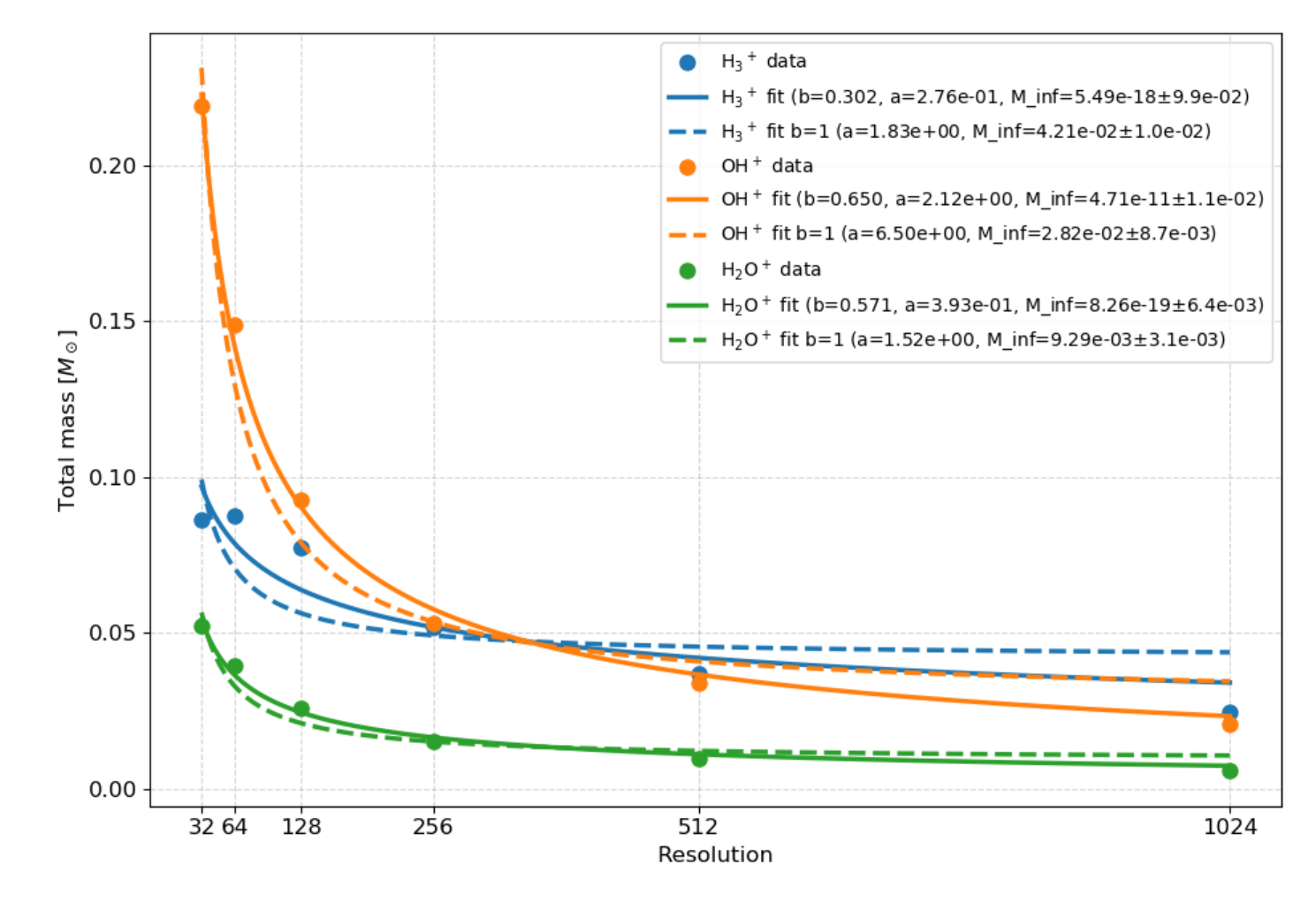}
		\caption{Convergence of trace molecule masses as a function of resolution. Simulation-derived masses (circles) are fitted with the function $M_\infty + a R^{-b}$ (lines).}
		\label{fig:GD32-1024}
	\end{center}
\end{figure*}

In the main text we discussed our requirement to use high resolution simulations for reliably estimating the abundance of trace molecules. Here we show the convergence of mass for trace molecules $\mathrm{OH^+}$, $\mathrm{H_2O^+}$ and H$_3^+$, as a function of resolution. Figure \ref{fig:GD32-1024} shows the total mass in units of $M_\odot$ at different resolutions $K$, fitted with the function $M_\infty + a R^{-b}$, where $M_\infty$, $a$ and $b$ are adjustable coefficients. Our simulations are run up to 60 Myr, while analysis shows that convergence occurs after only $\sim$30 Myr. Between 30-60 Myr, the simulation is in a quasi-steady state, where various quantities such as the CNM fraction, or the total masses of trace molecules are undulating around some mean value. Therefore, we use the mean of the time series between 30-60 Myr.

\begin{figure*}[]
    \subfigure[$K=32$] {\label{fig:2D_nH-T-nHWeighted-32}\includegraphics[scale=0.46]{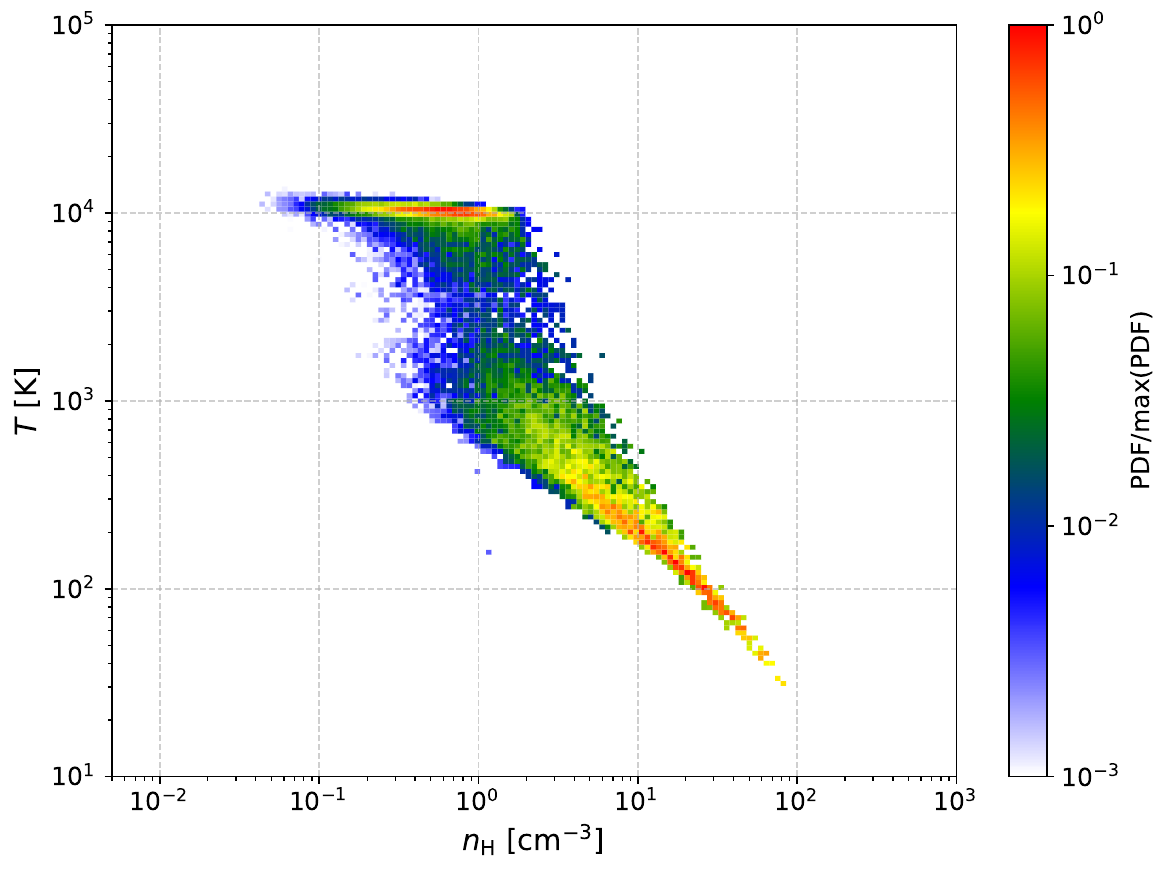}}
 	\subfigure[$K=64$] {\label{fig:2D_nH-T-nHWeighted-64}\includegraphics[scale=0.46]{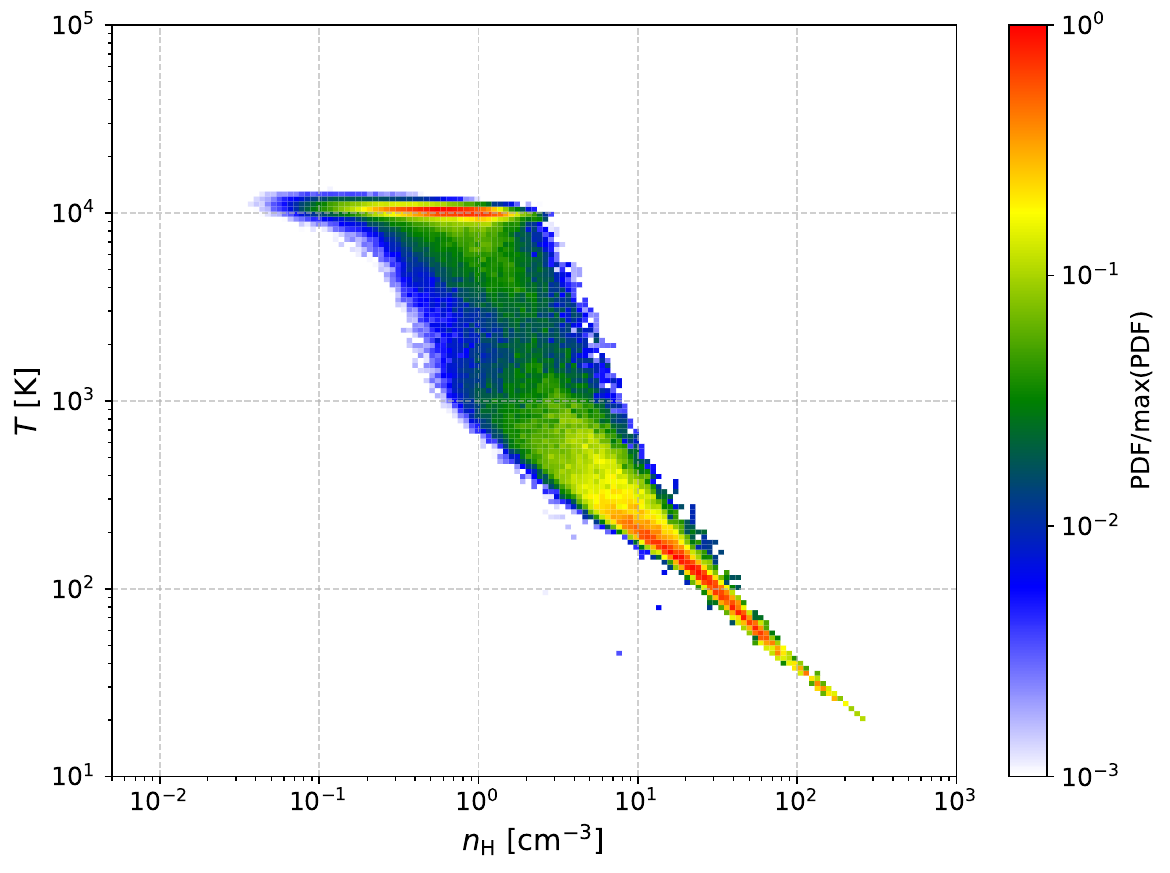}}
    \subfigure[$K=128$] {\label{fig:2D_nH-T-nHWeighted-128}\includegraphics[scale=0.46]{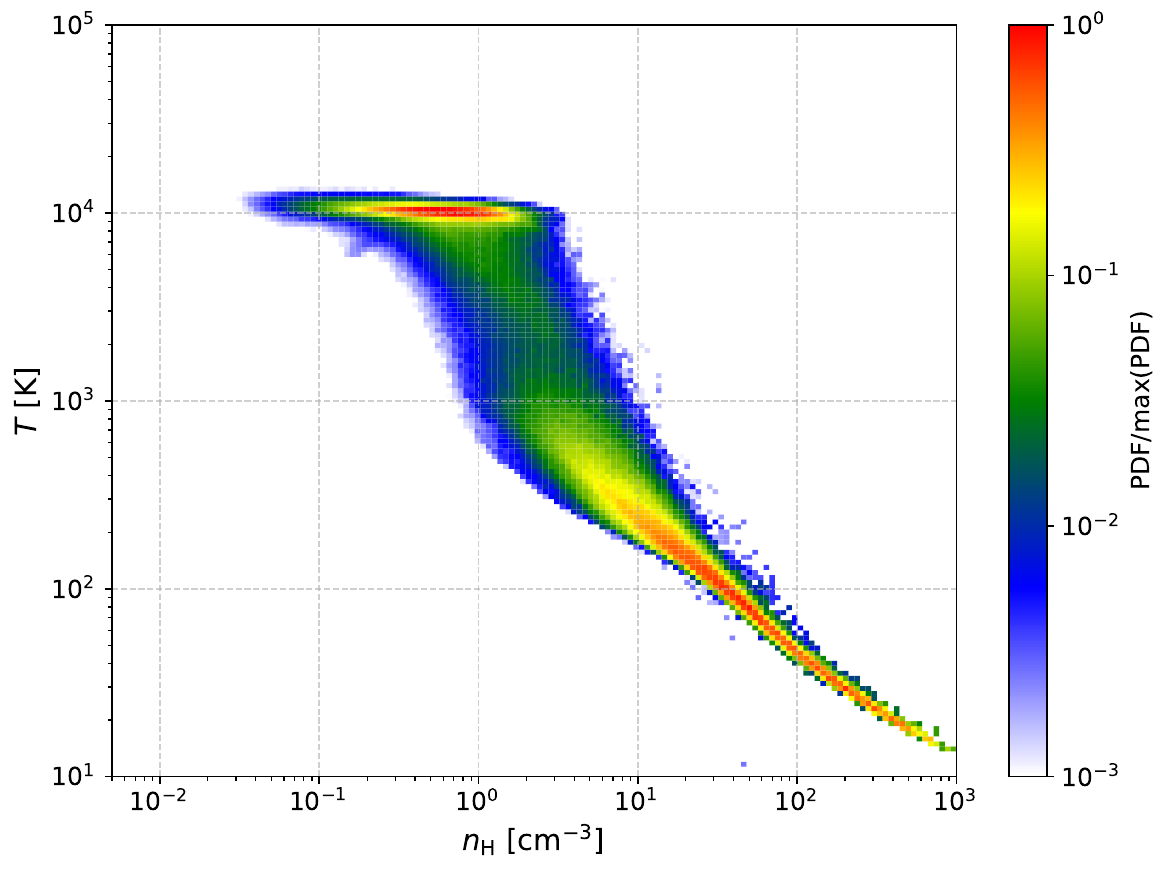}}
    \subfigure[$K=256$] {\label{fig:2D_nH-T-nHWeighted-256}\includegraphics[scale=0.46]{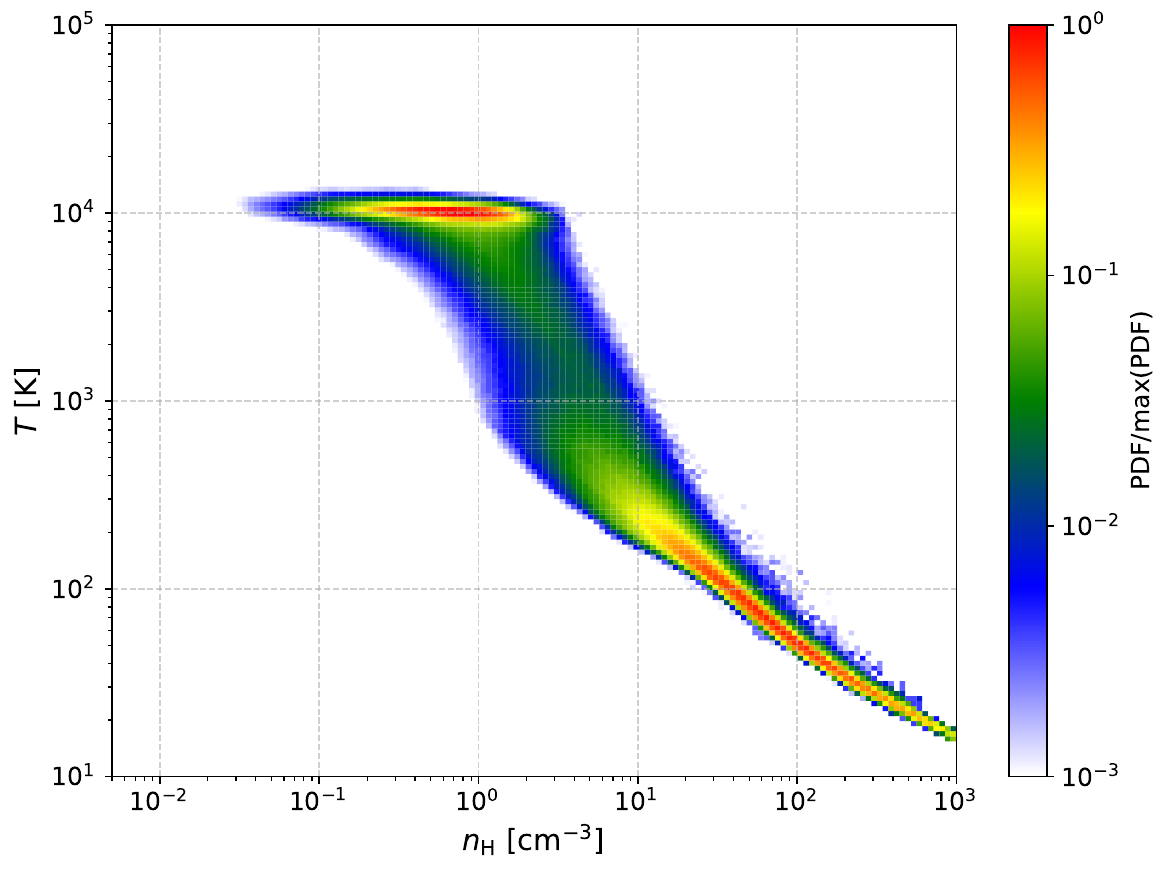}}
    \subfigure[$K=512$] {\label{fig:2D_nH-T-nHWeighted-512}\includegraphics[scale=0.46]{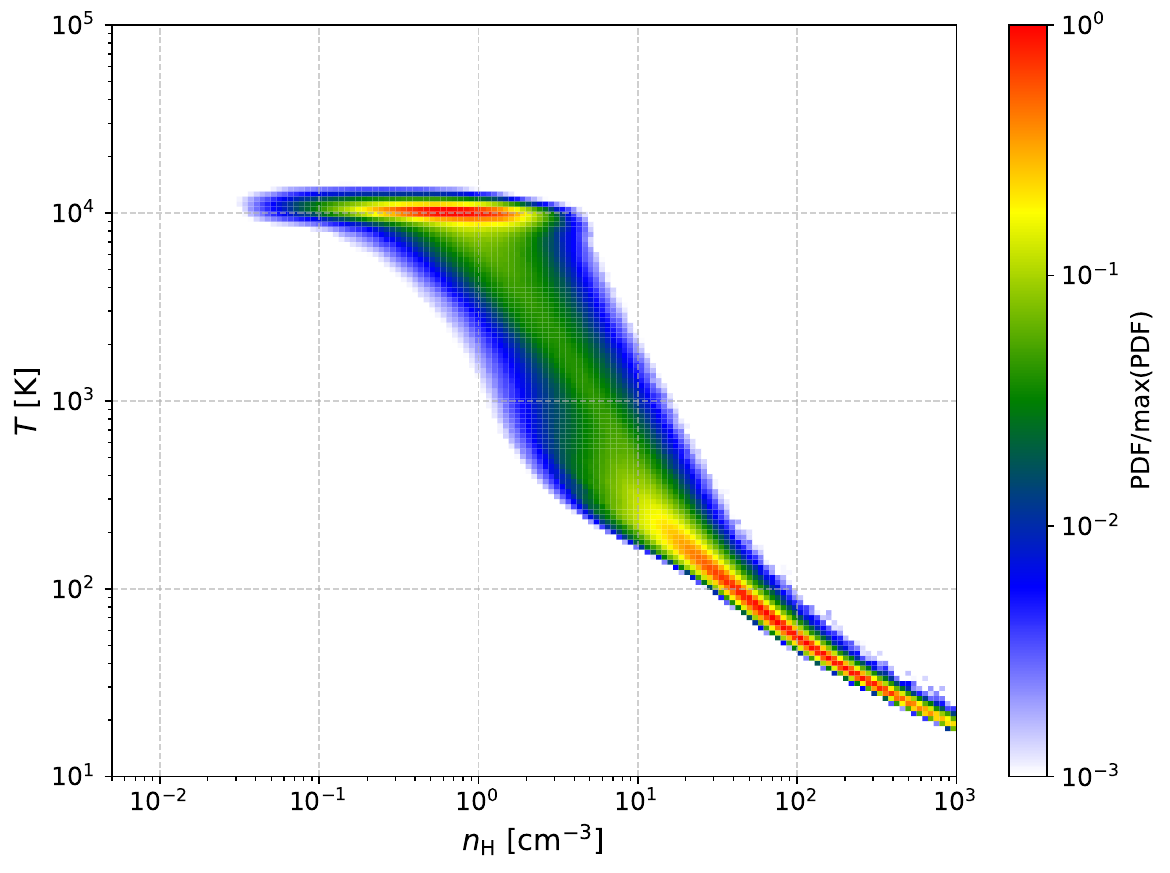}}
    \subfigure[$K=1024$] {\label{fig:2D_nH-T-nHWeighted-1024}\includegraphics[scale=0.46]{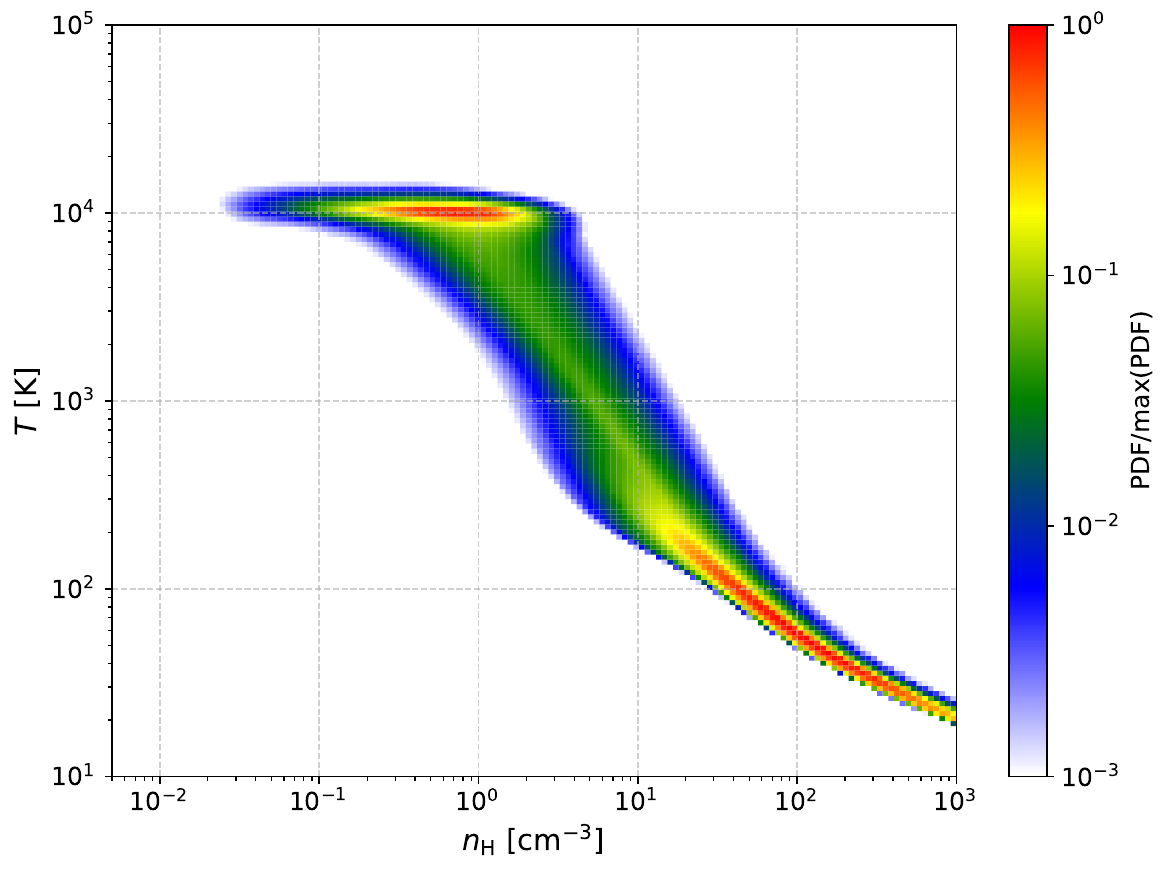}}

    \caption{2D mass-weighted histograms of temperature versus total density, across different mesh sizes ($K$).}
    \label{fig:2D-Temperature-Resolution-Convergence}
\end{figure*}

Mass generally decreases with resolution, as was previously also found in the predecessor study of \cite{GodardEtAl-2023} for the case of CH$^+$. For $\mathrm{OH^+}$ and $\mathrm{H_2O^+}$, which are formed predominantly in UNM, the lower the resolution, the more likely it becomes for interfaces to be smeared over a larger number of cells (numerical diffusion), which artificially creates more 'transition' gas, and therefore more mass. In contrast, H$_3^+$ mass has significant contributions from both UNM and CNM. However, at extremely low resolutions ($K<128$), high density/low temperature cells are not even resolved (Figure \ref{fig:2D-Temperature-Resolution-Convergence}), explaining the low resolution slope reversal (i.e, the reason for why the low resolution simulations produce less mass). In the other direction where the CNM is better resolved, the temperature is also better resolved (more low temperature cells = colder CNM). \rateref{-} and \rateref{12} respond to colder CNM by taking larger values, $x$(e) increasing the denominator in Equation \ref{eq:x_H3p} and $\mathrm{H_3^+}$ decreases.

\begin{figure*}[]
	\begin{center}
		\includegraphics[scale=0.7]{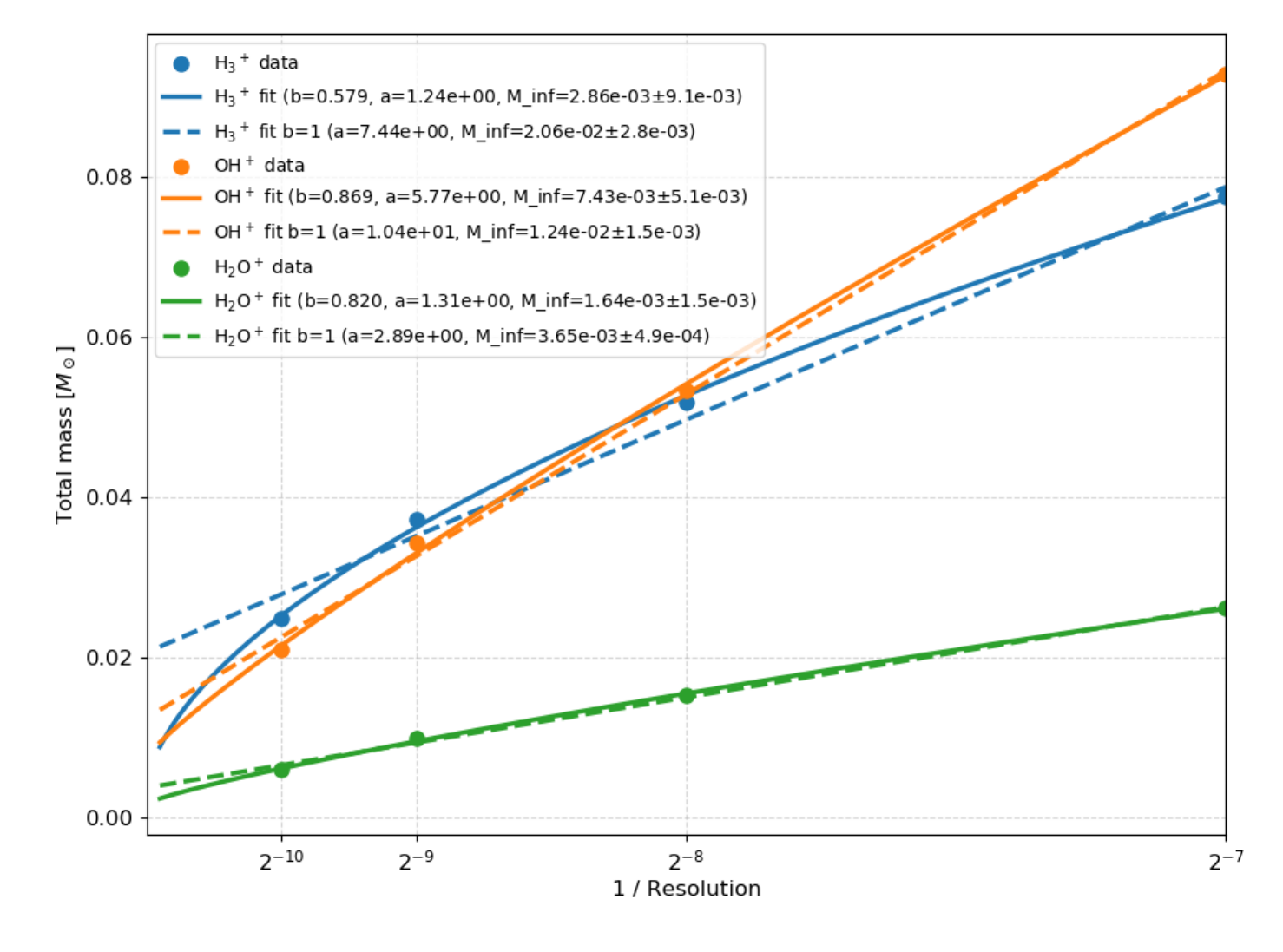}
		\caption{Similar to Figure \ref{fig:GD32-1024}, however the x-axis depicting 1/Resolution.}
		\label{fig:GD128-1024}
	\end{center}
\end{figure*}

Given the inadequacy of low resolution simulations, we constrict the mesh size to $K \ge 128$ in Figure \ref{fig:GD128-1024}. Here we depict the masses as a function of 1/Resolution, hence the fitting function intersects the $y$ axis at infinite resolution ($x$=0, $M(0)=M_\infty$), making the extrapolation visually intuitive.

We also explore two realizations: one in which the power-law coefficient $b$ is unconstrained, and one in which $b=1$. The latter is motivated by a fractal-geometry argument: if the trace-molecule mass resides on a structure of fractal dimension $D_\mathrm{f}$, the unresolved mass at resolution $K$ scales as $K^{-(3-D_\mathrm{f})}$, so $b = 3 - D_\mathrm{f}$. Figure \ref{fig:spatial_distribution} suggests that trace molecules may predominantly occupy filamentary or sheet-like structures, for which $D_\mathrm{f} \simeq 1$--$2$ and therefore $b \simeq 1$--$2$. We adopt the lower end, $b=1$ (corresponding to sheet-like structures, $D_\mathrm{f}\simeq 2$), which gives the more conservative estimate of $M_\infty$.
  
In the unconstrained case, $M_\infty$ converges at lower values of about  0.0029, 0.0074 and 0.0016 $M_\odot$ for $\mathrm{H_3^+}$, $\mathrm{OH^+}$ and $\mathrm{H_2O^+}$ respectively. In the case where $b=1$, the respective converged values are much higher: 0.021, 0.012 and 0.0037. In both cases these values are markedly lower than the corresponding values at $K=1024$, suggesting that even higher resolution is required for resolving these molecules. Taking the realization with $b=1$, $M_\infty/M_{1024}$ gives 0.87, 0.58 and 0.63 for $\mathrm{H_3^+}$, $\mathrm{OH^+}$ and $\mathrm{H_2O^+}$ respectively. Hence, for all three trace molecules, the mass at $K=1024$ converges to within a factor of $\sim 2$ of the asymptotic value, indicating that our highest-resolution simulation is reasonably close to the resolution required for full convergence.

Using unconstrained $b$, however, leads to lower ratios of 0.12, 0.35 and 0.28. $\mathrm{OH^+}$ and $\mathrm{H_2O^+}$ are still well converged within a reasonable factor of approximately 3, while the $\mathrm{H_3^+}$ mass at $K=1024$ is nearly an order of magnitude away from its infinite resolution counterpart. It may be that this difference arises in where $\mathrm{H_3^+}$ actually resides. Unlike $\mathrm{OH^+}$ and $\mathrm{H_2O^+}$ which live in shell-like transition layers around CNM, most $\mathrm{H_3^+}$ lives in diffuse CNM (with relatively little contribution from UNM or dense CNM). Various studies indicate that the fractal dimension of related molecular-clouds characteristically fulfills $D_\mathrm{f}>2.3$ \citep{SanchezEtAl-2005, FederrathEtAl-2009, RomanDuvalEtAl-2010}, hence, the low unconstrained $b$ value of $\mathrm{H_3^+}$ ($\sim 0.6$) may be grounded in reality. A CNM cloud may be physically large or small, yet even in the former case it may still contain unresolved internal thermal or density structures, requiring higher resolution for full convergence. Further studies will be required in order to quantitatively determine these sensitivities in the future. 

While we find the total trace-molecule mass to be close to convergence, in nearly all cases with the possible exception of only $\mathrm{H_3^+}$ at unconstrained $b$, this reveals no information about how the column ratio \emph{distributions} respond to increasing resolution. Does the shape, width or median of the distribution respond to resolution change, and what is the implication for the expected distribution at full convergence? In order to find out, we apply a similar analysis as in Figures \ref{fig:OHp_column_densities}-\ref{fig:H3p_column_densities}, Figure \ref{fig:1D-ModelVersusObservationAbundances} and Table \ref{tab:ratio_summary_statistics}, however at lower resolutions,  with $K=128,256,512$. Extremely low resolutions with $K<128$ are neglected, since as already discussed in Figure \ref{fig:2D-Temperature-Resolution-Convergence} and accompanying text, these simulations do not even resolve the CNM phase properly, and are thus highly inadequate for this particular investigation.

Our findings are summarized in Table \ref{tab:ratio_summary_resolution_dependent_statistics}, where instead of displaying many separate plots in such a sizable data set, we show a more compendious view, that includes only the median and the 25--75 percentile of the right-peak model theoretical distributions. As explained in the accompanying text to Table \ref{tab:ratio_summary_statistics}, the right-peak model is considered to be a more appropriate theoretical portrayer of the observations, given the observational bias towards large ratios (see Appendix \ref{Appendix:D} for details).

In this concise form, it is easy to see the effect of resolution. While the median decreases, as previously discussed (better resolution $\rightarrow$ sharper transitional gas interfaces $\rightarrow$ less trace molecules formed $\rightarrow$ trace/hydrogen column ratio diminishes), the distribution width and shape remains similar. One of the main conclusions of this work, namely that much of the variance in the observations can be explained just by multiphase turbulence ISM, even with a constant CRIR, remains valid regardless of resolution.

The primary implication for the distribution at full convergence therefore lies mainly in how the median changes. While we expect the width and shape of the distribution to remain the same -- the median will shift further left towards a lower ratio than in the $K=1024$ simulation. Given that the trace molecules' mass is converged by up to a factor of $\sim$2, in logarithmic scale the \emph{maximal} x-axis shift in Figures \ref{fig:OHp_column_densities}-\ref{fig:H3p_column_densities} shall be equivalent to a value of \emph{up to} 0.3 dex ($10^{0.3} \approx 2$). This may increase the gap between the model and observation median values, currently resulting from our highest resolution $K=1024$ model. However, as further discussed in Appendix \ref{Appendix:D}, the observations may also shift in the same sense, when observational limits are better constrained in the future. On the model side, it remains the task of future theoretical studies to better quantify the resolution-related aspects.

\begin{table*}
\centering
\caption{Summary statistics for the logarithmic column density ratios. The table lists the median and interquartile range (25--75 percentile) width of the \textbf{right-peak model} (see Figure \ref{fig:1D-ModelVersusObservationAbundances}), for different resolutions.}
\label{tab:ratio_summary_resolution_dependent_statistics}
\begin{tabular}{|l|ccc|ccc|ccc|}
\hline
 \multicolumn{1}{|c|}{Resolution} & \multicolumn{3}{c|}{$\log_{10}(N(\mathrm{\mathrm{OH}^+})/N(\mathrm{H}))$} & \multicolumn{3}{c|}{$\log_{10}(N(\mathrm{H}_2\mathrm{O}^+)/N(\mathrm{H}))$} & \multicolumn{3}{c|}{\text{\phantom{i}}$\log_{10}(N(\mathrm{\mathrm{H_3}^+})/N_\mathrm{H})$} \\
\cline{2-10}
\multicolumn{1}{|c|}{($K$)} & median & 25--75 percentile & $\Delta$ & median & 25--75 percentile & $\Delta$ & median & 25--75 percentile & $\Delta$ \\
\hline
1024 & $-8.25$ & $-8.40$--$-8.11$ & $0.29$ & $-8.84$ & $-9.03$--$-8.69$ & $0.34$ & $-7.72$ & $-8.21$--$-7.48$ & $0.73$\\
512  & $-8.05$ & $-8.20$--$-7.92$ & $0.28$  & $-8.63$ & $-8.82$--$-8.47$ & $0.35$ & $-7.55$ & $-8.01$--$-7.28$ & $0.73$\\
256  & $-7.85$ & $-8.18$--$-7.75$ & $0.33$  & $-8.43$ & $-8.61$--$-8.28$ & $0.33$ & $-7.38$ & $-7.84$--$-7.13$ & $0.71$\\
128  & $-7.56$ & $-7.69$--$-7.44$ & $0.25$  & $-8.15$ & $-8.32$--$-8.01$ & $0.31$ & $-7.33$ & $-7.79$--$-7.01$ & $0.78$\\

%1024 & $-8.25$ & $-8.40$--$-8.11$ & $0.29$ & $-8.84$ & $-9.03$--$-8.69$ & $0.34$ & $-7.70$ & $-8.20$--$-7.44$ & $0.76$\\
%512  & $-8.05$ & $-8.20$--$-7.92$ & $0.28$  & $-8.63$ & $-8.82$--$-8.47$ & $0.35$ & $-7.53$ & $-8.00$--$-7.25$ & $0.75$\\
%256  & $-7.85$ & $-8.18$--$-7.75$ & $0.33$  & $-8.43$ & $-8.61$--$-8.28$ & $0.33$ & $-7.35$ & $-7.84$--$-7.11$ & $0.73$\\
%128  & $-7.56$ & $-7.69$--$-7.44$ & $0.25$  & $-8.15$ & $-8.32$--$-8.01$ & $0.31$ & $-7.32$ & $-7.79$--$-6.98$ & $0.81$\\
\hline
\end{tabular}
\end{table*}

\section{Appendix D: Observational detection limits}
\label{Appendix:D}

The comparison of the column-density ratios with the model distributions is shaped by two distinct observational limitations, one affecting the molecular-ion column and one the hydrogen column, which we treat in turn. The first is the sensitivity of the molecular-ion detections.

The three tracers discussed in this paper are observed with different techniques and therefore carry different sensitivity limits. The oxygen-bearing ions $\mathrm{OH^+}$ and $\mathrm{H_2O^+}$ are detected in submillimeter absorption against bright continuum sources with \textit{Herschel}/HIFI \citep{IndrioloEtAl-2015}, where a component is registered as a detection at the $3\sigma$ level and the sensitivity of each sight line is set by the continuum signal-to-noise ratio. If absorption is shallower than the noise, an upper limit is essentially set by the noise. \cite{IndrioloEtAl-2015} tabulate the double-sideband continuum antenna temperature and the corresponding root-mean-square noise for every source and transition, from which the limiting optical depth follows as $\tau_{\rm lim}\simeq 3\,\sigma_{\rm rms}/T_{\rm cont}$ and the limiting column density from the transition strengths. Because $\mathrm{OH^+}$ is the strongest of these transitions it is detected in our sample in all velocity intervals, reaching values as low as $\sim 2\times 10^{12}\,\mathrm{cm^{-2}}$, whereas $\mathrm{H_2O^+}$ yields non-detections in several intervals and sets the more stringent floor, with the faintest reported columns of 3 $\times 10^{11}\,\mathrm{cm^{-2}}$. 

H$_3^+$, by contrast, is observed in near-infrared absorption toward comparatively nearby background stars. In the sample adopted here non-detections yield upper limits of the order of at least $10^{13}\,\mathrm{cm^{-2}}$, with a median value of around $3 \times 10^{13}\,\mathrm{cm^{-2}}$ \citep{ObolentsevaEtAl-2024,IndrioloEtAl-2025}, the exact value being governed by the continuum signal-to-noise ratio. Crucially, \citet{IndrioloEtAl-2015} note that H$_3^+$ absorption lines are at most a few percent deep, so that at low ionization rates the molecule is not produced in detectable abundances. $\mathrm{OH^+}$ and $\mathrm{H_2O^+}$ therefore remain observable in a regime in which H$_3^+$ is not.

Because the comparison in Figures \ref{fig:OHp_column_densities}-\ref{fig:H3p_column_densities} is expressed as a ratio to the hydrogen column, a hydrogen column detection maps to a horizontal limit on the ratio. Taking the above-mentioned characteristic trace limits for $\mathrm{OH^+}$, $\mathrm{H_2O^+}$ and $\mathrm{H_3^+}$, and adopting a representative hydrogen columns for the sample of $\sim 3 \times 10^{21}\,\mathrm{cm^{-2}}$ (for both $N$(H) and $N_\mathrm{H}$), the limit ratio floors yield approximately $\log(N(\mathrm{OH^+})/N(\mathrm{H}))\approx-9$, $\log(N(\mathrm{H_2O^+})/N(\mathrm{H}))\approx-10$ and $\log(N(\mathrm{H_3^+})/N(\mathrm{H}))\approx-8$, which are well reflected in Figures \ref{fig:OHp_column_densities}-\ref{fig:H3p_column_densities}.

The second, and even more dominant limitation in our sample, concerns the denominator rather than the molecular ion. For the $\mathrm{OH^+}$ and $\mathrm{H_2O^+}$ sight lines the atomic column $N(\mathrm{H})$ is obtained from H$\,$\textsc{i} 21 cm absorption toward the same continuum sources \citep{IndrioloEtAl-2015}, and because the 21 cm optical depth scales as $\tau\propto N(\mathrm{H\,\textsc{i}})/T_{\rm s}$, the line saturates toward the brightest sight lines, yielding only a lower limit on $N(\mathrm{H})$. This limitation applies to a larger fraction of the $\mathrm{OH^+}$ and $\mathrm{H_2O^+}$ sample than the molecular-ion non-detections. Unlike a molecular-ion non-detection, which is a purely horizontal upper limit at a measured $N(\mathrm{H})$, a saturated sight line is constrained in both coordinates of Figures \ref{fig:OHp_column_densities} and \ref{fig:H2Op_column_densities}. Such a point is at once a lower limit in $N(\mathrm{H})$ and an upper limit in the ratio $N(X)/N(\mathrm{H})$, and must be displaced upward and to the left of its plotted position.

A related and more systematic effect has already been discussed briefly in Section \ref{SS:TraceResponse}. It arises from the phase sensitivity of 21 cm absorption itself. Because the optical depth scales inversely with spin temperature, the absorption is weighted toward the cold neutral medium and is comparatively insensitive to warm or thermally unstable H\,\textsc{i}. The tabulated $N(\mathrm{H})$ values therefore trace mainly the CNM along each velocity interval and may omit a warm (WNM/UNM) contribution, which would further lower the true atomic column relative to that adopted here. Saturation and this cold-weighting effect act in union, in the same direction, both underestimating $N(\mathrm{H})$ and hence overestimating the observed ratios. Unlike saturation, cold-weighting however operates systematically on the whole sample.

The two aforementioned limitations bias the observed ratios reported in Table \ref{tab:ratio_summary_statistics} and Figure \ref{fig:1D-ModelVersusObservationAbundances} in the same sense. The molecular-ion detection floor removes low-ratio sight lines from the detected sample, while the combined effect of saturated $N(\mathrm{H})$ values and the inherent CNM-sensitivity of 21 cm absorption work to under-estimate the denominator and therefore over-estimate the ratio. Both effects raise the observed distribution relative to the underlying one and imply that the 'ground truth' distribution lies at systematically lower ratios, in the direction of the model rather than away from it. The H$_3^+$ sample is comparatively free of the second effect, its hydrogen columns being determined from direct ultraviolet measurements rather than saturated 21 cm absorption and its sight lines dominated by clean detections, so that its comparison with the model is actually the most direct of the three.

%\hlabel{lastpage}

\end{document}